\documentclass{jfm}

\usepackage{graphicx}
\usepackage{subcaption}   
\usepackage{newtxtext}
\usepackage{newtxmath}
\usepackage{natbib}
\usepackage{tabularx}  
\newcolumntype{C}{>{\centering\arraybackslash}X}  
\usepackage{circuitikz}
\usepackage{multirow}
\usepackage{hyperref}
\hypersetup{
    colorlinks = true,
    urlcolor   = blue,
    citecolor  = black,
}

\newcommand{\RomanNumeralCaps}[1]
\linenumbers
\usepackage{float}
\usepackage{xcolor} 
\usepackage{tikz}   

\definecolor{viridis0}{rgb}{0.267004, 0.004874, 0.329415}
\definecolor{viridis1}{rgb}{0.275191, 0.194905, 0.496005}
\definecolor{viridis2}{rgb}{0.212395, 0.359683, 0.55171}
\definecolor{viridis3}{rgb}{0.153364, 0.497, 0.557724}
\definecolor{viridis4}{rgb}{0.122312, 0.633153, 0.530398}
\definecolor{viridis5}{rgb}{0.288921, 0.758394, 0.428426}
\definecolor{viridis6}{rgb}{0.626579, 0.854645, 0.223353}
\definecolor{viridis7}{rgb}{0.8442608, 0.77023345, 0.1223456}

\newcommand{\markercircle}[1]{\tikz\fill[#1] (0,0) circle(2.5pt);}  
\newcommand{\markercross}[1]{\tikz\fill[#1] (0,0) -- (0.15/1.5,-0.3/1.5) -- (0.3/1.5,0) -- cycle;}  
\newcommand{\markerdiamond}[1]{\tikz\fill[#1] (0,0) -- (0.15/1.2,0.15/1.2) -- (0,0.3/1.2) -- (-0.15/1.2,0.15/1.2) -- cycle;}

\newcommand{\mycircuit}{
\begin{circuitikz}
\tikzstyle{every node}=[font=\LARGE]
\draw [short,ultra thick] (7.5/2,12.75/2) -- (7.75/2,12.5/2);
\draw [short,ultra thick] (7.75/2,12.5/2) -- (8/2,12.75/2);
\draw [short,ultra thick] (7.75/2,12.5/2) -- (7.75/2,12.25/2);
\end{circuitikz}
}

\newcommand{\mycircuitB}{
\begin{circuitikz}
\tikzstyle{every node}=[font=\LARGE]
\draw [short,ultra thick] (7.5/2,12.75/2) -- (7.75/2,13/2);
\draw [short,ultra thick] (8/2,12.75/2) -- (7.75/2,13/2);
\draw [short,ultra thick] (7.75/2,13/2) -- (7.75/2,13.25/2);
\end{circuitikz}
}

\title{Impact of freestream turbulence and thrust coefficient on wind turbine-generated wakes}

\author{Martin Bourhis\aff{1}
  \corresp{\email{m.bourhis@imperial.ac.uk}},
  Thomas Messmer \aff{2,3},
  Michael H\"olling  \aff{2,3}
 \and Oliver R.H. Buxton \aff{1}
 }

\affiliation{
\aff{1} Department of Aeronautics, Imperial College London, London SW7 2AZ, UK
\aff{2} Carl von Ossietzky Universit\"at Oldenburg, School of Mathematics and Science,  Institute of Physics, 26129 Oldenburg, Germany 
\aff{3} ForWind - Center for Wind Energy Research, K\"upkersweg 70, 26129 Oldenburg, Germany
}

\begin{document}

\maketitle

\begin{abstract}
This study investigates how variations in freestream turbulence (FST) and the thrust coefficient ($C_T$) influence wind turbine wakes. Wakes generated at $C_T \in \{0.5, 0.7,0.9\}$ are exposed to turbulent inflows with varying FST intensity ($1\% \lesssim TI_{\infty} \lesssim 11\%$) and integral length scale ($0.1 \lesssim {\cal L}_x/D \lesssim 2$, $D$ is the rotor diameter). For high-$TI_{\infty}$ inflows, a flow region within the wake is observed several diameters downstream, where a mean momentum deficit persists despite the turbulence intensity having already homogenised with the freestream, challenging traditional wake definitions. A ``turning point'' in the mean wake width evolution is identified, beyond which wakes spread at slower rates. Near-field ($x/D \lesssim 7$) wake growth rate increases with higher $TI_{\infty}$ and $C_T$, while far-field ($x/D \gtrsim 15$) wake growth rate decreases with higher $TI_{\infty}$—a finding with profound implications for wind turbine wake modelling that also bridges the gap with entrainment behaviours observed in bluff and porous body wakes exposed to FST. Increasing ${\cal L}_x$ delays wake recovery onset and reduces the mean wake width, with minimal effect on the spreading rate. Both $C_T$ and FST influence high- and low-frequency wake dynamics, with varying contributions in the near and far fields. For low-$TI_{\infty}$ and small-${\cal L}_x$ inflows, wake meandering is minimal, sensitive to $C_T$, and appears to be triggered by shear layer instabilities. Wake meandering is enhanced for high-$TI_{\infty}$ and large-${\cal L}_x$ inflows and is dominated by background turbulence. This emphasises the complex role of FST integral length scale: while increasing ${\cal L}_x$ amplifies meandering, it does not necessarily translate to larger mean wake width due to the concurrent suppression of entrainment rate.

\end{abstract}


\section{Introduction \label{Introduction} }

To meet the growing demand for electricity while minimising greenhouse gas emissions, the share of wind energy in the global energy mix is increasing significantly. This growth is accompanied by a notable rise in both the size and number of wind turbines. To optimise power generation within a given area, wind turbines are often arranged into large clusters, referred to as wind farms, particularly in offshore regions, where spatial constraints are minimised and winds are stronger, more consistent, and less turbulent than onshore, due to reduced surface roughness (\emph{e.g.} forests, buildings, mountains). Like any bluff body in a flow, a wind turbine generates a wake, which is conventionally characterised as a flow region of reduced velocity and increased turbulence compared to the freestream \citep{Pope2000,Vermeer2003}. This has two significant implications for wind farm operations, as, with the exception of the first row, the wakes of upstream turbines become the incoming flow for downstream turbines: (1) a reduction in the available power for downstream turbines, thereby reducing the total power output of $N$ turbines in a wind farm compared to $N$ independent turbines \citep{Barthelmie2007,Hansen2012}; and (2) increased turbulence in the incoming wind for downstream turbines, which compromises their structural integrity and leads to earlier and more frequent maintenance interventions \citep{Thomsen1999,Santelo2021}. Hence, a comprehensive understanding of wind turbine wake evolution, although not yet fully achieved, is a critical prerequisite for optimising both wind farm design and day-to-day operation. Accurately predicting key wake characteristics will enable the determination of the optimal number of turbines, their positioning, and real-time operations, to maximise the wind farm total power output whilst minimising the maintenance interventions, all while considering specific wind conditions and spatial constraints.

Historically, early studies on wakes were performed using basic geometries, such as cylinders, discs, or spheres, and were conducted mainly in non-turbulent flow conditions. These studies led to the development of the theoretical framework describing the evolution of axisymmetric wakes in laminar freestream conditions by \cite{Townsend1976} and \cite{George1989}. While the Townsend-George theory provides valuable insights for improving wind turbine wake modelling \citep{Neunaber2022}, wind turbine wakes are significantly more complex than canonical turbulent flows. Firstly, wind turbines are composed of several bodies of different sizes, including the blades, hub, nacelle and tower. As a result, their wakes exhibit multiple coherent structures—such as tip vortices, root vortices, vortex shedding from the tower and nacelle, and wake meandering—each with different characteristic time and length scales. A general and comprehensive overview of wind turbine wake aerodynamics can be found in \cite{Vermeer2003,PorteAgel2019}. As the wake develops, both the absolute and relative strengths of these structures vary, with some dominating in certain regions of the wake, while others dissipate and become negligible \citep{Biswas2024}. All dynamics, statistics, and ultimately the structure of the immediate wake directly behind a wind turbine are strongly influenced by its geometry \citep{Dong2023} and operating conditions, such as its tip-speed ratio ($\Lambda$) and thrust coefficient ($C_T$) \citep{Neunaber_2022,Biswas2024,Vahidi2025}. This first wake region is often defined in the literature as the near wake, starting immediately downstream of the rotor and extending coarsely up to a streamwise distance of $2D$ to $4D$, with $D$ being the rotor diameter \citep{Vermeer2003,Medici2006,Sorensen2015}. The far wake is located further downstream, where the influence of turbine-specific features diminishes, and the flow exhibits more universal characteristics compared to the near wake. In this region, the mean velocity profile is expected to approach a Gaussian distribution, turbine-specific flow structures, such as tip-vortices, have largely decayed, and wake meandering becomes the dominant dynamic feature \citep{PorteAgel2019,Dong2023}. At this stage, it is commonly accepted that a global wind turbine parameter, such as the thrust coefficient, is sufficient to characterise the wake with reasonable accuracy, thereby justifying the use of porous discs as wind turbine surrogates in experiments \citep{Neunaber2021,Vinnes2022,Vinnes2023}, as well as the concept of actuator discs in numerical simulations \citep{Blackmore2014_a,Hodgson2022,Hodgson2023,Vahidi2024,Vahidi2025}. It is important to emphasise that numerous definitions of the near and far wake exist in the literature; hence, these terms and the values provided above should be used with caution, as they sometimes refer to entirely different wake regions depending on the underlying physical definition. What is widely agreed upon, however, is that the length of the near wake is strongly influenced by the turbine's thrust coefficient and the turbulence characteristics of the incoming flow.

Secondly, the ambient flows in which wind turbines operate are far from being non-turbulent. Both the turbines and their wakes are exposed to complex and highly diverse turbulent environments, resulting from multiple interactions between the atmospheric boundary layer (ABL), the merging and mixing of turbulent wakes from upstream turbines, and the developing internal wind farm boundary layer \citep{Stevens2017}. Wind turbines and wind farms operate within the lowest portion of the ABL, which contains a wide range of turbulence length scales, from the Kolmogorov scale to scales on the order of the ABL's depth (several kilometres and multiple turbine diameters), as well as significantly varying turbulence intensity levels $TI_{\infty}$ \citep{PorteAgel2019}. Freestream turbulence intensities between 2\% and 15\% have been reported for offshore and onshore sites \citep{Pena2016,Shu2016,Argyle2018}, with even much higher levels at low wind speeds ($<5$~m.s$^{-1}$) \citep{Ren2018}. Moreover, wind turbine wakes are characterised by elevated turbulence intensity levels that can persist over considerable distances, particularly when the freestream turbulence intensity is low. Ultimately, as rotor size increases and optimal wind farm locations, such as the North Sea, become crowded, turbines are placed closer together, leading to higher wake-induced turbulence intensity in inflows for downstream units, as well as potential wind farms interactions \citep{Stevens2017}. As a result, wind turbine-generated wakes develop in highly diverse turbulent ambient flows, with different levels of turbulence intensity ($TI_{\infty}$) and integral length scale ($\cal L$). For ease of reading, freestream turbulence will be abbreviated as FST, and integral length scale as ILS throughout the text.

Typically, higher FST intensity is recognised to hasten the wake recovery. This is manifested by an accelerated transition of the velocity profiles toward self-similarity, a reduction in the maximum velocity deficit within the wake, and a larger wake width and growth rate \citep{Medici2006,Chamorro2009,Wu2012,Gambuzza2022}. Specifically, high $TI_{\infty}$ inflows enhance the early mixing of the highly three-dimensional, heterogeneous, and turbulent low-momentum flow in the near wake with the ambient flow, partly by accelerating the breakdown of the tip vortices, which restrict mass, momentum and energy exchanges between the wake and the background flow within the first few diameters downstream of the turbine \citep{Chamorro2009,Zhou2016}. Higher tip-speed ratios leads also to a faster breakdown of tip vortices \citep{Sorensen2015,Bayron2024,Biswas2024}. 

While a considerable body of literature exists on the influence of $TI_{\infty}$ on wind turbine wakes, the impact of FST length scales has only recently received attention, resulting in a limited number of studies. Furthermore, most wind tunnel experiments have focused on characterising the influence of $TI_{\infty}$ (often with unreported ${\cal L}$), as generating turbulent inflows with length scales representative of real-life operational conditions is experimentally challenging. As a result, the majority of the understanding regarding the impact of ILS on wind turbine wakes is derived from numerical approaches, especially Large Eddy Simulations (LES). \cite{Vahidi2024} conducted LES of wind turbine wakes under neutral atmospheric conditions, varying inflow turbulence scales by changing the boundary layer height while keeping $TI_{\infty}$ constant. They found that inflows with large turbulence length scales accelerate wake recovery and increase wake meandering. Similarly, \cite{Du2021} reported enhanced wake recovery under neutral atmospheric conditions compared to stable atmospheric conditions, attributing this to the presence of larger turbulent scales in the neutral boundary layer flow. In their LES of actuator disc wakes simulating a marine current turbine, \cite{Blackmore2014_a} found that increasing ${\cal L}$ led to a higher velocity deficit near the turbine, an enhanced wake recovery, and a wider wake. \cite{Hodgson2022,Hodgson2023} examined the impact of turbulent time scales on wind turbine wake recovery using sinusoidal inflow variations and synthetic turbulence fields, which were stressed and compressed to change the turbulent scale while maintaining the same flow structure and energy content. In both studies, they found that inflows with shorter time scales led to a faster breakdown of tip vortices, a shorter near-wake region, enhanced mixing, and quicker wake recovery, contrasting with the aforementioned studies. Conversely, longer time scales were associated with increased wake meandering. 

Recent advancements in the operation of active turbulence-generating grids have opened new opportunities for experimentally generating turbulence length scales representative of real-world wind turbine applications \citep{Hearst2017,Schottler2017,Neuhaus_2020_PRL}. In a recent study, \cite{Gambuzza2022} examined the development of wind turbine wakes subjected to turbulent inflows with different spectral contents and integral time scales, generated using an active grid. Inflows with large turbulent time scales (up to 10 times the convective time scale) were classified as non-Kolmogorov-like flows due to the presence of a gap in the turbulence spectra, which results from the artificial forcing of low frequencies through grid motion. In contrast, inflows with low integral time scales, which exhibited the canonical Kolmogorov energy distribution, were classified as Kolmogorov-like flows. Their study revealed that, for the same incoming turbulence intensity, wind turbine wakes exposed to non-Kolmogorov-like inflows with large temporal scales recover more slowly. In contrast, when exposed to Kolmogorov-like flows with smaller temporal scales, wake recovery begins closer to the turbine. However, it is not possible to ascertain whether these effects arise from the difference in turbulence time scales or the discontinuous distribution of the freestream turbulent energy.

Regarding the incorporation of FST effects into analytical wind turbine wake models, the wake growth rate -- defined as the slope of the linear wake expansion -- is typically assumed to be either constant \citep{Jensen1983}, a linear function of $TI_{\infty}$  \citep{Bastankhah2014,Niayifar2016,carbajo2018} or a power law function of both $TI_{\infty}$ and $C_T$ \citep{Ishihara2018}. These empirical correlations, often derived from LES data, are mostly valid for a single $C_T$ --  the one at which the LES have been performed --  or a narrow range, as well as for a limited range of FST conditions. Additionally, the influence of ${\cal L}$ is typically not accounted for in most wind turbine wake models. 

Importantly, recent studies examining the effect of FST on wakes of solid and porous bodies over large distances—porous discs being widely used as wind turbine surrogates in wind tunnel experiments due to their practicality, particularly at a wind farm scale —have provided a more comprehensive understanding of FST's role in wake development, revealing a more intricate interaction between the wake and the surrounding flow than what is typically reported for wind turbine wakes. In the near wake of a cylinder ($x/D \lesssim 10$), \cite{Kankanwadi2023,Buxton_Chen_2023,Chen2023,Chen_Buxton_2024} reported that both FST intensity and integral length scale act to enhance wake spreading and entrainment rates of mass, momentum and kinetic energy into the wake, compared to a non-turbulent background. However, an opposite trend was observed in the far wake ($x/D \gtrsim 30$) in \cite{Kankanwadi2020,Chen2023,Chen_Buxton_2024}: a reduction in the entrainment velocity and fluxes with increased FST intensity, with the ILS having a negligible influence. Similarly, \cite{Vinnes2023,Bourhis2024} reported a reduction in both wake growth and entrainment rates in the wake of porous discs for $x/D\gtrsim10$ with increased levels of FST intensity, while also highlighting the pivotal role played by the disc porosity and thrust coefficient. \cite{Kankanwadi2020} postulated that the reduction in the entrainment rate was caused by the suppression of small-scale nibbling-driven entrainment mechanisms in the presence of background turbulence intensity. In addition, \cite{Chen2023,Chen_Buxton_2024} identified a ``turning point'' in the wakes around $x/D\approx 15$, beyond which wakes spread at a significant slower rate, regardless of whether the background was turbulent or not. While FST intensity had little effect on entrainment velocity upstream of this point, it was found that beyond this transition region, the entrainment velocity significantly decreased as FST intensity increased.

Generally, these studies have emphasised that both FST intensity and ILS significantly influence the wakes of bluff and porous bodies, with their respective contributions and effects being functions of the streamwise distance from the object. Studies involving porous discs, for which $C_T$ can match values relevant to wind turbine applications, have further shown that these effects were contingent upon $C_T$. These findings raise important questions regarding the current body of knowledge on wind turbine wakes, as well as the validity of existing wake models. At first glance, there seems to be no clear reason why there wouldn't be a wake region, sufficiently far from the turbine, where the FST effects on wind turbine wakes are quantitatively similar to those observed in the far wakes of bluff and porous bodies. A primary limitation of the existing literature on wind turbine wakes lies in the limited wake region investigated, typically shorter than 10 turbine diameters, which aligns with a typical turbine-turbine spacing in wind farms, whereas wakes of bluff and porous bodies are examined over greater distances. Hence, it is plausible that wind turbine wakes have not been investigated over a sufficiently extended distance to observe FST effects analogous to those seen in the far wake of bluff and porous bodies. Therefore, exploring wind turbine wakes exposed to FST over greater distances may uncover a more qualitatively universal entrainment behaviour in the wakes of these different objects. While drawing connections in the entrainment behaviour in the presence of FST for various objects is valuable from a fundamental perspective, investigating wind turbine wakes over greater distances is also important in the context of farm-to-farm interaction, where adjacent wind farms may be separated by $O(10)\,\mathrm{km}$.

In summary, while fundamental to the optimisation of wind farms and, more broadly, to the enhanced deployment of wind energy technologies, the combined effects of $TI_{\infty}$, $\cal L$, and the turbine thrust coefficient $C_T$, on wakes remain insufficiently understood and modelled. The exploration of the \{$C_T$, $TI_{\infty}$, $\cal L$\} parameter space for a single turbine is relatively limited in the literature, and comparing results across different studies is challenging due to the absence of standardised lab-scale wind turbine models and consistent wind tunnel experimental conditions. The variations in turbine geometry, operating conditions (\emph{e.g.}, $C_T$, $\Lambda$) and FST conditions (\emph{e.g.}, Reynolds number, \{$TI_{\infty}$, $\cal L$\} combinations, turbulent energy spectral distribution) contribute to the observed scatter and, sometimes, divergent conclusions across different studies. To address this gap, we conducted an extensive parametric study of wind turbine wakes in high Reynolds number flows ($\text{Re}>10^5$), systematically examining the influence of the turbine thrust coefficient and FST conditions on the wake characteristics over streamwise distances of up to 20 turbine diameters. Specifically, the influence of 8 tailored FST ``flavours'' \{$TI_{\infty}$, $\cal L$\} on the wake behind a single turbine operating at 3 different $C_T$ is assessed, resulting in the evaluation of 24 different wind turbine wakes in the \{$C_T$, $TI_{\infty}$, $\mathcal{L}$\} parameter space. The remainder of the paper is structured as follows.  Section~\ref{Experimental method} introduces the details of the experimental setup, including the wind turbine characteristics, and the various \{$C_T$, $TI_{\infty}$, $\cal L$\} combinations examined. The major results are presented in \S~\ref{Results} where the influence of  \{$C_T$, $TI_{\infty}$, $\mathcal{L}$\} on time-averaged velocity and turbulence statistics (\S~\ref{subsec:Velodef}), wake width and growth rate (\S~\ref{subsec:WakeWidth}), wake-averaged quantities (\S~\ref{subsec:WakeAveraged}), near wake dynamics (\S~\ref{subsec:near wake dynamics}), and the statistics and dynamics of wake meandering (\S~\ref{subsec:wake meandering}) are thoroughly examined. The work is concluded in \S~\ref{Conclusion}.

\section{Experimental method \label{Experimental method}}

\subsection{Facility and experimental setup\label{Facility}}

\begin{figure}
    \centering
    \includegraphics[width=\textwidth]{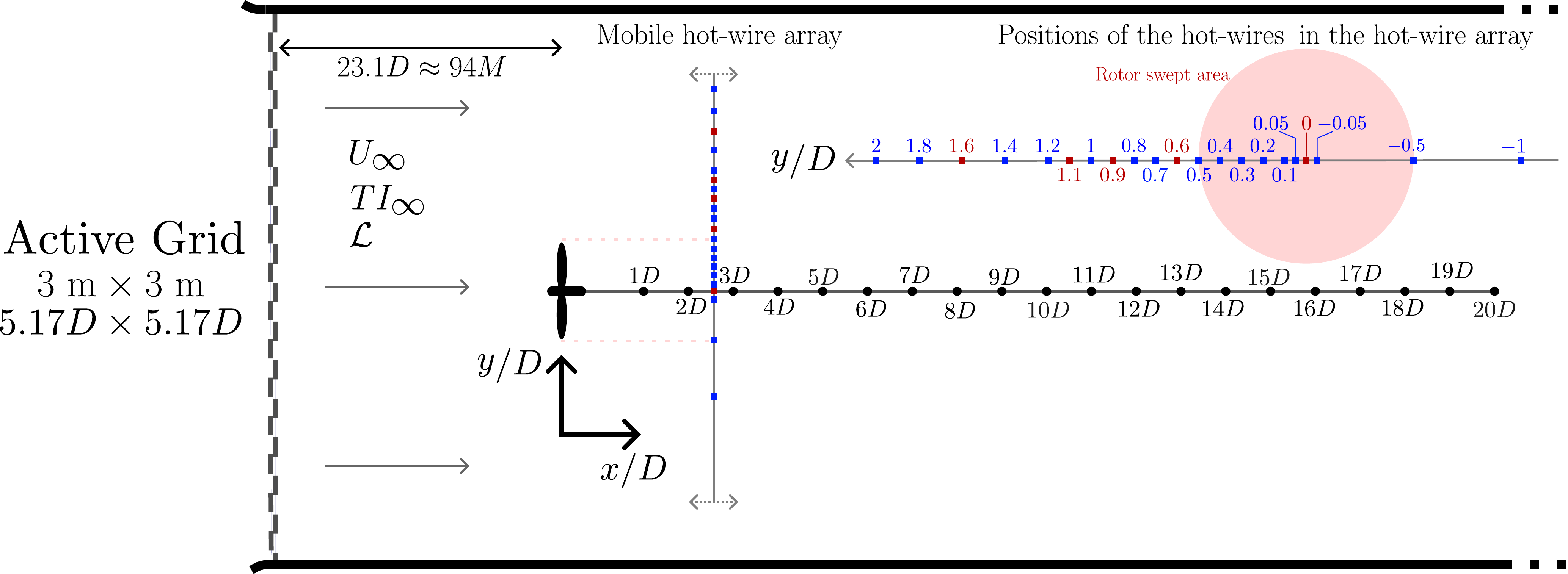}
    \caption{Schematic of the experimental setup, highlighting the positions of hot-wires along the horizontal line at hub height. Blue squares indicate hot-wires sampled at 6~kHz, while red squares represent hot-wires sampled at 20~kHz.}
    \label{Fig:Wind Tunnel}
\end{figure}

The experiments were performed in a large closed loop wind tunnel at the University of Oldenburg. The wind tunnel has a test section length of 30~m and a square cross-section of size $3~\text{m} \times 3~\text{m}$. The outlet of the contraction nozzle is equipped with an active turbulence-generating grid, allowing the generation of customised turbulent inflow conditions \citep{Kroger2018,Neuhaus_2020_PRL}. The grid has 80 individually controllable shafts with rhombic aluminium wings mounted on them, resulting in a mesh width of $M = 0.143~\text{m}$. The grid blockage of the flow is altered by changing the angle of the wings $\alpha(t)$, \emph{i.e.} the shaft angles, either dynamically (dynamic grid protocols) or statically (static grid protocols), with the latter resulting in a passive turbulence-generating grid. In dynamic protocols, the local grid blockage varies over time based on predefined shaft angle time series, while the global blockage, computed over the entire grid, remains constant at each instant. Further details on the transfer functions linking grid protocols to the resulting flows can be found in \cite{Kroger2018,Neuhaus2021}. The specific grid excitation protocols used in this study, along with the corresponding dynamics of the generated turbulence, will be presented in \S~\ref{FST flavours}

The wind turbine used in these experiments is the Model Wind Turbine Oldenburg 0.6 (MoWiTO 0.6), which has been employed in many experimental campaigns including wake measurements at different tip-speed ratios \citep{Neunaber_2022}, in yaw conditions \citep{Schottler_2018}, in floating-like conditions \citep{Messmer2024}, and tandem configuration \citep{Neunaber_2020}. The turbine's geometrical characteristics are detailed in \cite{Schottler_2016,Juchter_2022,Messmer2024}, and only the most relevant aspects are highlighted here. The turbine has three blades, a diameter of $D=0.58$~m, and a tower height of 0.45~m. The turbine is positioned at $23.1D \approx 94M$ from the active grid, and mounted on a platform such that the hub centre is $0.96$~m above the wind tunnel floor. The origin of the coordinate system used here is at the intersection between the rotation axis and the plane formed by the leading edges of the blades. The thrust acting on the turbine is measured using strain gauges attached on both the front and rear sides of the lower part of the tower, and set in a full bridge configuration. A real time closed-loop load control of the wind turbine enables maintenance of constant tip-speed ratio ($\Lambda = D \omega/2U_{\infty}$) and thrust coefficient ($C_T = T/0.5\rho (D/2)^2 U_{\infty}^2$) during the acquisition (further details on the control system can be found in \cite{Schottler_2016,Petrovic_2018,Juchter_2022}). 

The turbine was operated at three distinct thrust coefficients -- $C_T \approx 0.5$, $C_T \approx 0.7$, and $C_T \approx 0.9$ -- corresponding to three different tip-speed ratios -- $\Lambda \approx 1.7$, $\Lambda \approx 2.7$, and $\Lambda \approx 3.7$. The specific values of $C_T$ and $\Lambda$ for each FST case will be detailed in the following subsection. The incoming wind speed was $U_{\infty} \approx 6.5~\text{m.s}^{-1}$ for all FST cases, resulting in a diameter-based Reynolds number of $\textrm{Re} \approx 2.55 \times 10^5$. Although this global Reynolds number is lower than those typically experienced by large-scale wind turbines ($\textrm{Re} \sim 10^6 - 10^7$), which could lead to some differences in instantaneous flow characteristics \citep{McTavish2013,Bourhis2023}, \cite{Chamorro2012} showed that above $\text{Re} \approx 9.4 \times 10^4$, the main flow statistics (mean velocity, turbulence intensity, kinematic shear stress, and velocity skewness) become Reynolds number independent.

A motorised array consisting of 21 single hot-wire probes was used to horizontally scan the wake at hub-height ($z=0$) between $x/D = 1$ and $x/D = 20$ in steps of $1D$. The positions of the hot-wires are highlighted in figure~\ref{Fig:Wind Tunnel}. Half of the wake was scanned with high spatial resolution, while the other half was scanned sparsely and primarily used to ensure the turbine's alignment and the symmetry of the wake. Three hot-wires are positioned in close proximity around $y=0$ to capture the wake centreline with the highest possible spatial resolution, while the density of the hot-wires decreases progressively from the geometric centreline to the edge of the wind tunnel. Five hot-wires were operated using a StreamLine Pro CTA system  9091N0102 with 91C10 CTA (Constant Temperature Anemometer) modules from \emph{Dantec Dynamics} (see hot-wires highlighted in red in figure~\ref{Fig:Wind Tunnel}). The sampling rate was set to $f_s = 20$~kHz. All other probes were operated with multichannel 54N80 CTA modules, and sampled at a frequency of $f_s = 6$~kHz. Both systems were synchronised, and data were acquired for 240~s at each streamwise position, which is sufficiently long to ensure the convergence of all the quantities introduced and discussed in this paper. The hot-wires were calibrated twice a day. Differential pressure was measured using a Prandtl tube, and air density was computed based on the measured temperature, atmospheric pressure, and humidity. The hot-wire voltages were then converted into velocity using a fourth-order polynomial law. For each individual measurement point, the acquisition time was compared to the times at which the two calibrations were performed to compute a weighted average calibration, which was then used to convert the voltage signal to velocity. For each velocity profile, the mean of the time-averaged velocities measured by the two farthest hot-wires ($y/D = 1.8 - 2$) was used as the normalising measurement to account for the minor variations in the freestream velocity along the length of the tunnel.

\subsection{Freestream turbulence ``flavours'' \label{FST flavours}}

Eight different FST ``flavours'' were generated using tailored static and dynamic grid protocols. The first five columns of table~\ref{TABLE:FST} show the active grid operating parameters for the different FST cases. Images and movies showing the active grid in operation are available in the Supplementary Materials. With the exception of case S1, for which the grid was fully open ($\alpha =0^\circ$), the overall blockage of the grid to the flow remained the same for all FST cases. $\alpha$ is the angle of the wings relative to the incoming wind and is directly linked to the grid's blockage. In case S4, the wings alternated between being fully open and fully closed ($\alpha =90^\circ$), mimicking a regular grid pattern. In case S2 the wings were set to a constant angle of ($\alpha = 30^\circ$) along either the $\pm y$-axis or the $\pm z$-axis, ensuring a symmetrical distribution to prevent deflecting the flow towards any particular side of the wind tunnel. Finally in protocols M5, L3, L6, L7 and L8, the instantaneous local blockage was varied dynamically, with the angle standard deviation, $\sigma_{\alpha}$, and the angular speed of the wings $\Omega$ differing across the protocols. Here, $\Omega = \overline{ \left| \dot{\alpha} \right|}$ denotes the time-averaged absolute rotational rate of the grid's flapping wings, which oscillate back and forth around the mean angle $\overline{\alpha}$ during a dynamic protocol, without completing a full rotation, unlike in classical double random mode protocols \citep{Zheng2021}. However, at each instant, the global blockage, computed across the entire grid, remained identical across all protocols (both static and dynamic), so that the time-averaged angle over a grid cycle was similar ($\overline{\alpha} = 30^\circ$, again along either the $\pm y$-axis or $\pm z$-axis). The hot-wire acquisition system was synchronised with the grid protocol to ensure that data recording at each streamwise measurement station began at the start of a repeated grid cycle, thereby maintaining consistent inflow conditions across all stations. Each grid cycle lasted 5~min, providing sufficient time for the 4~min hot-wires measurement and for the hot wire cart to move between stations.

To characterise the turbulent inflows, the test section was left empty and the hot-wires were evenly spaced across the entire width of the wind tunnel (between $-1.4~\text{m} \leq y \leq 1.4~\text{m}$). The horizontal hot-wire line was positioned at $x/D = z/D = 0$, \emph{i.e.}, within the turbine's swept plane. In this paper, the velocity is decomposed using the Reynolds decomposition as $U(x,y,t) = \overline{U}(x,y,t) + u'(x,y,t)$, with the overline $\overline{\cdot}$ denoting a temporal average and the brackets $\langle \cdot \rangle$ denoting a spatial-average along the horizontal axis $y$. The various turbulent background flows were characterised in terms of turbulence intensity $TI_{\infty} = \sqrt{\overline{u'^{2}}}/U_{\infty}$, and integral length scales ${\cal L}_x$ and ${\cal L}_y$. The streamwise ILS, ${\cal L}_x$, gives an estimation of the turbulence length scale along the flow's main direction, while the spanwise ILS, ${\cal L}_y$, describes the lateral extent of the incoming flow coherent structures, which should be smaller than the wind tunnel width.

\begin{table}
  \begin{center}
\def~{\hphantom{0}}
\begin{tabularx}{\textwidth}{p{1.5cm}CCCCCCCCC}
            FST case & Grid protocol & $\overline{\alpha} (^\circ) $ & $\sigma_{\alpha} (^\circ)$ & $\Omega (^\circ.\textrm{s}^{-1}) $ &  $TI_{\infty}$ (\%) & ${\cal L}_x /D$ & ${\cal L}_y/D$ & ${\cal L}_x/{\cal L}_y$ \\[3pt]
      \hline 
     S1 \textcolor{viridis0}{+} & Static & $0^\circ$ & $0^\circ$ & - & $0.9 \pm 0.1$ & $0.09 \pm 0.01$  &  $0.13 \pm 0.01$ & 0.7   \\ 
     
     S2 \textcolor{viridis1}{\(\times\)} & Static & $30^\circ$ & $0^\circ$ & - & $1.8 \pm 0.1$ & $0.16 \pm 0.01$  &  $0.15 \pm 0.01$ & 0.9  \\
     
     S4  \markercircle{viridis2} & Static & $0^\circ$ or $90^\circ$ & $0^\circ$ & - & $4.5 \pm 0.4$ & $0.34 \pm 0.04$  &  $0.23 \pm 0.01$ & 1.5   \\
     
     M5  \textcolor{viridis3}{\(\blacksquare\)} & Dynamic & $30^\circ$ & $10^\circ$ & 152$^\circ.\textrm{s}^{-1}$  & $4.9 \pm 0.3$ & $1.17 \pm 0.26$  &  $0.72 \pm 0.06$ & 1.6  \\
     
     L3 {\markercross{viridis4}} & Dynamic & $30^\circ$ & $2.5^\circ$ & 10$^\circ.\textrm{s}^{-1}$  & $3.7 \pm 0.3$ & $1.76 \pm 0.17$  &  $1.09 \pm 0.10$ & 1.6  \\
     
     L6  {\markerdiamond{viridis5}} & Dynamic & $30^\circ$  & $5^\circ$  & 20$^\circ.\textrm{s}^{-1}$ & $5.8 \pm 0.4$ & $1.95 \pm 0.20$  &  $1.09 \pm 0.10$ & 1.8  \\
     
     L7 \textcolor{viridis6}{\mycircuit}  & Dynamic & $30^\circ$ & $10^\circ$  & 39$^\circ.\textrm{s}^{-1}$ & $8.8 \pm 0.4$ & $1.93 \pm 0.27$  &  $0.93 \pm 0.05$ & 2.1  \\
     
     L8 \textcolor{viridis7}{\mycircuitB} & Dynamic & $30^\circ$  &  $15^\circ$ & 61$^\circ.\textrm{s}^{-1}$ & $10.8 \pm 0.5$ & $2.00 \pm 0.29$  &  $0.90 \pm 0.04$ & 2.2    \\
\end{tabularx}
  \caption{Active grid operating parameters and corresponding freestream turbulence characteristics for the 8 turbulent inflows.}
  \label{TABLE:FST}
  \end{center}
\end{table}

\begin{figure}
    \centering
    \begin{subfigure}{0.48\textwidth}
        \centering
        \includegraphics[width=\linewidth]{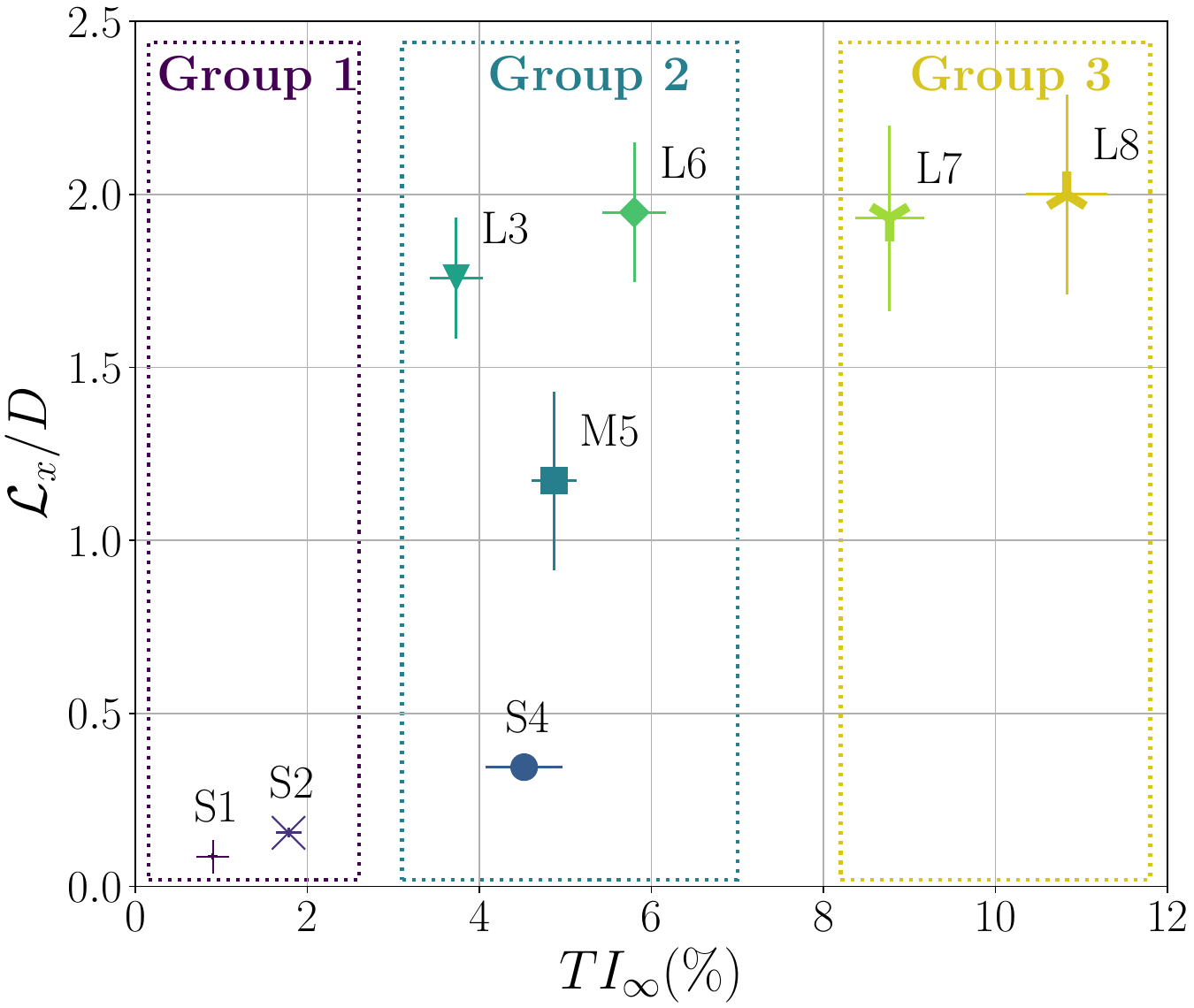}
        \subcaption{}
        \label{fig:figure1}
    \end{subfigure}
    \hfill
    \begin{subfigure}{0.48\textwidth}
        \centering
        \includegraphics[width=\linewidth]{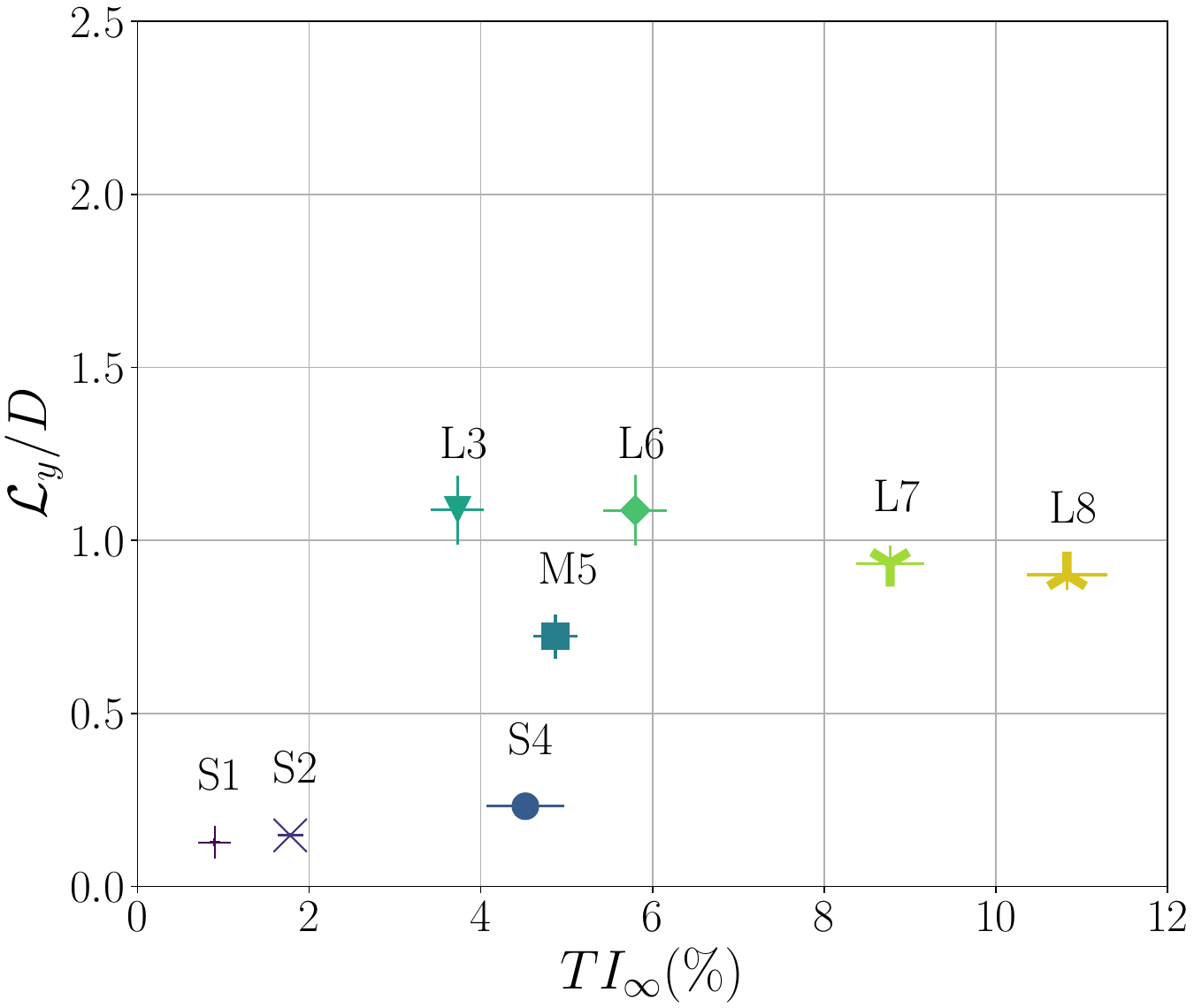}
        \subcaption{}
        \label{fig:figure2}
    \end{subfigure}
    \caption{Experimental envelope of freestream turbulence characteristics for the eight FST cases.}
    \label{Fig:FST_Flavours}
\end{figure}

${\cal L}_x$ is determined from the prior estimation of the integral time scale ${\cal T} = \int^{\tau_0}_0 R'(\tau) \mathrm{d}\tau$, and by assuming Taylor's hypothesis ${\cal L}_x = U_{\infty} {\cal T}$. Here, $R'(\tau)$ is the temporal correlation between $u'(x,y,t)$ and $u'(x,y,t+\tau)$ integrated up to $\tau_0$, where $R'(\tau_{0})=0.1$. Similarly, ${\cal L}_y = \int^{r_0}_0 R'(r) \mathrm{d}r$ is computed by integrating the spatial correlation $R'(r)$ between $u'(x,y,t)$ and $u'(x,y+r,t)$. Since the hot-wires were linearly spaced along the tunnel width, the spatial resolution for the integration $\mathrm{d}r$ is limited to the distance between each hot-wire, which is 14~cm. Moreover, for the hot-wires located near the walls, the range of $r$ for performing the correlation is maximum (\emph{e.g.}, 2.8~m for the probe closest to the wall, covering almost the entire width of the wind tunnel). This allows for the integration of $R'(r)$ up to the first zero-crossing ($R'(r_0)=0$) in all FST cases, particularly those characterised by a large spanwise ILS. In contrast, for the hot-wires located in the central region, the range of $r$ is smaller (\emph{e.g.}, 1.4 m for the centreline probe) and for cases with large ILS, the correlation function $R'(r)$ doesn't reach zero across the full range of $r$. Hence, for the sake of consistency, the integration is performed up to $r_{0}$, where $R'(r_{0})=0.1$ for all FST cases. Profiles of $TI_{\infty}$, ${\cal L}_x$ and ${\cal L}_y$ for all FST cases are shown in Appendix~\ref{AppendixA}, along with plots of the temporal $R'(\tau)$ and spatial $R'(r)$ correlation functions for selected FST cases and hot-wires, illustrating the aforementioned methodology for determining ${\cal L}_x$ and ${\cal L}_y$. The experimental envelopes of the 8 different turbulent inflows are obtained by averaging the profiles of $TI_{\infty}$, ${\cal L}_x$ and ${\cal L}_y$ for $y(\text{m}) \in \lbrack -1.2;1.2 \rbrack $ (\emph{i.e.}  $y/D \in \lbrack -2.1;2.1 \rbrack $). The three last columns of table~\ref{TABLE:FST} present the averaged quantities and their respective standard deviation, while a graphical representation of the different FST “flavours" $\{TI_{\infty}, {\cal L}_{x,y}\}$ is provided in figure~\ref{Fig:FST_Flavours}. 

A specific nomenclature has been established to classify the different FST cases, which are identified using a letter (S, M and L) and a number (from 1 to 8), and denoted as case “\#\#". The FST cases are first classified by their turbulence ILS: “S\#" for cases with small ILS (S1, S2, S4), “M\#" for medium ILS (M5), and case “L\#" for large ILS (L3, L6, L7, L8). Secondly, the number indicates the strength of the inflow turbulence intensity, $TI_{\infty}$, in hierarchical order, with case “\#1" representing the lowest $TI_{\infty} \approx 1\%$ (case S1), and case “\#8" the highest $TI_{\infty} \approx 11\%$ (case L8). For simplicity, the 8 FST cases are also categorised into 3 groups based on their turbulence intensity : Group 1 (low $TI_{\infty}$), Group 2 (medium $TI_{\infty}$), and Group 3 (high $TI_{\infty}$). 

The 8 FST cases encompass a broad spectrum of turbulence intensities, ranging from 1\% to 11\%, and streamwise integral length scales, spanning from $0.1$ to $2$ turbine diameters, which are representative of inflow conditions typically experienced by wind turbines \citep{Argyle2018,Burton2021}. Low-turbulence environments are typically encountered by offshore wind turbines, while higher turbulence levels are experienced by turbines in the central regions of wind farms or at onshore locations. The turbulence intensity levels notably fall within the classification defined in IEC standard 61400-1, Edition 3, that provide guidance for the design of wind farm layout \citep{international2019wind}. As can been seen in table~\ref{TABLE:FST} (last column) and figure~\ref{Fig:FST_Flavours}, the ILS in the streamwise direction, ${\cal L}_x$, is larger than in the spanwise direction ${\cal L}_y$, highlighting the stretching of the turbulent structures along the main flow direction. This is consistent with LES simulations of the flow within the  atmospheric boundary layer, with the ratio of ${\cal L}_x /{\cal L}_y$ being interestingly of the same order of magnitude for FST cases L\# \citep{Nandi2021,Thedin2023}. Although the range of ${\cal L}$ is limited compared to the size of the turbulent structures in the atmospheric boundary layer, which can exceed several times the diameter of wind turbines, the 8 FST cases capture specific ILS relative to the turbine's diameter. Cases S\# have an ILS smaller than the turbine radius (${\cal L}_x < D/2$), case M5 has an ILS on the order of the diameter (${\cal L}_x \simeq D$), and cases L\# have an ILS approximately twice the diameter (${\cal L}_x \simeq 2D$). In addition, it can be noticed that ${\cal L}_y$ is smaller than the tunnel width thereby reinforcing confidence in the method used to determine ${\cal L}_y$.

The grid protocols and FST cases were carefully designed and selected to separate the individual effects of the FST intensity $TI_{\infty}$, and its length scale ${\cal{L}}$, on turbine wakes, thus enabling the evaluation of their respective contribution to the wake recovery process, as previously studied with porous discs \citep{Bourhis2024} and solid bluff bodies \citep{Kankanwadi2020,Kankanwadi2023,Chen2023}. Hence, as shown in figure~\ref{Fig:FST_Flavours}, the four cases L\# provide a meaningful comparison of the impact of turbulence intensity at relatively fixed and large length scales. Cases S\# allow a similar comparison but at ILS smaller than the turbine's diameter. In addition, in Group 2, the variation in turbulence integral length scale is \emph{a priori} more pronounced than the changes in turbulence intensity, allowing for the potential isolation of the effect of ${\cal L}$ at a relatively constant $TI_{\infty}$.

The FST power spectra are presented in Appendix~\ref{AppendixA}. For all FST cases, a constant $\mbox{-5/3}$ slope, characteristic of fully developed turbulence can be recognised in the spectrum \citep{Kolmogorov1191}. Interestingly, and unlike the inflow conditions with large integral scales examined by \cite{Gambuzza2022} (also generated with an active grid), no spectral gap between low and high frequencies is observed, even for L\# cases, for which low frequencies contain more energy. As reported by \cite{Neuhaus_2020_PRL}, this suggests that the turbine is sufficiently far from the grid so that enough energy has been transferred from the large scales to the smallest scales through the energy cascade, resulting in the merging of the power spectra into a single large cascading process. Therefore, all FST cases are Kolmogorov-like flows, as defined by \cite{Gambuzza2022}, with significantly large inertial subranges for cases L\# (around 2 decades).

\subsection{Wind turbine operating points}
The time-averaged thrust coefficients, and the corresponding tip-speed ratios, for all FST cases are presented in table~\ref{Table:CT}. It's important to highlight that the values of $C_T$ presented in the table represent the total load acting on the turbine, including the nacelle, tower, and blades. In a previous study, \cite{Neunaber_2020} estimated that the tower and nacelle of the MoWito 0.6, without the blades, account for 17\% of the total measured thrust. Similarly to the analysis conducted by \cite{Juchter_2022}, using the same wind turbine, the standard deviations $\sigma_{C_T}$ and $\sigma_{\Lambda}$ for each $\{C_T, \text{FST}\}$ combination are defined as the standard deviation of the time-averaged values of ten consecutive time series of equal duration. For all combinations $\sigma_{C_T} \leq 0.03$ and $\sigma_{\Lambda} \leq 0.05$. 

For each operating point, denoted as low-$C_T$, medium-$C_T$, and high-$C_T$, we computed the mean values, along with their respective standard deviations, based on the time-averaged values across the 8 FST cases (see the two last rows of table~\ref{Table:CT}). As observed, for each operating point, the variations in $C_T$ and $\Lambda$ across the different FST cases are negligible, suggesting that the turbine's time-averaged thrust seems to be unaffected by both the FST intensity $TI_{\infty}$ and integral length scale ${\cal L}_x$. The effects of the inflow turbulence characteristics on the thrust coefficient of turbines are inconsistently reported in the literature. While studies by \cite{Gambuzza2021,Gambuzza2022} present results that align with our observations using their wind turbine model, other investigations on tidal turbines report contrasting trends. For instance, some studies report that $C_T$ decreases with increasing $TI_{\infty}$ \citep{Mycek2014,Blackmore_2016,Vinod2019}, but increases with the turbulence ILS \citep{Blackmore_2016}. \cite{Gambuzza2021} suggested that these conflicting trends may be attributed to the variations in turbine design or differences in the frequency content of the incoming turbulent flows. For solid and porous bodies, which are commonly used as wind turbine surrogates in wind tunnel or numerical studies, it has been shown that drag increases with turbulence intensity \citep{Blackmore2014,Rind2012,Jafari2018}, but decreases with larger longitudinal ILS \citep{Jafari2018}, which contrasts also with the trends observed for turbines. The seminal work of \cite{bearman1983} summarises these varied observations, stating that “Sometimes FST acts to increase drag, sometimes to decrease it, and at other times it has no effect at all". Ultimately, in our experiments, which cover a wide range of $TI_{\infty}$ and ${\cal L}$ for a unique wind turbine and well-characterised incoming flows, our observations indicate that $C_T$ is unaffected by the turbulence characteristics of the incoming flow.

\begin{table}
  \begin{center}
\def~{\hphantom{0}}
\begin{tabularx}{\textwidth}{p{1.3cm}||CC |CC |CC}
    FST case & \multicolumn{2}{c|}{Low-$C_T$} & \multicolumn{2}{c|}{Medium-$C_T$} & \multicolumn{2}{c}{High-$C_T$} \\[3pt]
             & $C_T$ & $\Lambda$  & $C_T$ & $\Lambda$ & $C_T$ & $\Lambda$ \\
    S1  & 0.47 & 1.63 & 0.67 & 2.56  & 0.89 & 3.78 \\  
    S2  & 0.46 & 1.70 & 0.62 & 2.85  & 0.89 & 3.56 \\
    S4  & 0.48 & 1.73 & 0.71 & 2.71  & 0.92 & 3.68 \\
    M5  & 0.48 & 1.70 & 0.69 & 2.70  & 0.89 & 3.67 \\
    L3  & 0.48 & 1.73 & 0.67 & 2.66  & 0.89 & 3.80 \\
    L6  & 0.49 & 1.76 & 0.70 & 2.69  & 0.88 & 3.60 \\
    L7  & 0.49 & 1.71 & 0.69 & 2.60  & 0.91 & 3.68 \\
    L8  & 0.47 & 1.70 & 0.69 & 2.61  & 0.91 & 3.68 \\
    \hline
    \multicolumn{7}{c}{Averaged values and standard deviations across the 8 FST cases} \\
    & \multicolumn{2}{c|}{$C_{T,{\text{low}}} \pm \sigma_{C_T} = 0.48 \pm 0.01$}   &
    \multicolumn{2}{c|}{$C_{T,{\text{med}}}\pm \sigma_{C_T}   = 0.68 \pm 0.03$}     & 
    \multicolumn{2}{c}{$C_{T,{\text{high}}}\pm \sigma_{C_T}  = 0.90 \pm 0.01$} \\
     & \multicolumn{2}{c|}{$\Lambda_{T,{\text{low}}} \pm \sigma_{\Lambda} = 1.71 \pm 0.04$}   &
    \multicolumn{2}{c|}{$\Lambda_{T,{\text{med}}} \pm \sigma_{\Lambda}  = 2.67 \pm 0.09$}     & 
    \multicolumn{2}{c}{$\Lambda_{T,{\text{high}}} \pm \sigma_{\Lambda}  = 3.68 \pm 0.08$}
\end{tabularx}
  \caption{Time-averaged operating thrust coefficients and tip-speed ratios for the wind turbine across the 8 FST conditions.}
  \label{Table:CT}
  \end{center}
\end{table}

\section{Results \label{Results}}

\subsection{Profiles of turbulence statistics \label{subsec:Velodef}}

\begin{figure}
    \centering
    \includegraphics[width=\linewidth]{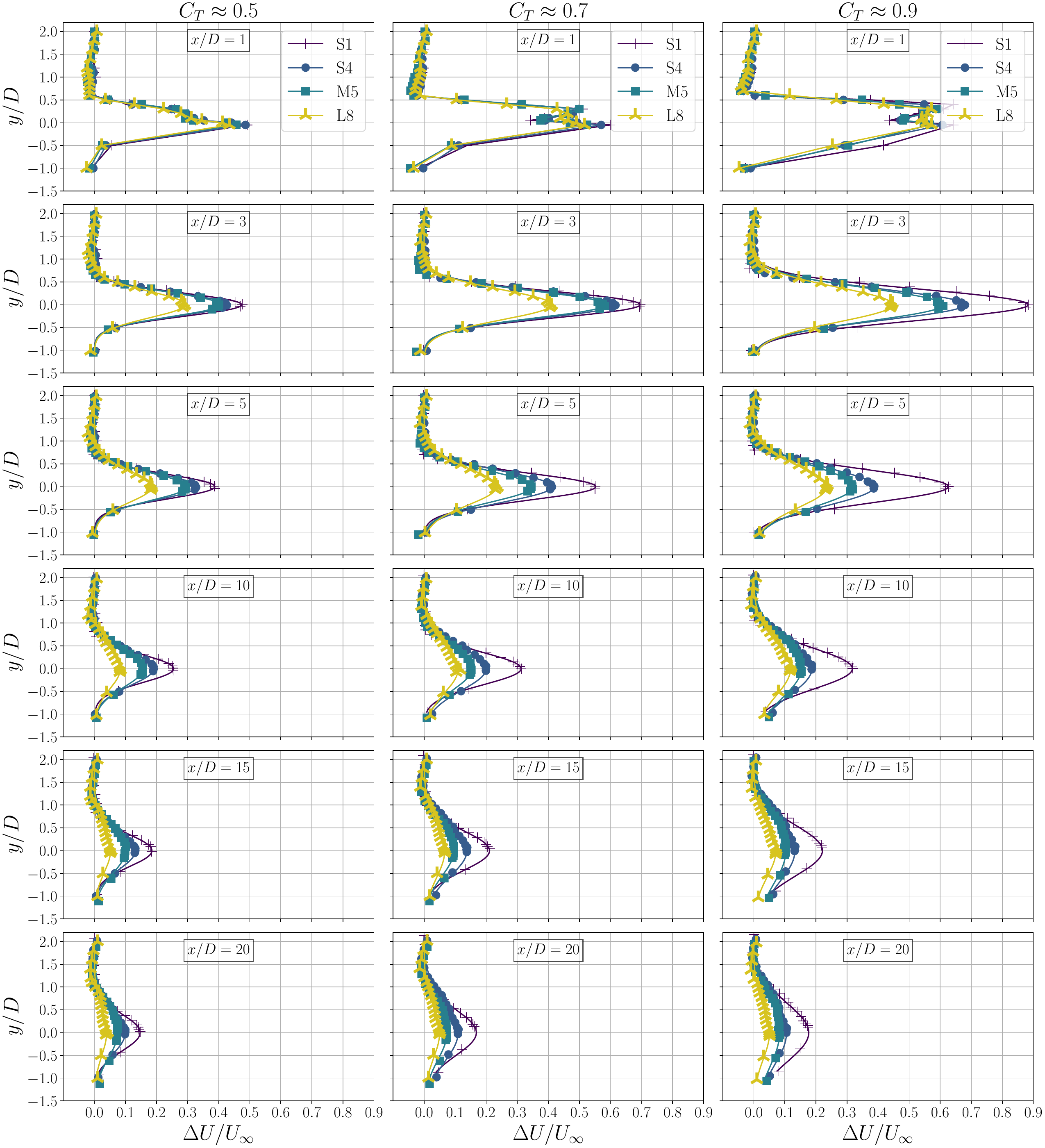}
    \caption{Time-averaged velocity deficit profiles.}
    \label{Fig:VelocityDeficitProfiles}
\end{figure}

\subsubsection{Time-averaged velocity deficit}

The time-averaged velocity deficit profiles ($\Delta U/U_{\infty}$, where $\Delta U = U_{\infty}(x) - \overline{U}(x,y)$) at five different streamwise locations are shown in figure~\ref{Fig:VelocityDeficitProfiles} for FST cases S1, S4, M5, and L8. Firstly, multiple inflection points in $\Delta U$ are observed at $x/D = 1$, highlighting the multiscale nature of the initial wake \citep{Baj2017,Biswas2024}. At this measurement station, the velocity profiles for each $C_T$ are nearly identical across all FST cases, as expected from the relationship between $\Delta U$ and $C_T$ described by momentum theory ($C_T \sim \int \overline{U} \Delta U r\textrm{d}r$). As the wake develops, the time-averaged velocity deficit profiles evolve towards a Gaussian-like shape for all $\{C_T, \text{FST}\}$ combinations, which is the typical far-wake shape assumed in standard wake models \citep{Bastankhah2014}. Although slight deviations from an ideal Gaussian profile are observed near the wake edge, within the shear layer where tip-vortices may persist, the velocity profiles follow a Gaussian-like shape by $x/D \approx 2$ (not shown here for brevity). Higher $TI_{\infty}$ accelerates the transition to a Gaussian shape by smoothing this steep velocity gradient and promoting the early dissipation of tip-vortices. Interestingly, although top-hat shapes \citep{Jensen1983,Frandsen2006} are commonly used to model the near wake velocity deficit (and occasionally the entire wake), none of the test cases exhibit such profiles at the streamwise locations examined. Instead, the profiles at $x/D=1$ for $C_T\approx 0.7$ and $C_T \approx0.9$ more closely resemble double Gaussian distributions  \citep{Keane2016,Keane2021}, especially for low-$TI_{\infty}$ cases (see S1). As a result, even more elaborate wake models like super-Gaussian models \citep{Shapiro2019,Vahidi2022}, which assume a transition from a top-hat profile in the near wake to a Gaussian distribution in the far wake, fail to accurately model the behaviour of the entire wake with a single function due to inadequate near-wake modelling. 

Considering the effect of FST, significant differences in both the magnitude and width of the profiles are observed for all $C_T$ after just a few diameters (see profiles at $x/D=3$), underscoring the strong and immediate impact of FST, with $TI_{\infty}$ seemingly having a greater impact on the velocity deficit profiles than ${\cal L}$. In addition, more pronounced differences in the $\Delta U/U_{\infty}$ profiles between the FST cases are observed as $C_T$ increases. Far downstream of the turbine, the presence of a strong turbulent background seems to disrupt the Gaussian distribution in particular for $C_T\approx0.5$ (see the velocity profiles for L8 at $x/D \geq 15$ ). The weak momentum deficit, compared to the strength of the turbulent background, might not be sufficient to ``sustain" a Gaussian distribution, with strong turbulent events potentially disrupting a self-similar evolution of the wake \citep{Rind2012}. 

At the farthest measurement station ($x/D=20$), a velocity deficit profile is still noticeable for all $\{C_T, \text{FST}\}$ combinations, emphasising the considerable distance needed for the velocity to fully recover. In the lowest $TI_{\infty}$ case S1, the initial differences in the width and magnitude of $\Delta U/U_{\infty}$ profiles, driven by changes in the turbine operating point ($C_T$), propagate throughout the entire measurement domain, with noticeable difference between the three $C_T$ cases persisting even at $x/D=20$. However, as $TI_{\infty}$ increases, the $C_T$-induced differences become less pronounced, and at $x/D=20$ the variations in $\Delta U/U_{\infty}$ between the three $C_T$ cases are barely discernible for high-$TI_{\infty}$ cases, like case L8. The presence of an inflow with a strong background turbulence intensity tends to mitigate the influence of $C_T$ on $\Delta U/U_{\infty}$, with the effect of $C_T$ on the wake being limited to a shorter distance downstream of the turbine. Finally, for the current  $\{C_T, TI_{\infty},{\cal L}\}$ dataset, variations in $\Delta U/U_{\infty}$ due to modifications in the FST conditions are generally more pronounced than those resulting from changes in $C_T$ , especially when considering the two highest $C_T$ cases. At low FST intensity, the influence of variations in $C_T$ on the wake persists over a considerable distance, whereas at high FST intensity, the effects of varying $C_T$ are progressively attenuated.

\subsubsection{Turbulence intensity and kinetic energy}
\begin{figure}
    \centering
    \includegraphics[width=\linewidth]{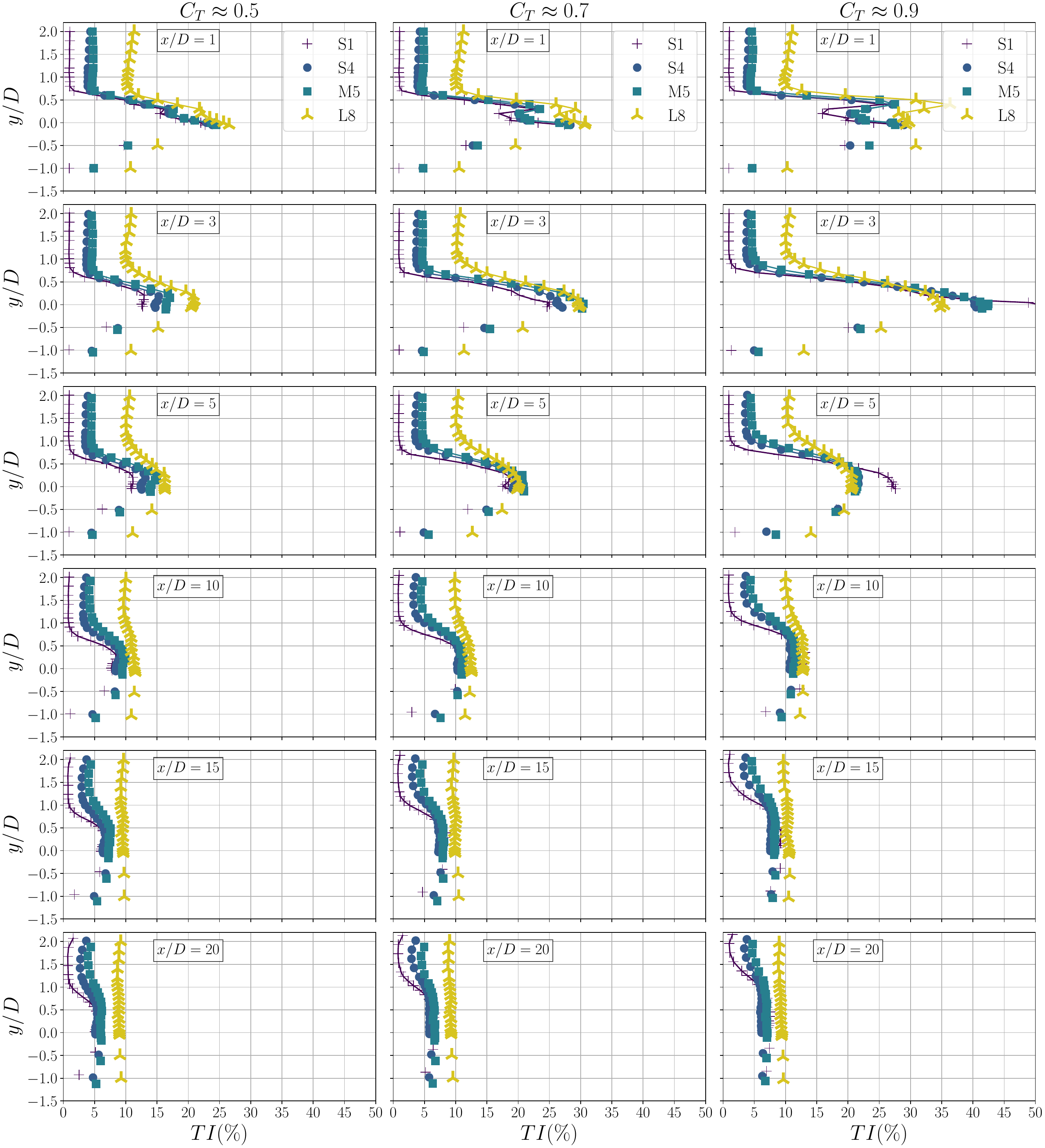}
    \caption{Turbulence intensity profiles.}
    \label{Fig:TIProfiles}
\end{figure}

\begin{figure}
    \centering
    \includegraphics[width=\linewidth]{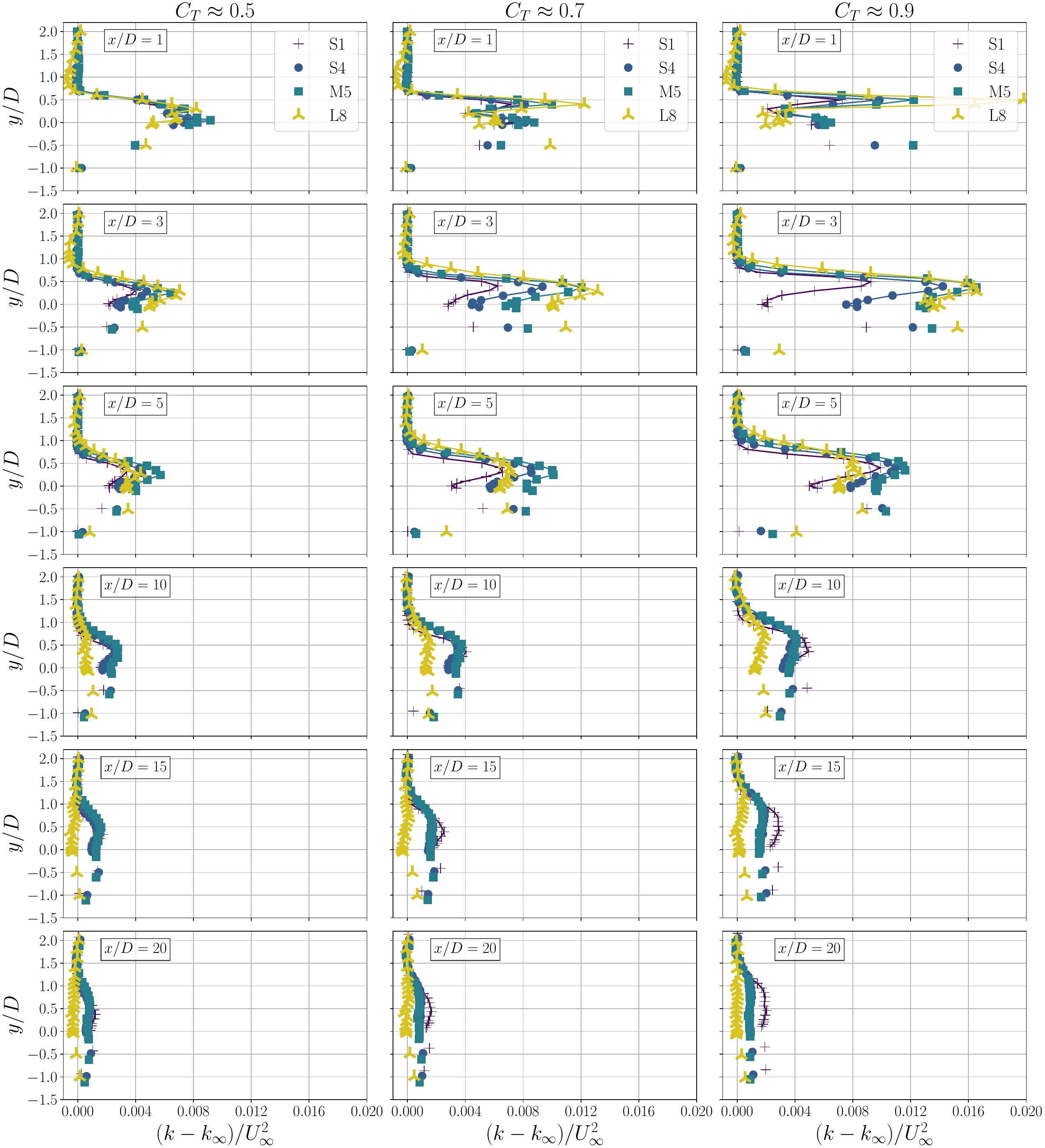}
    \caption{Turbulence-added kinetic energy profiles.}
    \label{Fig:TKEProfiles}
\end{figure}

The profiles of turbulence intensity ($TI = \sqrt{\overline{u'^{2}}}/\overline{U}$), and turbine-added turbulence kinetic energy ($\Delta k = (k-k_{\infty})/U_{\infty}^2$) are shown in figures~\ref{Fig:TIProfiles} \& \ref{Fig:TKEProfiles}. It is worth noting that the turbulence kinetic energy (TKE) here refers only to its streamwise component, representing only a fraction of the total TKE. To account for the gradual decay of the background turbulence kinetic energy along the wind tunnel length, inherent to grid-generated turbulence, $k_{\infty}$ is computed at each streamwise position by averaging the TKE measurements from the two hot-wires furthest from the centreline, \emph{i.e.}  $k_{\infty }=\langle k(x,y/D=1.8,2)\rangle $ , while $k = 0.5 \overline{u'^{2}}$.

At the first measurement station ($x/D=1$), the figures show that both $TI$ and $\Delta k$ increase as $C_T$ and $TI_{\infty}$ increase. For all \{$C_T$, FST\} combinations, a peak of turbine-added TKE is observed in the rotor tip shear layer. The magnitude of this peak increases with both $C_T$ and $TI_{\infty}$, with more pronounced differences observed among the FST cases at higher $C_T$. As the wake develops, the tip shear layer expands both inward and outward from the wake, leading to a decrease in the TKE peak and an initial increase in the centreline TKE, which then progressively decays as the wake mixes with the ambient flow. A similar initial build-up of the centreline $TI$, fuelled notably by the high shear tip-region, is also observed before it begins to decay. While a net saddle shape for the TKE profiles, characteristic of self-similarity for axisymmetric wakes, persists for the entire measurement domain for low $TI_{\infty}$ cases, the presence of a strong turbulent background disrupts this typical TKE saddle shape (see case L8), mirroring the breakdown of self-similar Gaussian velocity deficit profiles observed for high-$TI_{\infty}$ cases. Interestingly, \cite{Rind2012} reported a comparable behaviour in the far wake of a solid disc ($x/D \geq 65$). They observed that, in a weakly turbulent inflow, the far wake exhibited self-similarity, whereas the presence of a strong turbulent background suppressed any possibility for far-wake self-similarity. This was notably emphasised in the turbulence normal stresses, which deviated from their typical self-similar cross-stream behaviour (the saddle shape), gradually becoming constant along the radial axis and aligning with freestream stress levels 

The rate at which $TI$ and $\Delta k$ decay in the near wake increases as both $C_T$ and $TI_{\infty}$ increase. Close to the turbine ($x/D \leq 3$),  the following trend is observed: $\Delta k_{\textrm{L8}}\geq\Delta k_{\textrm{M5}}\geq \Delta k_{\textrm{S4}}\geq \Delta k_{\textrm{S1}}$. However, at measurement stations located farther downstream ($x/D \gtrsim15$), the inequality is reversed, such that $\Delta k_{\textrm{L8}} \leq\Delta k_{\textrm{M5,S4}}\leq \Delta k_{\textrm{S1}}$, indicating a faster mixing of the low-TKE background flow with the high-TKE wake flow as $TI_{\infty}$ increases. The faster decay and homogenisation of $TI$ and TKE in the near wake as $TI_{\infty}$ increases likely result from enhanced turbulence mixing between the near wake and the freestream due to both higher background turbulence intensity and an initially higher $\Delta k$. At $x/D = 3$, the turbulence intensity exhibits a distinct behaviour depending on the thrust coefficient. For  $C_T=0.5$ and $C_T=0.7$, the inequality $TI_{\textrm{L8}} \geq TI_{\textrm{M5}}\geq  TI_{\textrm{S4}} \geq TI_{\textrm{S1}}$ aligns with the augmentation of $TI_{\infty}$. However,  for $C_T = 0.9$, the opposite trend is observed, with $TI_{\textrm{L8}} \leq TI_{\textrm{M5,S4}} \leq TI_{\textrm{S1}}$ at the same downstream location. This reversal is likely attributable to the elevated turbulence intensity levels in the wakes at high $C_T $, which result in a steeper $TI$ gradient between the wake and the freestream, thereby accelerating $TI$ mixing and decay within the first initial diameters downstream of the turbine. $TI$ homogenisation in the near-wake region is further amplified at high-$TI_{\infty}$. Given the definition of turbulence intensity, the rate at which $TI$ decreases is influenced by both the recovery of the velocity deficit and the dissipation of the turbulence kinetic energy, processes that evolve at different rates, thereby leading to the non-identical evolution of $\Delta k$ and $TI$. 

Interestingly, for the highest $TI_{\infty}$ case (L8) and all three $C_T$, it can be observed that sufficiently far from the turbine ($x/D \gtrsim 15$), $TI$ in the wake becomes indistinguishable from the background turbulence intensity $TI_{\infty}$, with quasi-flat profiles. Similarly, the turbine-added TKE approaches $\Delta k \approx 0$, aligning with the observations of \cite{Rind2012}. Hence, while a wake defined by the presence of a momentum deficit still exists at $x/D=20$, as evidenced by the persistence of a small time-averaged velocity deficit $\Delta U$, a wake based on an increased level of turbulence kinetic energy vanishes earlier in the presence of a strong background turbulence intensity, as these quantities have effectively homogenised with the ambient flow. The faster recovery of $TI$ and $\Delta k$  emphasises that the background TKE may be entrained and mixed into the wake at a faster rate than the momentum. This aligns with the results of \cite{Buxton_Chen_2023}, who reported an increased efficiency of the entrainment of TKE compared to the entrainment of mass and momentum in the near wake of a cylinder exposed to FST ($x/D \lesssim 10$). It results in a scenario where the wind turbine wake is characterised by a flow region with solely reduced mean momentum, rather than one with both lower mean momentum and higher mean TKE.

\begin{figure}
    \centering
    \includegraphics[width=\linewidth]{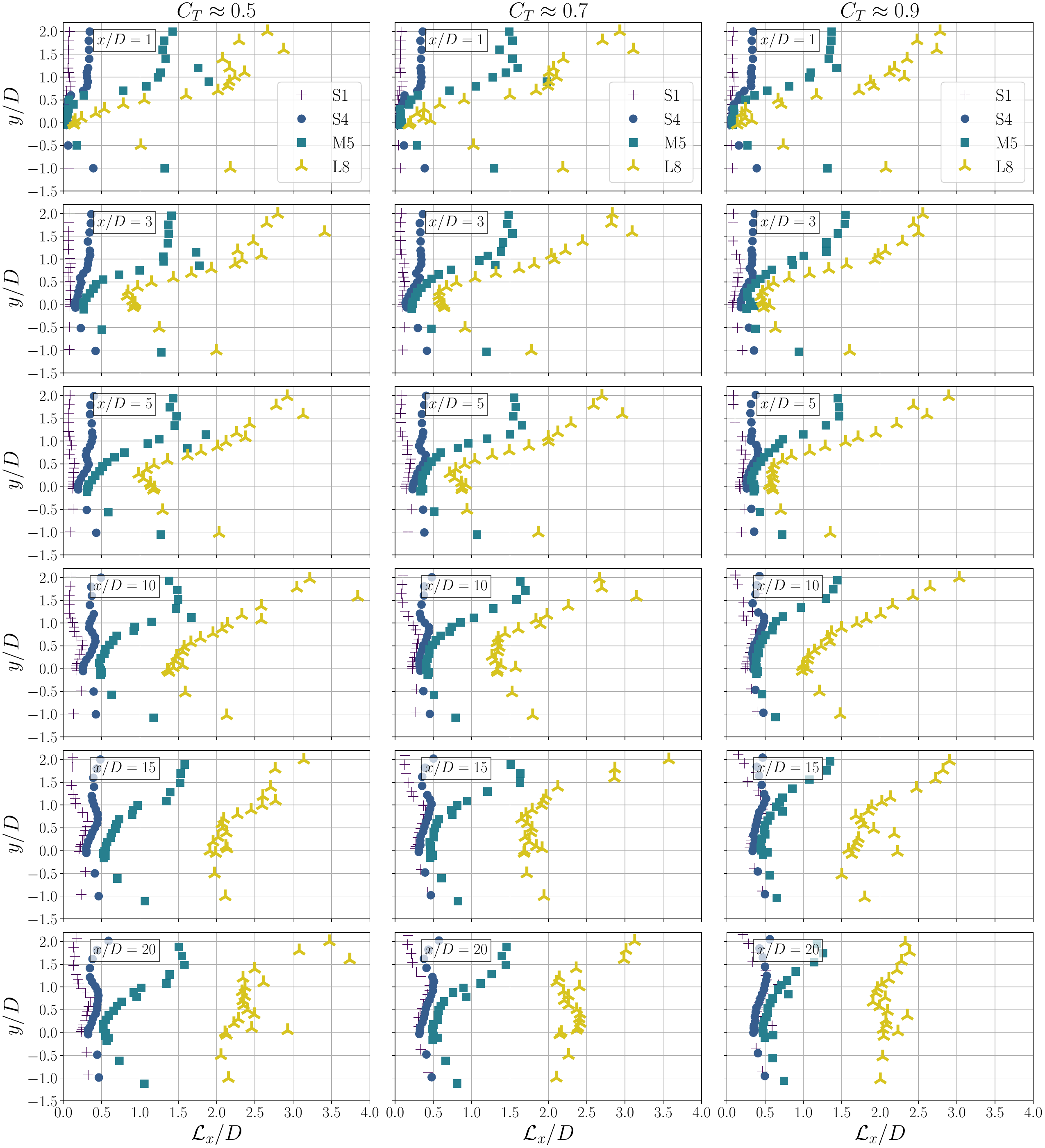}
    \caption{Streamwise turbulence integral length scale profiles.}
    \label{Fig:LProfiles}
\end{figure}

\subsubsection{Turbulence integral length scale}

Figure~\ref{Fig:LProfiles} shows the profiles of streamwise ILS, ${\cal L}_{x}$, for the FST cases S1, S4, M5 and L8. For reference, the inflows ILS are respectively ${\cal L}_{x}<D/2$ for S1 and S4, ${\cal L}_{x}\sim D$ for M5, and ${\cal L}_{x}\sim 2D$ for L8. The ILS in the wakes are computed in a similar manner to the background turbulence ILS introduced in \S~\ref{Experimental method}, as ${\cal L}_{x} = \overline{U}\int^{\tau_0}_0 R'(\tau)d\tau$ and $R'(\tau_0) = 0.1$.

At $x/D=1$, similarly to the velocity deficit, an ILS deficit is observed in the near wake for all \{$C_T$, FST\} combinations, with ${\cal L}_{x}(x/D=1)$ being smaller than the background ILS. As suggested by \cite{Chamorro2012_ILS}, the rotor acts as a high-pass filter, damping the largest scales of the incoming turbulent flow. The effect is notably more pronounced for M\# and L\# cases, where steep gradients of ${\cal L}_x$ between the wake and the background flow are observed in the profiles, than for S\# cases, which exhibit nearly flat profiles. Nonetheless, even for the S\# cases, the ILS immediately downstream of the turbine is slightly smaller than the background ILS. As $C_T$ increases, the profiles at $x/D=1$ become progressively less steep around $y/D=0$, suggesting that with increasing $C_T$ (and $\Lambda$), the streamtube within which large structures are filtered expands radially, a trend that is further emphasised in the widest ${\cal L}_x$ profiles at higher $C_T$.

Interestingly, further downstream of the turbine the behaviour of ${\cal L}_{x}$ in the wakes varies strongly depending on the FST integral length scale. Focusing on cases S1 and S4, it can be observed that for $x/D \gtrsim 3$, ${\cal L}_x$ in the turbine wakes exceeds that in the freestream. In particular, the largest coherent structures are located in the tip-shear layer, as indicated by the presence of of a peak in ${\cal L}_x$ in the profiles around $y/D \approx 0.6$, which becomes more pronounced and shifts radially as the wake develops. These structures resulting from tip-shear layer instabilities, grow in size as $x$ increases and as the thrust coefficient increases, with ${\cal L}_{x,\text{Low-}C_T}\leq{\cal L}_{x,\text{Med-}C_T}\leq{\cal L}_{x,\text{High-} C_T}$ for the entire measurement range. At the farthest measurement station, the ILS in the wake is approximately one rotor radius for $C_T\approx0.9$, which remains small compared to the background ILS in cases M\# and L\#. A similar behaviour has been observed in the tip-shear layers of porous bodies above a porosity threshold, with small vortices being initially being confined in the unstable shear layer, generating then larger structures as the shear layer develops \citep{Cicollin2024,Bourhis2024}. 

Shear layer instability, similar to the vortex shedding of bluff bodies, is one of the two mechanisms proposed in the literature as the possible origin for the onset of wake meandering, the other being the presence of large eddies in the freestream (see \cite{Yang2019} for a general review on wake meandering). The second mechanism forms the basis of the Dynamic Wake Meandering (DWM) model introduced by \cite{Larsen2008}, which assumes that wind turbine wakes are advected passively by the large-scale eddies in the freestream, with these eddies driving the wake's large-scale motions. While this topic will be discussed in greater detail in \S~\ref{subsec:wake meandering}, it is important to note at his stage that, for the small ${\cal L}_x$ FST cases S\#, the largest structures in the wind turbine wakes originate from the tip-shear layer, grow with streamwise distance, and are influenced by the turbine operating point $C_T$, \emph{i.e} its porosity, exhibiting characteristics similar to the classical vortex shedding behaviour of porous and bluff bodies.

In contrast, for M\# and L\# cases, the ILS within the wakes remains smaller than the background ILS throughout the entire measurement domain, with no peak in ${\cal L}_x$ observed in the rotor tip-shear layer. Hence, the high-pass filter behaviour of the turbine on the incoming flow structures persists over a longer streamwise extent. Ultimately, the ILS exhibits a form of recovery for these FST cases, with ${\cal L}_x$ in the wake progressively aligning with that of the freestream, similar to the recovery of the velocity deficit. Moreover, higher background ILS  leads to a faster recovery of ${\cal L}_x$ in the wake, similar to the evolution of $TI$ and $k$ discussed previously. Interestingly, in contrast also to the S\# cases, we observe that ${\cal L}_{x,\text{Low-}C_T}\geq{\cal L}_{x,\text{Med-}C_T}\geq{\cal L}_{x,\text{High-} C_T}$ in the wake for the entire measurement range, meaning that ${\cal L}_x$ recovers faster at low thrust coefficient. This can be qualitatively interpreted by drawing a simple comparison with porous discs: at high porosity, and consequently low thrust coefficient, larger structures pass through the disc, \emph{i.e} the turbine, without being filtered, leading to a slightly larger ${\cal L}_x$ in the wakes at low $C_T$. This analysis holds true only for the FST cases with a sufficiently large ILS, for which the turbine-induced eddies are smaller than the background eddies. 

In summary, for S\# FST cases, the large scale structures in the wakes are primarily turbine-induced coherent structures and their evolution is therefore strongly influenced by the turbine's operating point. The structures originating in the tip-shear layer grow and propagate toward the background flow, resulting in a larger ILS in the wake compared to the ambient flow. On the other hand, for M\# and L\# FST cases, the ILS within the wake recovers similarly to $TI$ and $\Delta U$,  progressively increasing to adjust with that of the background turbulence,  with higher background ILS leading to a faster wake ILS recovery. Larger meandering of the wake, vortex-vortex interaction between the FST and the wake, or the entrainment of large structures from the background flow into the wake, could explain the progressive adjustment of the ILS. For the range of ILS examined, distinct and clear interactions between the background flow and the wake is observed: at small ILS, the growth of structures in the wake is primarily driven by turbine-induced structures, while for larger ILS values, the growth of wake structures is predominantly fuelled by the background flow. 

\subsection{Wake width and wake growth rate \label{subsec:WakeWidth}}

\subsubsection{Wake width based on the time-averaged velocity deficit $\Delta U$}

\begin{figure}
    \centering
    \includegraphics[width=0.6\linewidth]{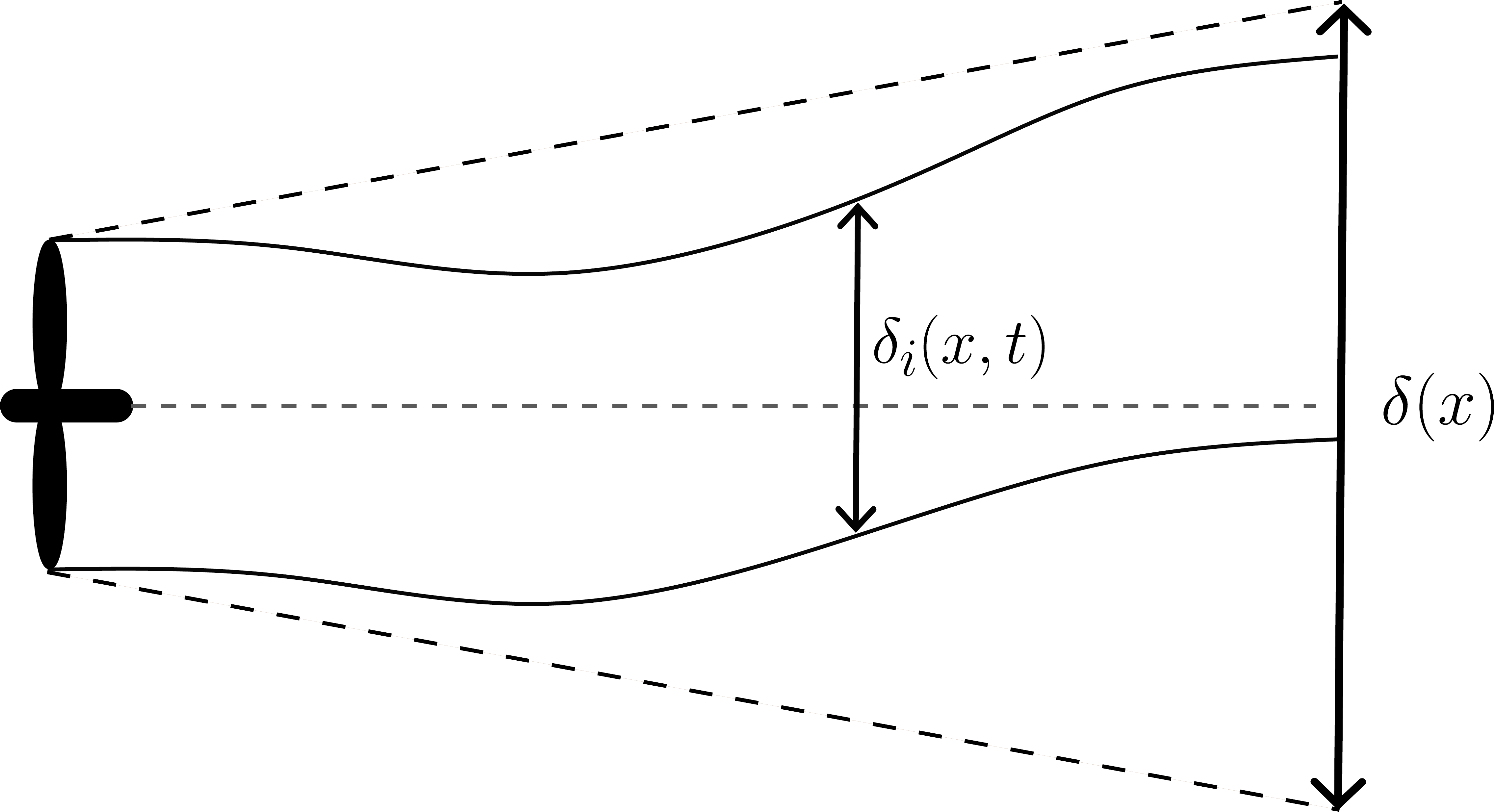}
    \caption{Sketch illustrating the contribution of wake meandering to the time-averaged wake width $\delta(x)$. $\delta_i(x,t)$ is the instantaneous wake width, determined from a single snapshot of the wake. Reproduced, with permission, from \cite{Kankanwadi2023}.}
    \label{fig:cartoon}
\end{figure}

\begin{figure}
    \centering
    \includegraphics[width=\linewidth]{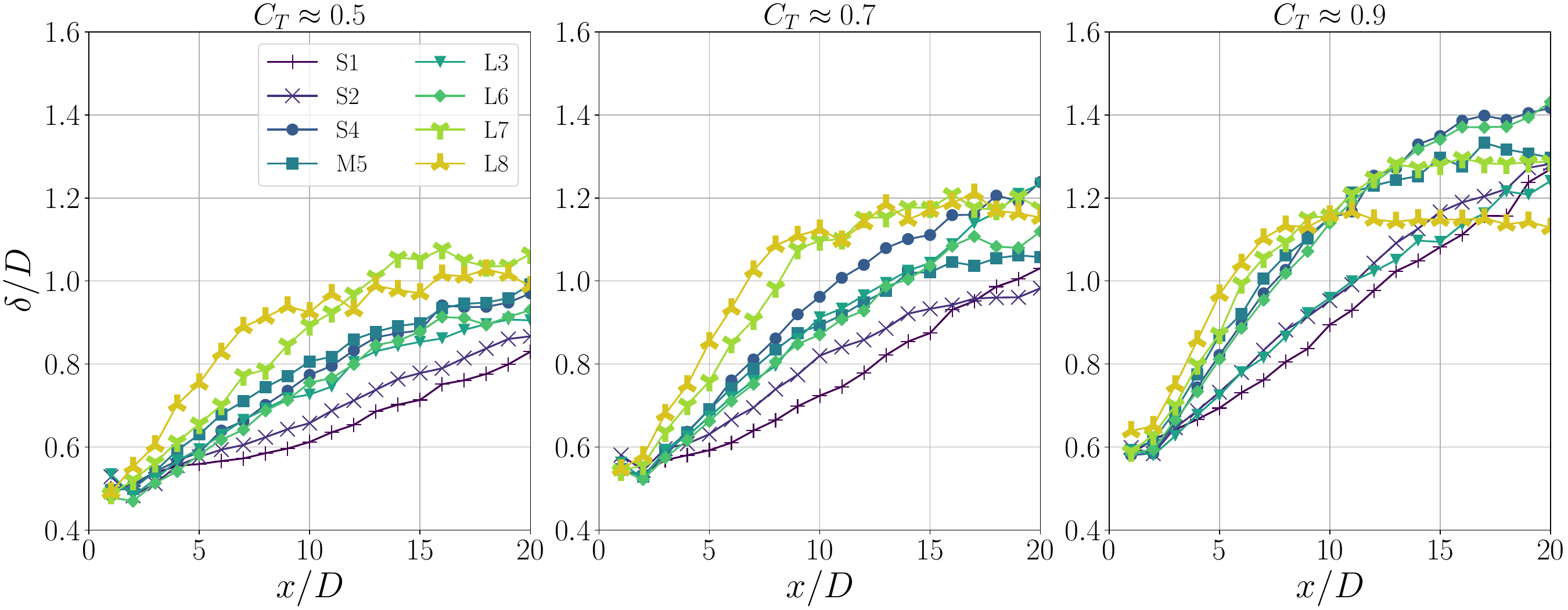}
    \caption{Streamwise evolution of the wake width $\delta(x)$.}
    \label{Fig:wakewidth}
\end{figure}

Figure~\ref{Fig:wakewidth} shows the streamwise evolution of the time-averaged wake width, $\delta(x)$, for all 24 \{$C_T$, FST\} combinations. At a given streamwise location $x$, the time-averaged wake width is defined as the radial location $r$ at which the time-averaged velocity deficit is $10\%$ that of the maximum, \emph{i.e.} $\delta(x)$ such that $\Delta U(x,y=\delta) = 0.1\Delta U_{max}(x)$, where $\Delta U_{max}$ is the maximum velocity deficit. It is important to note that, as introduced by \cite{Kankanwadi2023}, the time-averaged wake width $\delta(x)$ captures both the growth of the instantaneous wake through the entrainment of background flow and the lateral motion of the wake induced by wake meandering. The sketch in Figure~\ref{fig:cartoon} illustrates the contribution of wake meandering to $\delta(x)$. While we analyse this metric—widely used in empirical wind turbine wake models—without distinguishing the two contributions in that section, we will further discuss the influence of wake meandering in \S~\ref{subsec:wake meandering}.

At first glance, the most striking observation from this figure is the presence of a turning point in the slope of $\delta(x)$ occurring after several diameters—particularly for wakes developing in a highly turbulent background—beyond which the wake width grows significantly more slowly than it does farther upstream. All wakes initially grow linearly with streamwise distance, consistent with the evolution widely reported for wind turbine wakes \citep[e.g.][]{Schumann2013,Bastankhah2014} and two-dimensional bluff body wakes \citep{Eames2011,Kankanwadi2023,Chen2023} in the initial development region. Moreover, since the seminal work of \cite{Jensen1983}, the assumption of linear wake growth has been central to most used analytical wake models (see the paper from \cite{Wang2024} for a general review on wind turbine wakes modelling). The wake width $\delta$ is typically modelled as $\delta = \lambda x + \beta$, where $\lambda = \text{d}\delta/\text{d}x$ represents the wake growth rate, which is generally assumed to be constant with respect to $x$. However, it appears that this assumption is valid only up to a certain distance from the turbine, which depends on $C_T$ and the background turbulence. Indeed, with the exception of the fully open grid case (S1), where the wakes grow nearly linearly throughout the entire measurement range, a change in the slope of the wake width is observed for all FST cases, becoming progressively more abrupt and occurring closer to the turbine as $TI_{\infty}$ increases. For instance, for the highest-$TI_{\infty}$ case L8, a plateau evolution of the wake width begins at $x/D \approx 7$. For the medium-$TI_{\infty}$ cases (Group 2), the transition to a plateau shape is more gradual and occurs farther downstream in the wake. This change in the wake growth rate occurs for all three $C_T$, though it is more pronounced at larger $C_T$. Interestingly, this change in slope appears to coincide with the position in the wake where both $TI$ and TKE have nearly homogenised. The absence of a significant TKE and $TI$ gradient between the wake and the ambient flow may result in less efficient mixing and turbulent diffusion, ultimately leading to a reduced wake expansion rate. 

Focusing on the wake region in relative proximity to the turbine  ($x/D \lesssim 7$), we observe a clear increase in $\delta$ as both $C_T$ and $TI_{\infty}$ increase. Specifically, FST case L8 yields the widest wake, S1 the narrowest, and the evolution of $\delta$ follows the order $\delta_{\textrm{Group 3}} \geq \delta_{\textrm{Group 2}} \geq \delta_{\textrm{Group 1}}$. Similarly, we observe that $\delta_{\textrm{High-$C_T$}} \geq \delta_{\textrm{Med-$C_T$}} \geq \delta_{\textrm{Low-$C_T$}}$, a trend that holds true for all measurement stations and FST cases. Focusing on Group 2 FST cases, it can be observed that for $C_T = 0.5$, $\delta_{\textrm{S4,M5}} \geq \delta_{\textrm{L3,L6}}$ across the entire measurement range, while for $C_T = 0.7$ and $C_T=0.9$, this inequality holds up to $x/D \approx 10$. Since $TI_{\infty}$ is higher in case L6 compared to cases S4 and M5, one might \emph{a priori} anticipate that $\delta_{\textrm{S4,M5}} \leq \delta_{\textrm{L6}}$. However, the ILS are significantly larger in cases L3 and L6 (${\cal L}_x \sim 2D$) compared to case S4 (${\cal L}_x \sim 0.5D$) and M5 (${\cal L}_x \sim 1D$). Therefore, we may conclude that the FST integral length scale of the inflow contributes to a reduction in the time-averaged wake width, at least in the initial wake region. Interestingly, \cite{Kankanwadi2023} reported a reduction in the time-averaged instantaneous near wake width ($\overline{\delta_i}$) downstream of a cylinder for inflows with a large ILS, as well as an initial decrease in the time-averaged width wake ($\delta$) for $x/D \lesssim 3$. However, further downstream ($x/D \gtrsim 4$), they found that wake meandering, enhanced by larger ILS, became sufficiently pronounced to compensate for the reduction in the instantaneous wake width, such that the time-averaged wake width remained larger for high ILS inflows compared to the non-turbulent background. In our wind turbine experiments, at first glance wake meandering appears not to compensate for the decrease in instantaneous wake width induced by larger ILS. While further analysis on the influence of the large motions of the wake will be provided in \S~\ref{subsec:wake meandering}, it is already noteworthy to observe certain similarities in the FST effects on both bluff body and wind turbine wake widths.  

Further downstream, for Group 3 FST cases, a clear plateau in the evolution of $\delta$ is observed for all three $C_T$. For Group 2, a slowdown in wake width growth is evident, while for Group 1, the evolution remains nearly linear across the entire measurement range. Moreover, it can be noted that, for $C_T=0.9$, the far wake is wider for the lowest $TI_{\infty}$ case S1 than for the highest $TI_{\infty}$ case L8. It can be hypothesised that a similar result would be observed for the two lower $C_T$ if the measurement range was extended. This contrasts with the current state of the art on wind turbine wakes, which assumes an increase in wake width with increasing $TI_{\infty}$ throughout the entire wake. However, interestingly, it aligns with recent findings on porous disc wakes, where larger far wakes were observed for low or non-turbulent inflows compared to high $TI_{\infty}$ ones \citep{Vinnes2023,Bourhis2024}.

\begin{figure}
    \centering
    \begin{subfigure}[t]{0.48\linewidth}
        \centering
        \includegraphics[width=\linewidth]{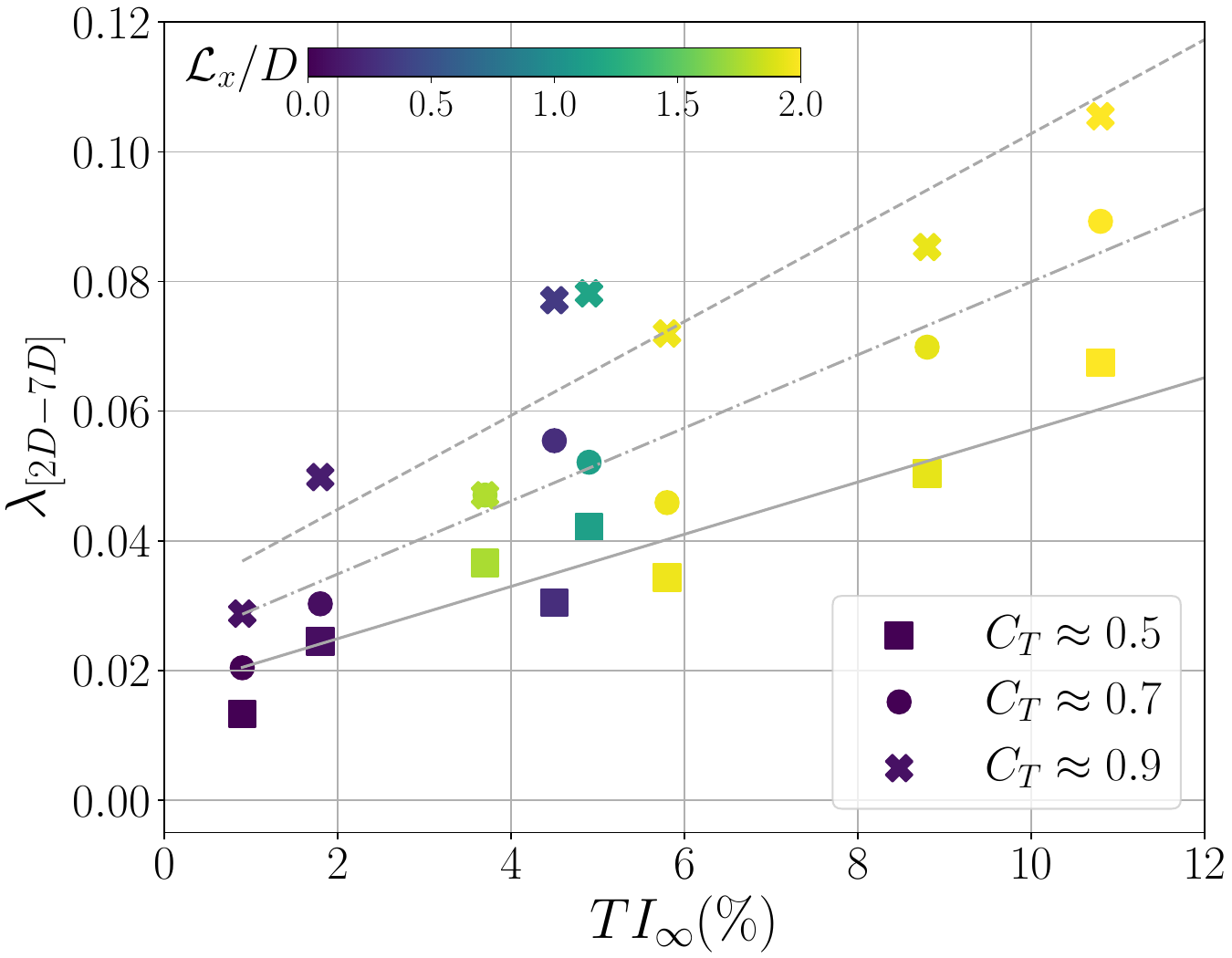}
        \caption{}
        \label{fig:WakeGrowthRate1}
    \end{subfigure}%
    \hfill
    \begin{subfigure}[t]{0.48\linewidth}
        \centering
        \includegraphics[width=\linewidth]{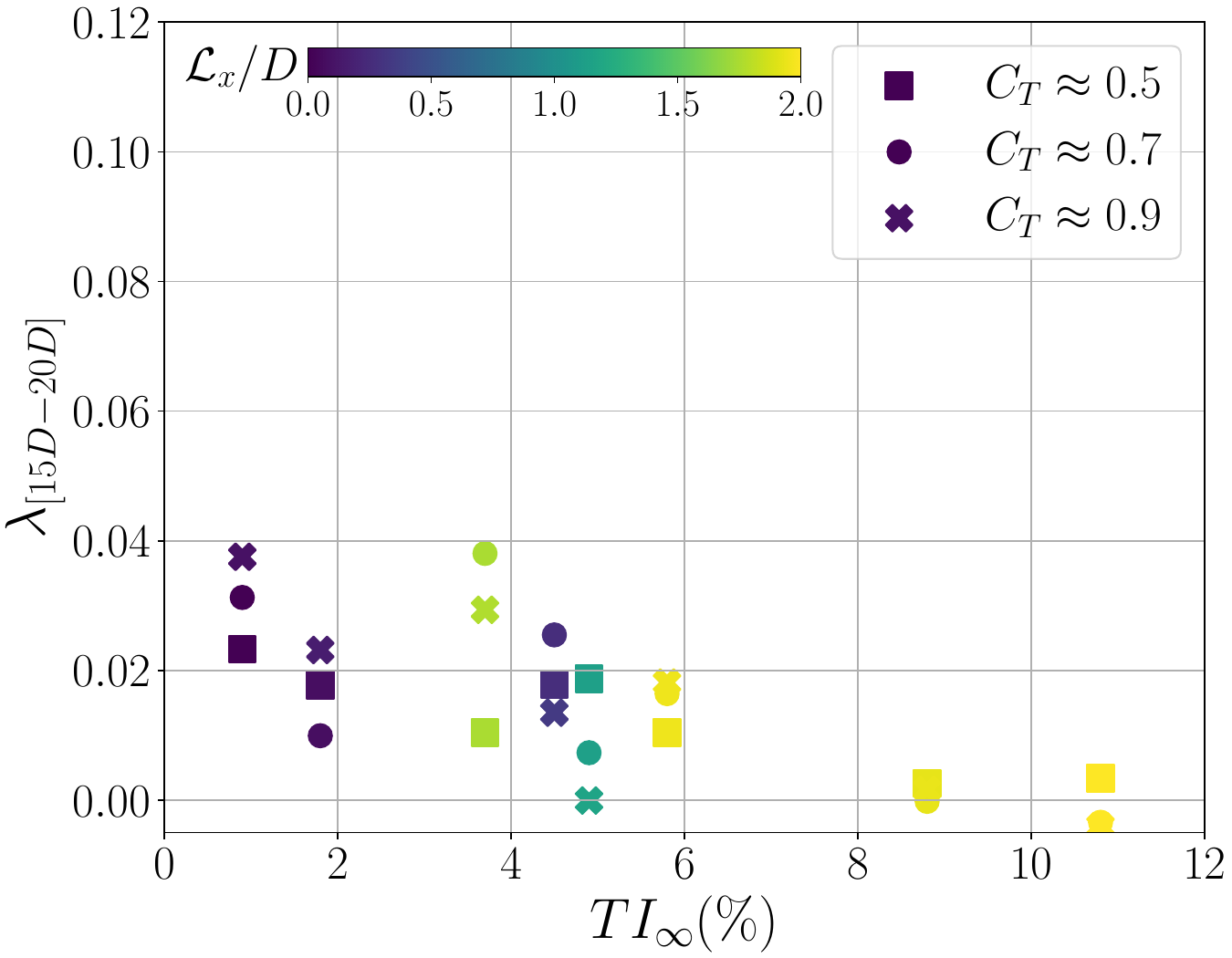}
        \caption{}
        \label{fig:WakeGrowthRate2}
    \end{subfigure}
    \caption{Wake growth rates calculated from $\lambda = \overline{\text{d}\delta/\text{d}x}$ for (\subref{fig:WakeGrowthRate1}) $2 \leq x/D \leq 7$, and (\subref{fig:WakeGrowthRate2}) $15 \leq x/D \leq 20$.}
    \label{Fig:wakegrowthrate}
\end{figure}

Figures~\ref{fig:WakeGrowthRate1} \& \ref{fig:WakeGrowthRate2} present the 24 wake growth rates $\lambda$, calculated for two distinct intervals: one in the ``near wake'', calculated from the gradient of $\delta(x)$ for $2 \leq x/D \leq 7$, and one in the ``far wake'', calculated from the gradient of $\delta(x)$ for $15 \leq x/D \leq 20$. In the near wake, we observe that $\lambda$ increases with both $C_T$ and $TI_{\infty}$. In this wake region, the variations in $\lambda$ with $C_T$ are more pronounced at high $TI_{\infty}$ than at low $TI_{\infty}$, with also larger differences between FST cases at higher $C_T$. While ${\cal L}_x$ clearly affects the wake width $\delta$, its effect on $\lambda$ appears secondary compared to $TI_{\infty}$ and $C_T$. However, in the far wake $x/D \gtrsim 15 $, we observe the opposite trend: a decrease in $\lambda$ as $TI_{\infty}$ increases.

Importantly, this change in how FST influences wind turbine wakes spreading aligns with the entrainment behaviours reported for bluff and porous bodies (\emph{cf.} \S~\ref{Introduction}). As a reminder, both FST intensity and ILS have been found to enhance the entrainment and wake growth rates in the near wake of cylinders, whereas, in the far wake, FST intensity suppresses entrainment \citep{Kankanwadi2020,Kankanwadi2023,Chen2023,Chen_Buxton_2024}. Moreover, \cite{Chen2023,Chen_Buxton_2024} identified a ``crossover location'' in cylinder wakes, located approximately $15D$ downstream (for the specific Reynolds number and FST ``flavour'' parameter space studied), beyond which the wakes spread at a slower rate. Similarly, porous-disc-generated wakes exposed to FST have been found to be wider in the near wake, but both wake growth and entrainment rates in the far wake decrease as FST intensity increases \citep{Vinnes2023,Bourhis2024}. The similar variation in wind turbine wake growth rates exposed to FST, along with the presence of a similar turning point in the wake width streamwise evolution further highlights the similarities between the entrainment behaviours observed in wind turbine and bluff body wakes.

\subsubsection{Modelling of the wake growth rate $\lambda$}

Meaningful conclusions can be drawn by comparing the $\lambda$ dataset obtained for the near wake (figure~\ref{fig:WakeGrowthRate1}) with standard wake growth rate models. In early studies, within the framework of the top-hat Jensen model, the wake growth rate was assumed to be constant. Specifically, \cite{Jensen1983} suggested $\lambda=0.1$ in his seminal work, while subsequent studies recommended $\lambda=0.075$ for onshore wind turbines, and $\lambda=0.03 - 0.05$ for offshore ones \citep{Katic1987,Mortensen2007,Barthelmie2006,Barthelmie2010}. It is noteworthy that the analytical models by \cite{Jensen1983} and \cite{Katic1987}, which assume a constant wake growth rate, are implemented in many major commercial wind turbine and wind farm software packages \citep{Triantafyllou2021}. Although these values typically fall within the range observed in our experiments, assuming a constant wake growth is a crude assumption, potentially leading to significant inaccuracies  in the preliminary assessment of a wind farm's potential. 

In recent decades, more elaborate wake growth rate models have emerged, typically derived by fitting linear or power laws to $\lambda$ datasets obtained through LES or experiments. Several studies have reported a linear relationship between the wake growth rate and FST intensity, as $\lambda = a TI_{\infty} + b$, with varying $\{a,b\}$ coefficients. \cite{Niayifar2016} proposed that $a = 0.38$ and $b=0.004$ (LES), \cite{carbajo2018} suggested $a = 0.35$ and $b=0$ (LIDAR measurement of full-scale turbines), and \cite{Cheng2019} derived  $a = 0.223$ and $b=0.022$ (LES). It is important to note that while these authors acknowledged that $\lambda$ is a function of $C_T$, these empirical relationships were derived for a unique $C_T$ (close to $C_T \approx 0.8$ for all three studies), and for a limited range of FST conditions, resulting in a relatively narrow examination of the \{$C_T$, $TI_{\infty}$, ${\cal L}_x$\} parameter space. Fitting the equation $\lambda = a TI_{\infty} + b$ to our entire $\lambda$ dataset results in a poor fit ($R^2=0.621$ with $a=0.592$ and $b=0.022$), due to the strong influence of $C_T$ on $\lambda$. However, when fitting each $C_T$ dataset individually, the fits improve significantly, supporting the idea of a linear relationship between $TI_{\infty}$ and $\lambda$. For instance, fitting the $\lambda$ dataset for $C_T \approx 0.5$ yields $a=0.47$ and $b=0.013$ ($R^2 = 0.907$); for $C_T \approx 0.7$ the fit gives $a=0.62$ and $b=0.019$ ($R^2 = 0.929$); and for  $C_T \approx 0.9$, the fit results in $a=0.68$ and $b=0.033$ ($R^2 = 0.850$). Drawing inspiration from these models, we fitted the following equation over the entire $\lambda$ dataset: $\lambda = a C_TTI_{\infty} + b C_T + c{\cal L}_x$.  The fitting coefficients are $a = 0.83$, $b = 0.032$, and $c \sim 10^{-11}$, with a coefficient of determination $R^2 = 0.914$ (see the grey lines on Figure~\ref{fig:WakeGrowthRate1}). This empirical model provides a reasonable approximation of the wake growth rate, while further highlighting the negligible influence of ${\cal L}_x$ on the wake growth rate compared to $TI_{\infty}$ and $C_T$.

Recently, \cite{Ishihara2018} proposed the following empirical power law wake growth rate model based on LES data $\lambda = 0.11 C_T^{1.07} TI_{\infty}^{0.2}$, which yields a poor fit with the current dataset ($R^2 = 0.343$). Taking inspiration from this model, we fitted the following equation over the entire dataset $\lambda=aC_T^{\alpha_1}TI_{\infty}^{\alpha_2}{\cal L}^{\alpha_3}$. This yielded a fairly high coefficient of determination $R^2 = 0.923$, with the fit resulting model $\lambda = 0.372C_T^{0.99}TI_{\infty}^{0.53}$ ($\alpha_3 \sim  10^{-11}$). While the power exponent for the thrust coefficient nearly aligns with that proposed by \cite{Ishihara2018}, our findings suggest a stronger influence of $TI_{\infty}$. It is worth noting that, although their study explored a wide range of $C_T$, it considered only two levels of $TI_{\infty}$, potentially limiting the model's sensitivity to this parameter 

With the current dataset, it is challenging to determine which model fits better, as the differences in $R^2$ are small and subject to uncertainty. Nevertheless, both empirical models provide a fairly good approximation of $\lambda$ across a wide range of thrust coefficients and freestream turbulence conditions. They outperform most models in the literature, which are typically designed for a single thrust coefficient and based on a significantly smaller number of data points. In addition, we found that wake growth rate models assuming a linear expansion of the wake are only valid up to a certain streamwise distance from the turbine, which decreases as $TI_{\infty}$ increases.

\subsubsection{Wake width based on the turbine-added TKE $\Delta k$}

\begin{figure}
    \centering
    \includegraphics[width=\linewidth]{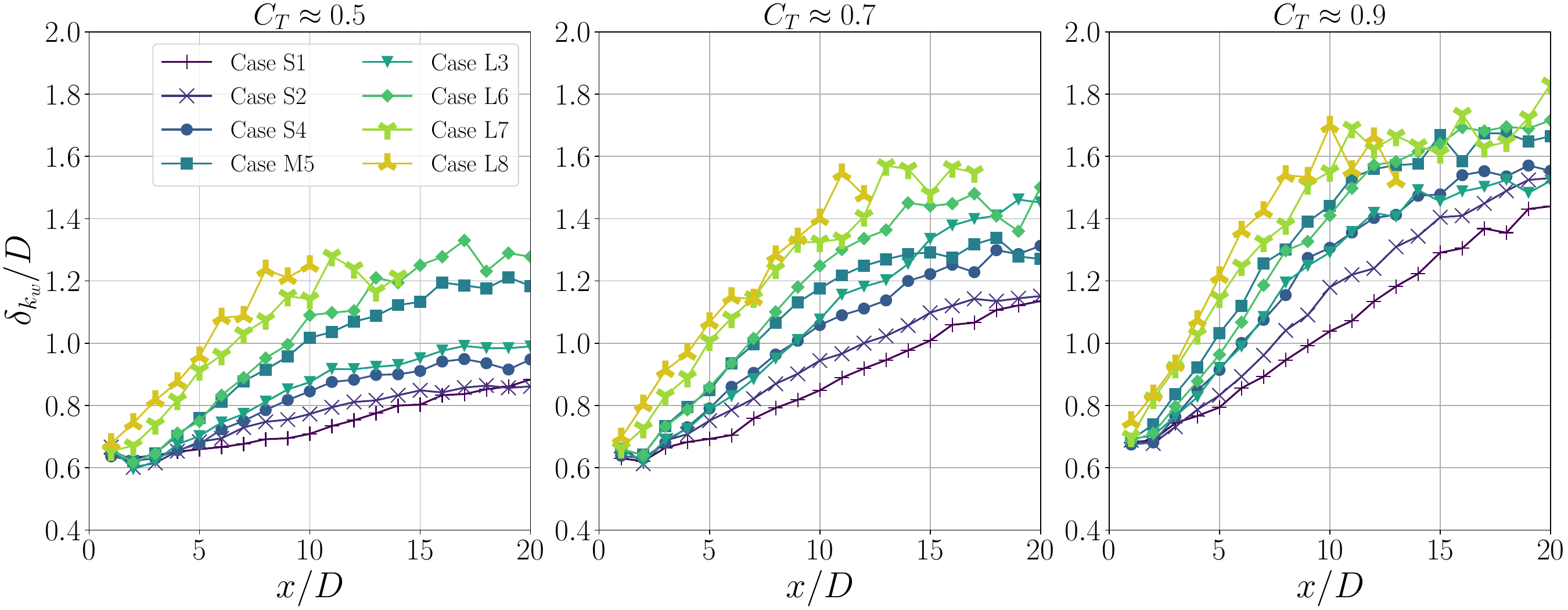}
    \caption{Wake width defined by the turbine-added turbulence kinetic energy $\Delta k$.}
    \label{Fig:wakewidth_tke}
\end{figure}

\begin{figure}
    \centering
    \begin{subfigure}[t]{0.48\linewidth}
        \centering
        \includegraphics[width=\linewidth]{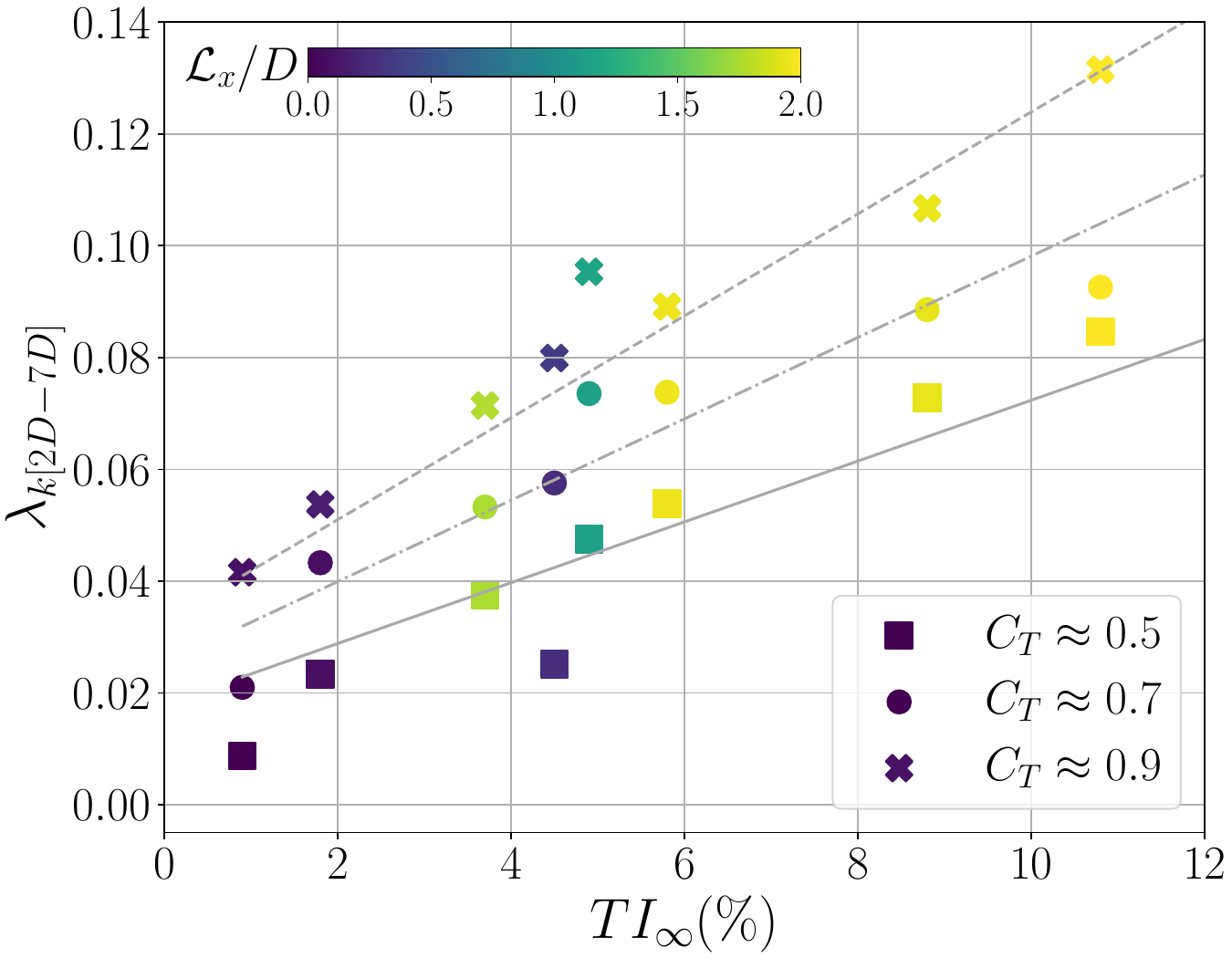}
        \caption{}
        \label{fig:wakegrowthrate_tke1}
    \end{subfigure}%
    \hfill
    \begin{subfigure}[t]{0.48\linewidth}
        \centering
        \includegraphics[width=\linewidth]{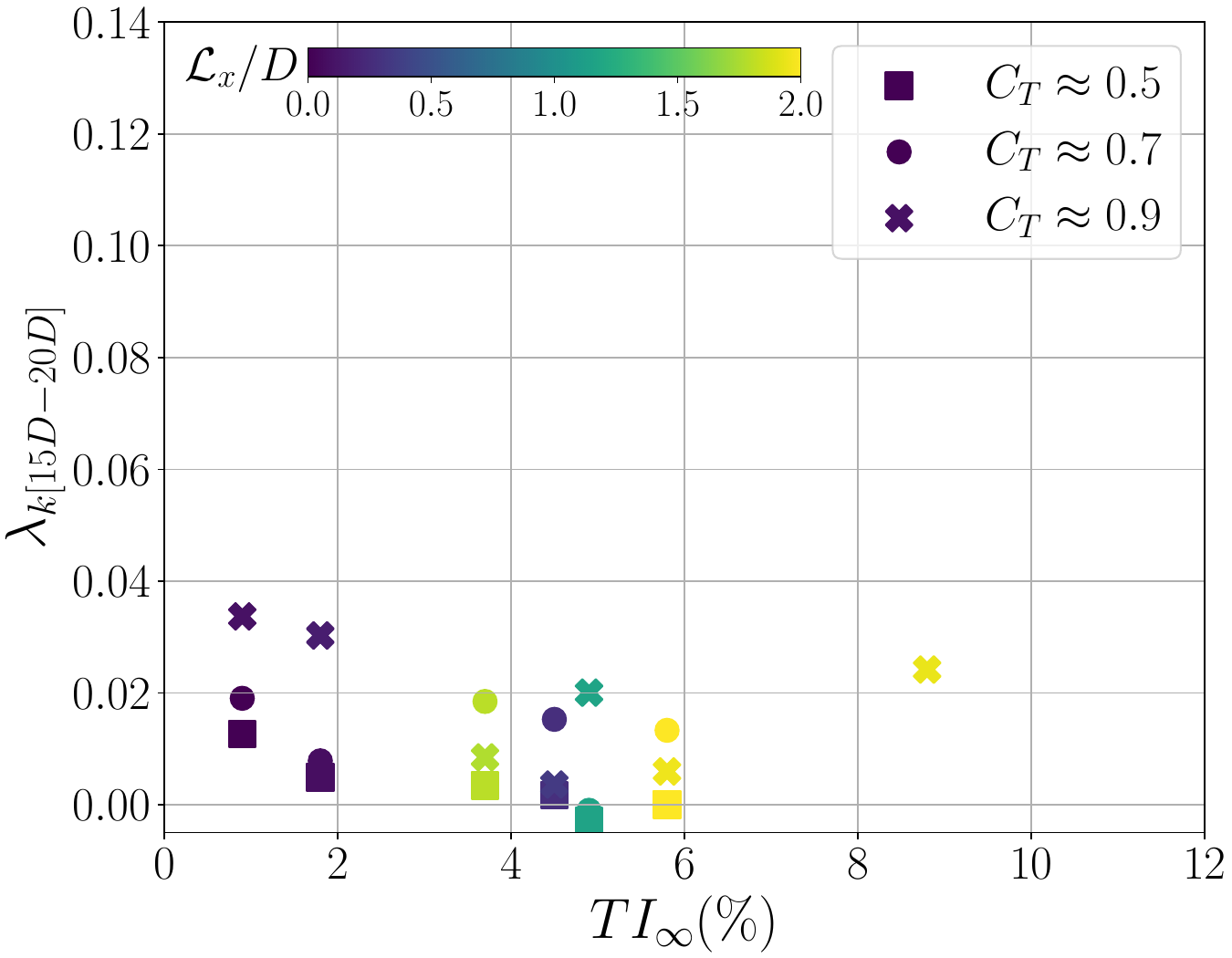}
        \caption{}
        \label{fig:wakegrowthrate_tke2}
    \end{subfigure}
    \caption{Wake growth rates calculated from $\lambda_{k} = \overline{\text{d}\delta_{k_w}/\text{d}x}$ for (\subref{fig:WakeGrowthRate1}) $2 \leq x/D \leq 7$, and (\subref{fig:WakeGrowthRate2}) $15 \leq x/D \leq 20$.}
    \label{fig:wakegrowthrate_tke}
\end{figure}

Similarly to the definition of the wake width based on the time-averaged velocity deficit $\delta$, we define a characteristic wake width $\delta_{k_w}$ based on the turbulence kinetic energy profiles. At a given streamwise location $x$, $\delta_{k_w}$ is defined as the radial location $r$ at which $\Delta k$ is $10\%$ that of the maximum. Moreover, analogous to the wake growth rate $\lambda$, we computed the expansion rate of the TKE profiles, denoted as $\lambda_k$, based on the mean gradient of $\delta_{k_w}(x)$ over two intervals: one within the near wake ($2 \leq x/D \leq 7$), and another within the far wake ($15 \leq x/D \leq 20$). Figures~\ref{Fig:wakewidth_tke} \& \ref{fig:wakegrowthrate_tke} present the streamwise evolution of the wake width, $\delta_{k_w}(x)$, and the TKE-based expansion rates, $\lambda_k$, for all 24 \{$C_T$, FST\} combinations.

At first glance, the behaviour of $\delta_{k_w}(x)$ closely mirrors that of $\delta(x)$. Specifically, $\delta_{k_w}(x)$ initially exhibits an initial near-linear increase before reaching a plateau, with larger $TI_{\infty}$ leading to a steeper change in the slope of $\delta_{k_w}(x)$. One can note that for the two highest $TI_{\infty}$ cases (L7 and L8), the wake width based on TKE cannot be defined beyond a certain position, as the profiles become nearly flat (see figure~\ref{Fig:TKEProfiles}). Similar to $\delta$, $\delta_{k_w}$ increases as both $C_T$ and $TI_{\infty}$ increase, with $\delta_{k_w,\textrm{High-$C_T$}} \geq\delta_{k_w,\textrm{Med-$C_T$}} \geq \delta_{k_w,\textrm{Low-$C_T$}}$ and $\delta_{k_w,\textrm{Group 3}} \geq \delta_{k_w,\textrm{Group 2}} \geq \delta_{k_w,\textrm{Group 1}}$, while the integral length scale ${\cal L}_x$ seems to have only a secondary influence on $\delta_{k_w}$. Moreover, the scattering of points for $\lambda_{k_w}$ in figure~\ref{fig:wakegrowthrate_tke1} closely aligns with the distribution observed for $\lambda$ in figure~\ref{fig:WakeGrowthRate1}. A fit was performed over the entire $\lambda_{k_w[2D-7D]}$ dataset using a model analogous to that employed for $\delta$, resulting in the following equation : $\lambda_{k_w} = 0.92C_TTI_{\infty} + 0.037 C_T + 0.004{\cal L}$, with a coefficient of determination of $R^2=0.931$ (see the grey lines in figure~\ref{fig:wakegrowthrate_tke}).

In all 24 combinations, the wake width defined by TKE is consistently larger than that based on the velocity deficit, $\delta_{k_w}(x) \geq \delta(x)$, and expands at a faster rate $\lambda_{k_w}(x) \geq \lambda(x)$, with more pronounced differences at high-$TI_{\infty}$ and large $C_T$. These observations, also reported in the wakes of porous discs \citep{Lingkan2023}, further emphasise the fast mixing of the TKE in the wake with the background TKE. Moreover, the deceleration in $\delta_{k_w}$ growth closely aligns with that of $\delta$, and occurs when a shallower gradient of TKE between the wake and the ambient flow is observed, highlighting the leading role of the TKE and $TI$ differences in driving wake expansion. Once the gradient of TKE becomes small, as is observed in the far wake for high-$TI_{\infty}$ cases, both the TKE and the velocity deficit expansion rates are reduced, explaining the smaller values of $\lambda_{k_w}(x)$ and $\lambda(x)$ for Group 3 FST cases compared to Group 1 FST cases for $15 \leq x/D \leq 20$ (figure~\ref{fig:wakegrowthrate_tke2}). Hence, the presence of background turbulence intensity accelerates the recovery of both TKE and velocity; however, TKE recovers more rapidly, eventually reaching a quasi-uniform state. Once TKE becomes uniform, the recovery of the velocity deficit slows significantly as it propagates through a region of nearly homogeneous TKE, reducing momentum and mass transfer from the background flow into the wake and limiting further wake recovery.

\subsection{Evolution of wake-averaged quantities \label{subsec:WakeAveraged}}

\subsubsection{Wake velocity recovery}

\begin{figure}
    \centering
    \begin{subfigure}{0.9\textwidth}
         \centering
            \includegraphics[width=\linewidth]{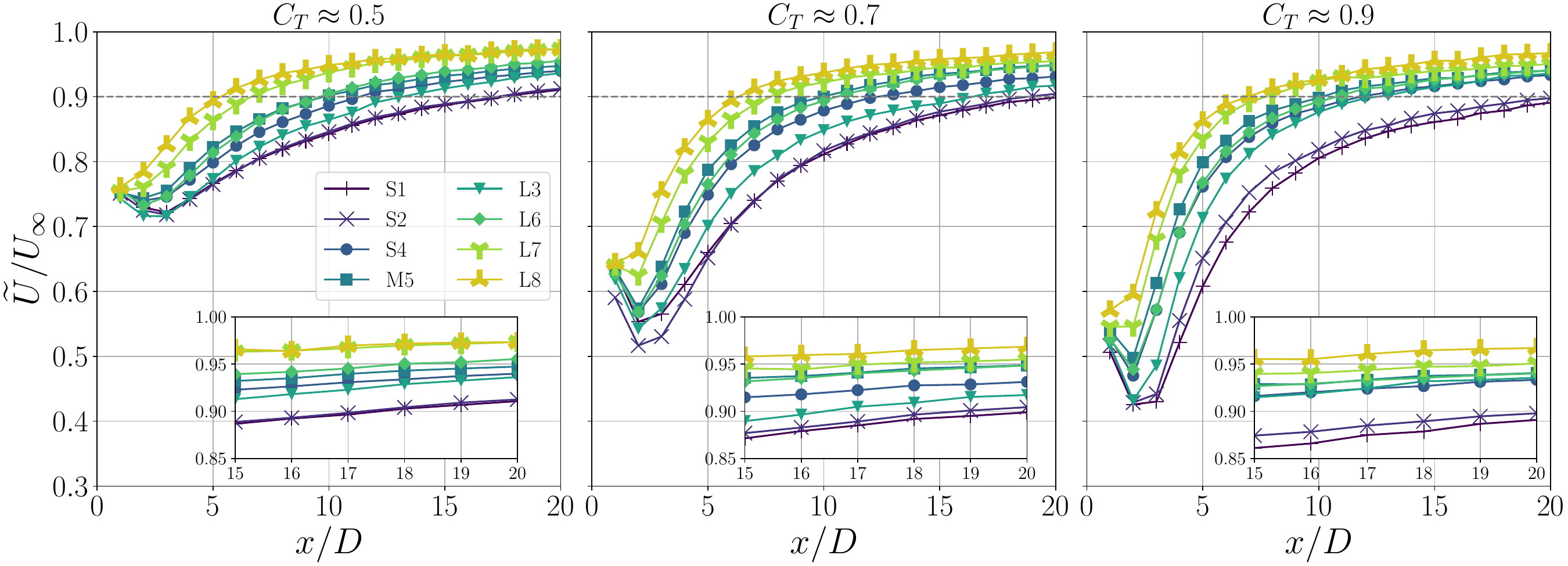}
            \subcaption{}
            \label{fig:velocity_tilde}
    \end{subfigure} \\
    \begin{subfigure}{0.9\textwidth}
         \centering
            \includegraphics[width=\linewidth]{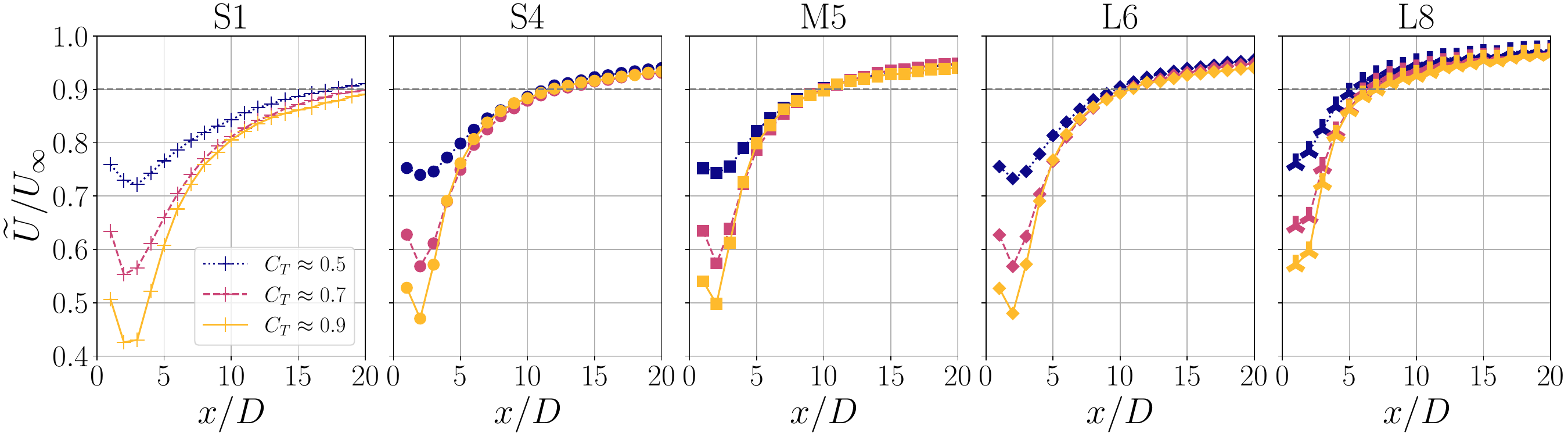}
            \subcaption{}
             \label{fig:velocity_tilde2}
    \end{subfigure} \\
     \begin{subfigure}{0.9\textwidth}
         \centering
            \includegraphics[width=\linewidth]{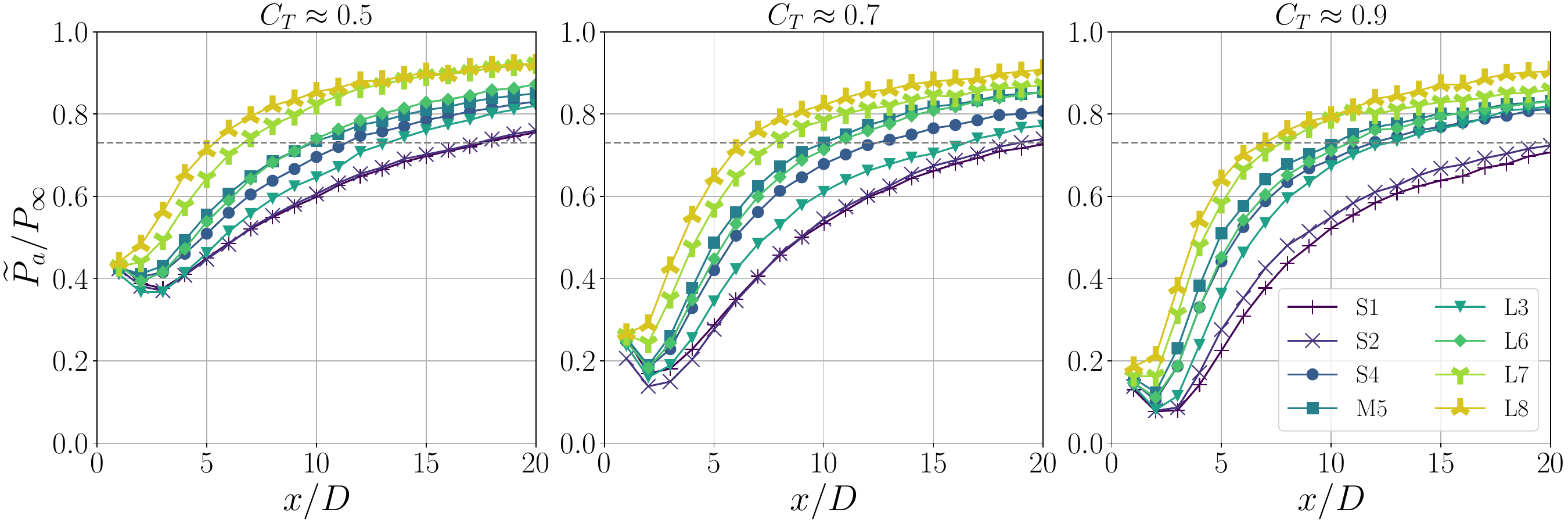}
            \subcaption{}
             \label{fig:puissance_tilde}
    \end{subfigure} \\
    \begin{subfigure}{0.9\textwidth}
         \centering
            \includegraphics[width=\linewidth]{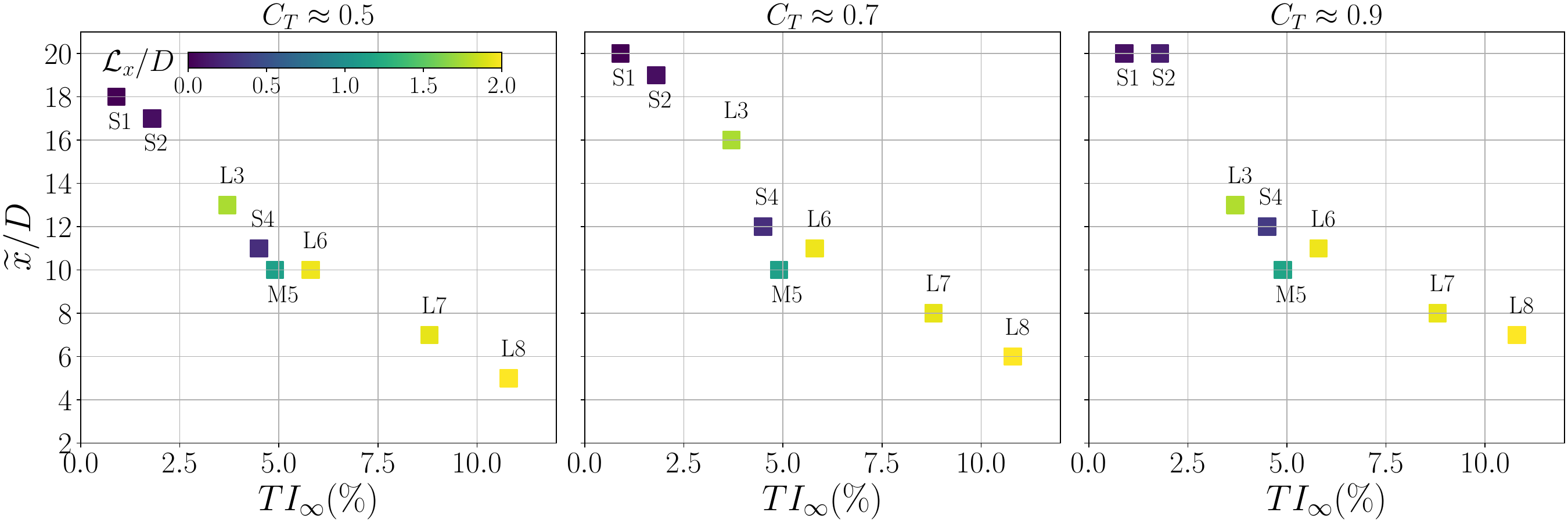}
            \subcaption{}
            \label{fig:x_tilde}
    \end{subfigure} 
    \caption{Streamwise evolution of the wake velocity recovery, $\widetilde{U}/U_{\infty}$ ((\subref{fig:velocity_tilde}) \& (\subref{fig:velocity_tilde2})), and wake-averaged available power, $\widetilde{P}_a/P_{\infty}$ (\subref{fig:puissance_tilde}). A focus is placed on the influence of FST on $\widetilde{U}$ in panel  (\subref{fig:velocity_tilde}), and of $C_T$ in panel (\subref{fig:velocity_tilde2}). Panel (\subref{fig:x_tilde}) shows the streamwise distance, $\widetilde{x}$, required for the wake to recover to $\widetilde{U} = 0.9 U_{\infty}$, for all $\{C_T, \textrm{FST}\}$ combinations.}
    \label{fig:Mean_Tilde}
\end{figure}

The influence of FST on the wake velocity recovery is assessed similarly to \cite{Li2022,Messmer2024,Wei2024}, by averaging radially the time-averaged velocity from the centre of the wake to the wake width $\delta(x)$, \emph{i.e} $\widetilde{U}(x) = \langle \overline{U}(x,y) \rangle_{\delta(x)}$. $\widetilde{U}(x)$ is the mean time-averaged velocity within a streamtube of width $\delta(x)$, and serves as an indicator of wake recovery, with $\widetilde{U}/U_{\infty}$ approaching 1 indicating a higher degree of velocity recovery. $\widetilde{U}(x)$ is also directly related to the available power for a virtual downstream turbine operating within the wake of an upstream turbine, as $\widetilde{P}_a/P_{\infty} \propto (\widetilde{U}/U_{\infty})^3$. Analogously to the parameter $\overline{L}$ introduced by \cite{Wei2024} to characterise the streamwise extent of the wake based on $\widetilde{U}$, we define the parameter $\widetilde{x}$ as the distance required for $\widetilde{U}$ to recover to $\widetilde{U}(\widetilde{x}) = 0.9U_{\infty}$, \emph{i.e.} for the available power to recover to $\widetilde{P}_a(\widetilde{x}) = 0.9^3P_{\infty} = 72\% P_{\infty}$. These quantities are of particular interest for optimising the positioning of wind turbines within a wind farm, as the power generated by a turbine located downstream of others is given $P \propto C_p\widetilde{P}_a$ with $C_p \leq 0.59$ \citep{Sorensen2016}. Therefore, having a good understanding of the distance inter-turbines required to maintain sufficiently high available power for downstream turbines -- depending on the upstream turbine $C_T$, and the FST conditions-- is essential to optimising turbine placement and control strategies. 

Figure~\ref{fig:Mean_Tilde} shows the streamwise evolution of $\widetilde{U}$, $\widetilde{P}_a$ and $\widetilde{x}$ for all 24 \{$C_T$, FST\} combinations. As expected, higher momentum is extracted from the flow as $C_T$ increases, leading then to lower wake-averaged velocity  $\widetilde{U}$ (figures~\ref{fig:velocity_tilde} \&\ref{fig:velocity_tilde2}), and lower wake-averaged available mechanical power $\widetilde{P}_a$ immediately downstream of the turbine (figure~\ref{fig:puissance_tilde}). As the wake develops, high-momentum fluid from the surrounding flow is entrained and mixes with the low-momentum flow within the wake, causing $\widetilde{U}$ and $\widetilde{P}_a$ to approach the freestream wind velocity $U_{\infty}$ and mechanical power $P_{\infty}$. In the following analysis, we will focus on the streamwise evolution of  $\widetilde{U}$ for convenience, though all observations made for $\widetilde{U}$ equally apply to $\widetilde{P}_a$, with the effects being more pronounced for the latter due to the cubic relationship.

Firstly, for all three $C_T$, the wake-averaged velocity initially increases rapidly in the near wake  ($3 \lesssim x/D \lesssim 7$) before transitioning to a slower recovery rate as the wake develops. While, the recovery in the near wake is intense for high-$TI_{\infty}$ cases (Group 3), as evidenced by the steep slopes of  $\widetilde{U}$ and $\widetilde{P}$, low-$TI_{\infty}$ cases (Group 1) exhibit a more gradual wake recovery over the measurement domain. For both Group 1 and Group 2 cases, an initial reduction in $\widetilde{U}/U_{\infty}$ is observed for $x/D \leq 2 $, which is related to pressure recovery. Similarly to porous discs, as $C_T$ increases, flow blockage by the turbine intensifies, leading to a greater pressure drop and a more significant reduction in  $\widetilde{U}/U_{\infty}$. As $TI_{\infty}$ increases, this reduction progressively diminishes, indicating a faster pressure recovery due to enhanced momentum mixing just behind the turbine. 

For all $C_T$, the main differences between the FST cases are generated within the near wake and then propagate throughout the entire measurement range. The location of the maximum difference in $\widetilde{U}/U_{\infty}$ between the lowest-$TI_{\infty}$ (S1) and highest-$TI_{\infty}$ (L8) cases moves closer to the turbine as $C_T$ increases, occurring at $x/D = 6$ for $C_T\approx0.5$, $x/D = 4$ for $C_T\approx0.7$,  and $x/D = 3$ for $C_T\approx0.9$, with these differences progressively diminishing downstream. At the farthest measurement station, greater differences are observed between the FST cases at $C_T \approx 0.9$. As $TI_{\infty}$ increases, the recovery of $\widetilde{U}/U_{\infty}$ in the near wake is enhanced with a clear difference between Group 1 FST cases (low-$TI_{\infty}$ cases), Group 2 cases (medium-$TI_{\infty}$ cases) and Group 3 (high-$TI_{\infty}$ cases). Indeed, for all $C_T$ and $x/D$, $(\widetilde{U}/U_{\infty})_{\text{Group 3}} \geq (\widetilde{U}/U_{\infty})_{\text{Group 2}} \geq (\widetilde{U}/U_{\infty})_{\text{Group 1}}$. However, when focusing more in detail on Group 2 cases, the evolution of $\widetilde{U}/U_{\infty}$ doesn't follow the hierarchical order of $TI_{\infty}$. Specifically, for all three $C_T$, there exists a streamwise range where $(\widetilde{U}/U_{\infty})_{\text{M5}} \geq (\widetilde{U}/U_{\infty})_{\text{L6}}$. At $C_T \approx 0.5$, $(\widetilde{U}/U_{\infty})_{\text{M5}} \geq (\widetilde{U}/U_{\infty})_{\text{L6}}$ up to $x/D \approx 10$, whereas at $C_T \approx 0.7$ and $C_T \approx 0.9$  this relationship persists up to $x/D \approx 15$, beyond which the two curves converge and become indistinguishable. In addition, for $C_T\approx0.9$, it is observed that  $(\widetilde{U}/U_{\infty})_{\text{L6}}$ is nearly identical to $(\widetilde{U}/U_{\infty})_{\text{S4}}$ in the near wake. Given that $ {\cal L}_{x,\text{L6}} \geq {\cal L}_{x,\text{M5}} \geq {\cal L}_{x,\text{S4}} $, the preceding comparisons suggest that the integral length scale ${\cal L}_x$, has an effect opposite to that of  $TI_{\infty}$, namely it slows down the recovery of the velocity deficit in the near wake. Focusing now on the FST case L3, it is noteworthy that in the near wake, $(\widetilde{U}/U_{\infty})_{\text{L3}}$ is closer to that of Group 1 FST cases (S2) than to Group 2 FST cases (S4), despite the fact that, \emph{a priori}, $TI_{\infty,\mathrm{L3}}$ is closer to $TI_{\infty,\mathrm{S4}}$ than $TI_{\infty,\mathrm{S2}}$. This observation further supports the observation that in the near wake ${\cal L}_x$ slows down the velocity recovery, though its influence is secondary to that of $TI_{\infty}$ and $C_T$. Moreover, it aligns with the findings of \cite{Gambuzza2022, Hodgson2022,Hodgson2023}, who showed that while higher FST intensity enhances velocity recovery, the FST integral length scale tends to inhibit wake recovery in the near wake.

\begin{table}
    \centering
\begin{tabularx}{\textwidth}{p{1.7cm}|| C C C}   
 &$C_T \approx 0.5$& $C_T \approx 0.7$ & $C_T \approx 0.9$ \\
$5D$ &  58\% &  124\% & 185\% \\ 
$10D$ & 42\% &  54\% &  52\% \\
$15D$ & 29\%&   33\% &  37\% \\
$20D$ & 22\%&   25\% &  28\% \\
        \end{tabularx}
        \caption{Quantification of wake-recovery enhancement due to FST, expressed as the percentage increase in available power in the wake, $\widetilde{P}_{a}$, at $x/D=[5 ,10,15,20]$, through a comparison of FST cases S1 and L8: $(\widetilde{P}_{a,L8} -\widetilde{P}_{a,S1})/\widetilde{P}_{a,S1}(\%)$. }
        \label{tab:power_percent}
\end{table}

Figure~\ref{fig:velocity_tilde2} highlights the influence of $C_T$ across 5 different FST cases. Greater differences in $(\widetilde{U}/U_{\infty})$ between the three $C_T$-cases are observed both in the near wake and at the last measurement station as $TI_{\infty}$ decreases. An increase in $TI_{\infty}$ mitigates the discrepancies observed in the wake-averaged velocity resulting from changes in the turbine's operating point, whereas varying ${\cal L}_x$ appears to have a negligible effect (see FST cases S4, M5 and L6 in figure~\ref{fig:velocity_tilde2}). In the near wake, the recovery slope of $\widetilde{U}$ is strongly influenced by $C_T$. However, further downstream, $TI_{\infty}$ has a greater impact on $\widetilde{U}$ than $C_T$, as evidenced by the similar slopes of $\widetilde{U}$ across the three $C_T$ values for all FST cases. This suggests that while the influence of $C_T$ is primarily confined to the near wake, FST exerts a prolonged effect on the wake evolution.

As shown in figure~\ref{fig:x_tilde}, the wake length variable $\widetilde{x}$ diminishes dramatically as $TI_{\infty}$ increases. Even though the wake length slightly increases with $C_T$, $\widetilde{x}$ is significantly more sensitive to variations in $TI_{\infty}$ than to variations in $C_T$ or ${\cal L}_x$. For the wind turbine operating at $C_T \approx 0.9$ in the lowest $TI_{\infty}$ case (S1), even after 20D, the wake-averaged available power has not yet reached 72\% of the freestream available power $P_{\infty}$. In contrast, for the highest turbulence intensity case (L8), $\widetilde{x} \leq 7$ for all values of $C_T$.  This underscores that when the background flow is highly turbulent, the role of $C_T$ in driving the velocity recovery process and homogenising the velocity field becomes secondary, whereas in the case of a weak turbulent background, the imprint left by $C_T$ on the wake persists. 

This faster wake recovery is also reflected in the wake-averaged available power $\widetilde{P}_a$. Table~\ref{tab:power_percent} shows the percentage increase in $\widetilde{P}_a$ at various streamwise positions due to higher levels of $TI_{\infty}$, with a comparison between cases S1 and L8. At 5D, $\widetilde{P}_a$ is enhanced by 185\% for L8 compared to S1 when the turbine operates at $C_T=0.9$. Although the difference decreases as $C_T$ decreases, it remains significant. Even at the furthest measurement station ($x/D = 20$), notable differences in $\widetilde{x}$, $\widetilde{U}$, and $\widetilde{P}_a$ between the FST cases persist, although they tend to diminish. Therefore, an inflow with high levels of turbulence intensity appears advantageous, as $C_T$ seems unaffected by $TI_{\infty}$ (\emph{cf.} table~\ref{Table:CT}), and it allows the wake-averaged quantities to recover more quickly, resulting in larger available power for turbines operating within the wake of an upstream turbine.

\subsubsection{Wake TKE recovery}

\begin{figure}
    \centering
    \begin{subfigure}{0.9\textwidth}
         \centering
            \includegraphics[width=\linewidth]{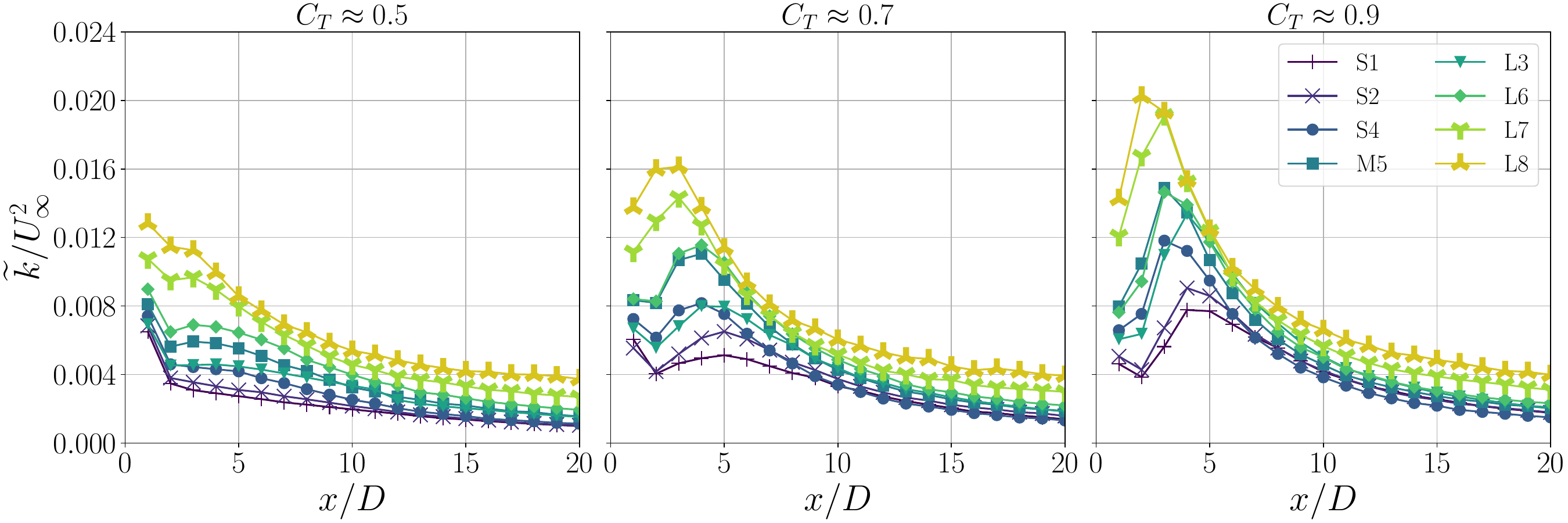}
            \subcaption{\label{fig:ktilde}}
    \end{subfigure} \\
     \begin{subfigure}{0.9\textwidth}
         \centering
            \includegraphics[width=\linewidth]{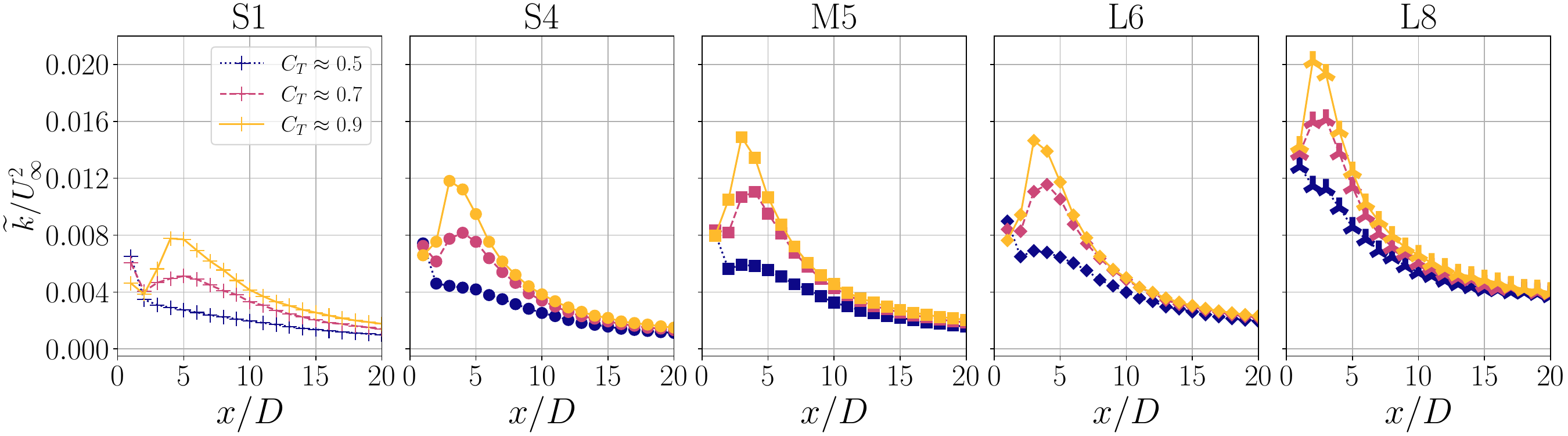}
            \subcaption{\label{fig:ktildeCT}}
    \end{subfigure} \\
    \begin{subfigure}{0.9\textwidth}
         \centering
            \includegraphics[width=\linewidth]{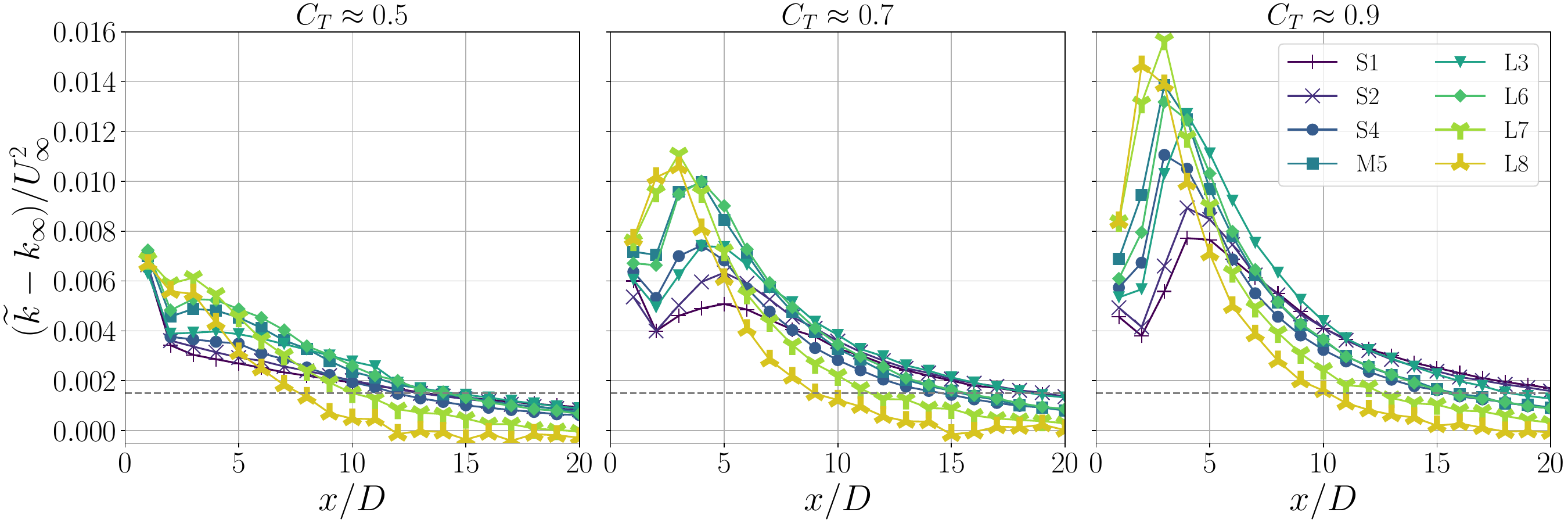}
            \subcaption{\label{fig:Deltaktilde}}
    \end{subfigure} \\
    \begin{subfigure}{0.9\textwidth}
         \centering
            \includegraphics[width=\linewidth]{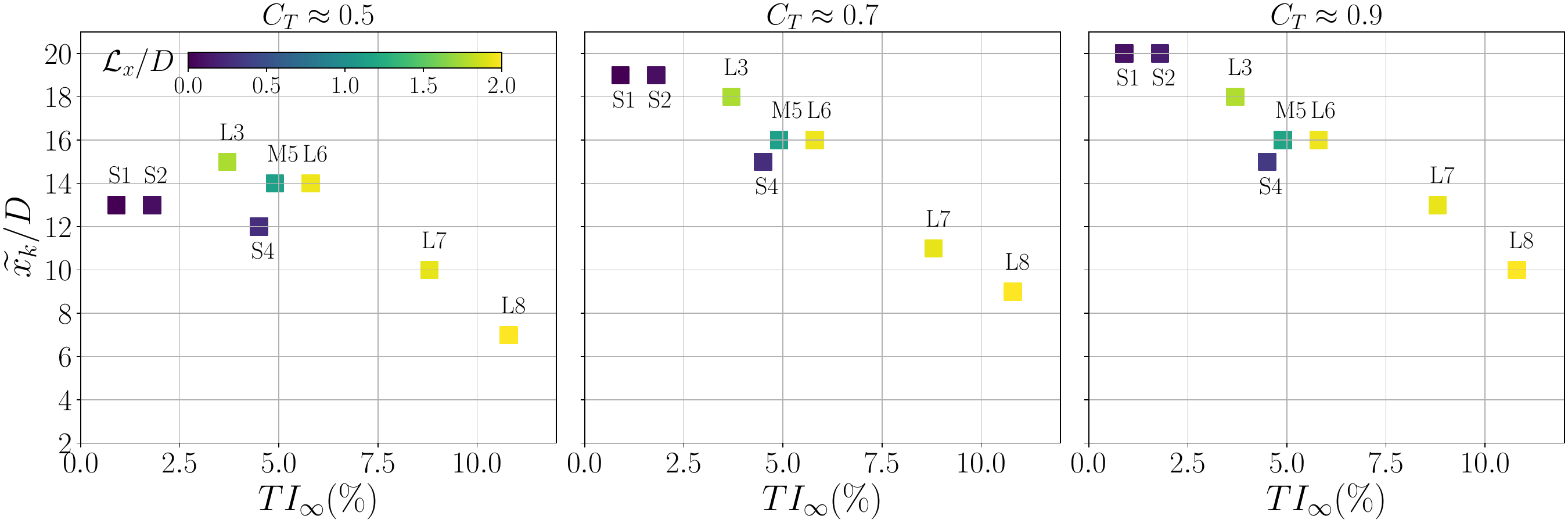}
            \subcaption{\label{fig:xtilde_k}}
    \end{subfigure}
    \caption{Streamwise evolution of the wake-averaged TKE, $\widetilde{k}/U_{\infty}^2$ ((\subref{fig:ktilde}) \& (\subref{fig:ktildeCT})), and turbine-added wake-averaged TKE, $(\widetilde{k} - k_{\infty})/U_{\infty}^2$ (\subref{fig:Deltaktilde}). A focus is placed on the influence of FST on $\widetilde{k}/U_{\infty}^2$ in panel  (\subref{fig:ktilde}), and of $C_T$ in panel (\subref{fig:ktildeCT}). Panel (\subref{fig:xtilde_k}) shows the streamwise distance, $\widetilde{x}_k$, required for the wale to recover to $\Delta \widetilde{k}(\widetilde{x}_k)/U_{\infty}^2  = 0.0015$, for all $\{C_T, \textrm{FST}\}$ combinations.}
    \label{fig:TKE_tilde}
\end{figure}

Similarly to $\widetilde{U}$, we define the wake-averaged turbulence kinetic energy  $\widetilde{k}(x) = \langle \overline{k}(x,y) \rangle_{\delta(x)}$. Since the background TKE, $k_{\infty}$, varies across the FST cases, the recovery of the TKE within the wakes is assessed using the turbine-induced wake-averaged TKE, $\Delta \widetilde{k}/U_{\infty}^2=(\widetilde{k} - k_{\infty})/U_{\infty}^{2}$, with $\Delta \widetilde{k}$ approaching 0 indicating a higher degree of wake-TKE recovery. We also introduce $\widetilde{x}_k$, the TKE-based wake length, defined as the streamwise distance at which $\Delta \widetilde{k}(\widetilde{x}_k)/U_{\infty}^2  = 0.0015$. For reference, this threshold corresponds to a difference of approximately 4\% between the background turbulence intensity and the wake-averaged turbulence intensity. Figure~\ref{fig:TKE_tilde} presents the streamwise evolution of  $\widetilde{k}(x)/U_{\infty}^2$ and $\Delta \widetilde{k}(x)=(\widetilde{k} - k_{\infty})/U_{\infty}^{2}$, along with the TKE-based wake length $\widetilde{x}_k$  for all 24 $\{C_T, \text{FST}\}$ combinations.

Firstly, both $\widetilde{k}$ and $\Delta \widetilde{k}$ initially increase as $C_T$ increases; however, some differences are observed in their evolution further downstream. For $C_T = 0.5$, both quantities begin to decay immediately downstream of the turbine. In contrast, for $C_T = 0.7$ and $C_T = 0.9$, $\widetilde{k}$ and $\Delta{\widetilde{k}}$ initially increase before decaying in all FST cases, likely due to the stronger tip-shear layer, which enhances turbulence production behind the rotor  (see figure~\ref{Fig:TKEProfiles}). As $C_T$ and $TI_{\infty}$ increase, the streamwise location where $\widetilde{k}$ and $\Delta \widetilde{k}$ begin to decay shifts closer to the turbine, reflecting the reduced persistence of the shear layer and the faster onset of wake recovery. Additionally, in case L3, the decay of $\widetilde{k}$ and $\Delta \widetilde{k}$ occurs further downstream compared to cases with similar $TI_{\infty}$ but smaller ${\cal L}_x$ (S4 and M5), suggesting that inflows with larger ${\cal L}_x$ delay the onset of the TKE recovery. Larger differences in $\widetilde{k}$ between the different $C_T$ cases are observed at the farthest measurement stations for low-$TI_{\infty}$ cases (figure~\ref{fig:ktildeCT}), highlighting the prolonged influence of the turbine operating point on the TKE recovery in weakly turbulent ambient conditions.

For all $C_T$, the maximum $\Delta \widetilde{k}$ follows the order: $\text{max}(\Delta\widetilde{k}/U_{\infty}^2)_{\text{Group 3 }} \geq \text{max}(\Delta\widetilde{k}/U_{\infty}^2)_{\text{Group 2 }} \geq \text{max}(\Delta\widetilde{k}/U_{\infty}^2)_{\text{Group 1}}$, although it does not strictly follow the hierarchical order of increasing $TI_{\infty}$, due to the secondary influence of $C_T$ and ${\cal L}_x$. The recovery rate for $\Delta \widetilde{k}$ is largely a function of the ambient turbulence intensity $TI_{\infty}$ (\emph{i.e} the ambient TKE $k_{\infty}$), with the strongest recovery seen for case L8, and the weakest for case S1. As previously observed in the TKE profiles, $\Delta \widetilde{k}\approx0$ for case L8 at $x/D \gtrsim 15$ across all three $C_T$ cases, driven by enhanced mixing resulting from both higher ambient turbulence and a greater initial $\Delta \widetilde{k}$. Focusing on L\# cases, we observe that for all three $C_T$, after a few diameters downstream, $(\Delta\widetilde{k}/U_{\infty}^2)$ follows a hierarchical order corresponding to decreasing $TI_{\infty}$, \emph{i.e}  $(\Delta\widetilde{k}/U_{\infty}^2)_{\text{L8 }} \geq (\Delta\widetilde{k}/U_{\infty}^2)_{\text{L7}} \geq (\Delta\widetilde{k}/U_{\infty}^2)_{\text{L6}} \geq (\Delta\widetilde{k}/U_{\infty}^2)_{\text{L3}}$. A similar trend is observed for S\# cases, where $(\Delta\widetilde{k}/U_{\infty}^2)_{\text{S4 }} \geq (\Delta\widetilde{k}/U_{\infty}^2)_{\text{S1,S2}}$ at a certain location downstream of the turbine. These observations clearly illustrate that $TI_{\infty}$ enhances the recovery of the turbine-induced TKE within the wake. This effect becomes more pronounced at higher $C_T$, where the increased initial $\Delta \widetilde{k}$ accelerates TKE homogenisation, resulting in the complete dissipation of the TKE-based wake for the highest $TI_{\infty}$ case. However, the recovery rate also depends on the local turbine-induced TKE $\Delta \widetilde{k}$, which varies as the wake expands, thereby modifying the local recovery rate. Hence, while the TKE recovery rate is hastened in the early wake for high $TI_{\infty}$ and high $C_T$ cases due to both higher $k_{\infty}$ and higher $\Delta \widetilde{k}$ immediately downstream of the turbine, the initial recovery rate for weakly turbulent background and low $C_T$ cases is diminished due to lower $k_{\infty}$ and $\Delta \widetilde{k}$, resulting in a more gradual recovery.  

Focusing on Group 2 FST cases -- specifically cases L6 ({\markerdiamond{viridis5}}) , M5 (\textcolor{viridis3}{\(\blacksquare\)}), and S4 (\markercircle{viridis2}) --it can be observed that for all $C_T$, $(\Delta\widetilde{k}/U_{\infty}^2)_{\text{L6 }} \geq (\Delta\widetilde{k}/U_{\infty}^2)_{\text{M5}} \geq (\Delta\widetilde{k}/U_{\infty}^2)_{\text{S4}}$ for $x/D \gtrsim 5$ (figure~\ref{fig:Deltaktilde}). This trend may seem counter-intuitive, as one might expect the opposite given the earlier conclusion that higher $TI_{\infty}$ promotes a faster decay of the turbine-induced TKE. However, given that ${\cal L}_{x,\text{L6} } \approx 2{\cal L}_{x,\text{M5}}\approx 4{\cal L}_{x,\text{S4}}$, this suggests that for inflows with nearly identical $TI_{\infty}$, a larger turbulence integral length scale ${\cal L}_x$ slows the recovery of the TKE in the near wake. Additionally, for $C_T \approx 0.9$, the largest $\Delta\widetilde{k}/U_{\infty}^2$ within the range $5 \leq x/D \leq 11$ is observed for the FST case L3, characterised by a large ${\cal L}_x$ and a small $TI_{\infty}$. This is also highlighted in figure~\ref{fig:ktilde}, where at $C_T\approx 0.5$,  the wake-averaged TKE, $\widetilde{k}$, does not strictly follow the hierarchical order of increasing $TI_{\infty}$ for group 2 FST cases. Specifically, when comparing cases L3 ({\markercross{viridis4}}) and S4 (\markercircle{viridis2}), it is seen that $(\widetilde{k}/U_{\infty}^2)_{\text{L3 }} \geq (\widetilde{k}/U_{\infty}^2)_{\text{S4 }} $, particularly for $x/D \leq 10$, despite $TI_{\infty,\text{L3}} \leq TI_{\infty,\text{S4}}$. Given that ${\cal L}_{x,\text{L3} } \approx 4{\cal L}_{x,\text{S4}}$, this further reinforces the role of a larger turbulence integral length scale ${\cal L}_x$ in slowing the mixing of the TKE and its recovery in the near wake. 

Finally, when comparing the near wake length, $\widetilde{x}_k$ (figure~\ref{fig:xtilde_k}), we observe that as $C_T$ increases, the near wake length also increases, primarily due to the greater turbine-induced TKE, which requires a longer distance for recovery. Conversely, the near wake length decreases significantly as $TI_{\infty}$ increases. Comparing Group 2 cases (L3, S4, M5, and L6), it appears that increasing the turbulence integral length scale increases the near wake length. Indeed, for all $C_T$, $\widetilde{x}_{k,\textrm{S4}}$ is smaller than that of the other cases in the group, while $\widetilde{x}_{k,\textrm{L3}}$ is higher. In fact, $\widetilde{x}_{k,\textrm{L3}}$ is also greater than $\widetilde{x}_{k,\textrm{S1}}$ and $\widetilde{x}_{k,\textrm{S2}}$ at $C_T \approx 0.5$ and $C_T \approx 0.7$. This further emphasises that inflows with large integral length scale slows down wake recovery,

\begin{figure}
    \centering
    \includegraphics[width=0.6\linewidth]{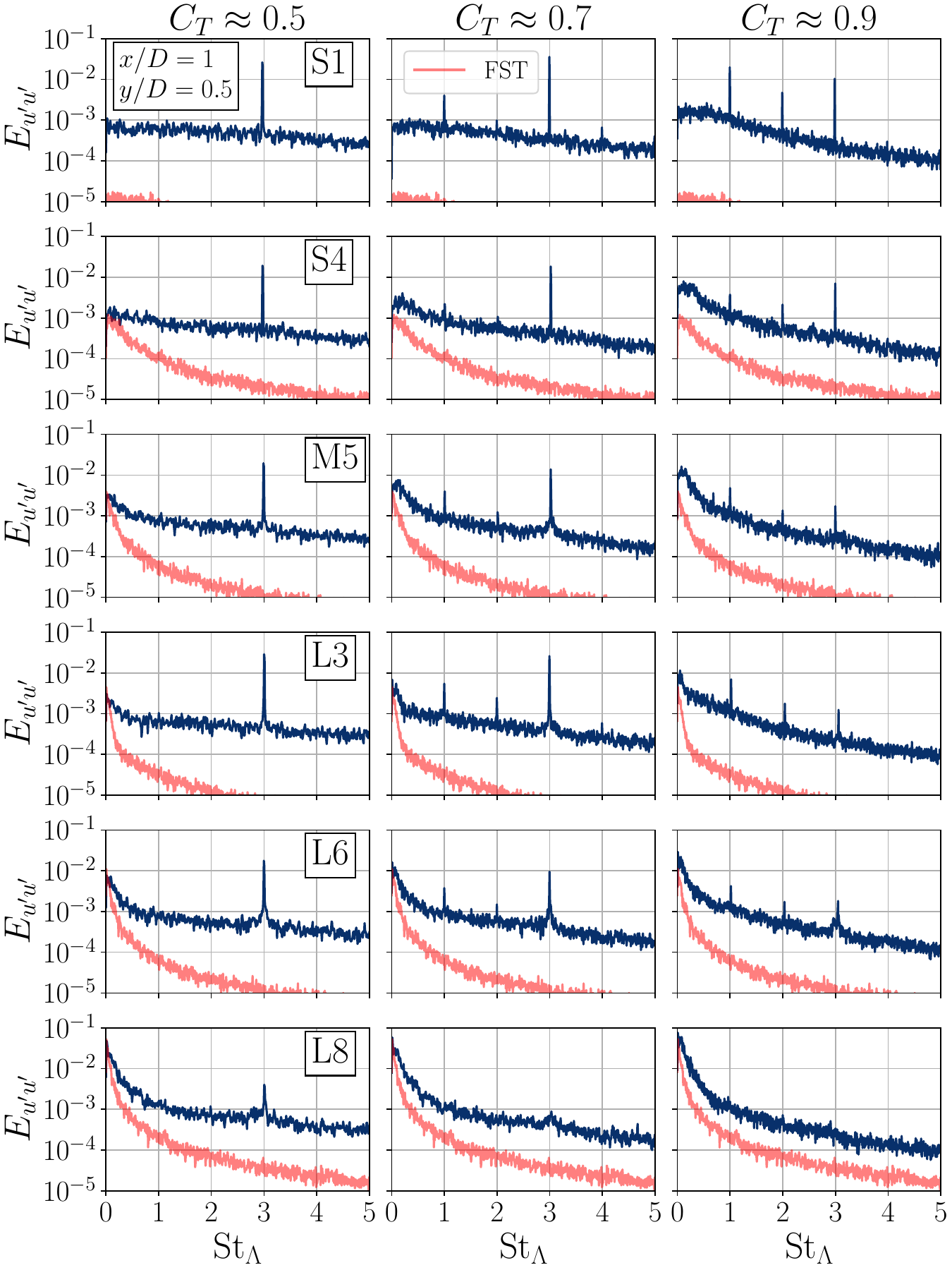}
    \caption{Power spectra of the fluctuating velocity $E_{u'u'}$ as a function of the Strouhal number $\textrm{St}_{\Lambda} = fL_c/U_{\infty} = f/f_r$. Spectra are computed in the tip-shear layer and at the first measurement station, $\{x/D,y/D\} = \{1,0.5\}$. FST power spectra, measured without the turbine at $\{x/D , y/D\} = \{0,0\}$, are shown in {\textcolor{red}{red}}. }
    \label{fig:PSD_XD1_YD0p5}
\end{figure}

\begin{figure}
    \centering
    \begin{subfigure}{0.49\textwidth}
         \centering
            \includegraphics[width=\linewidth]{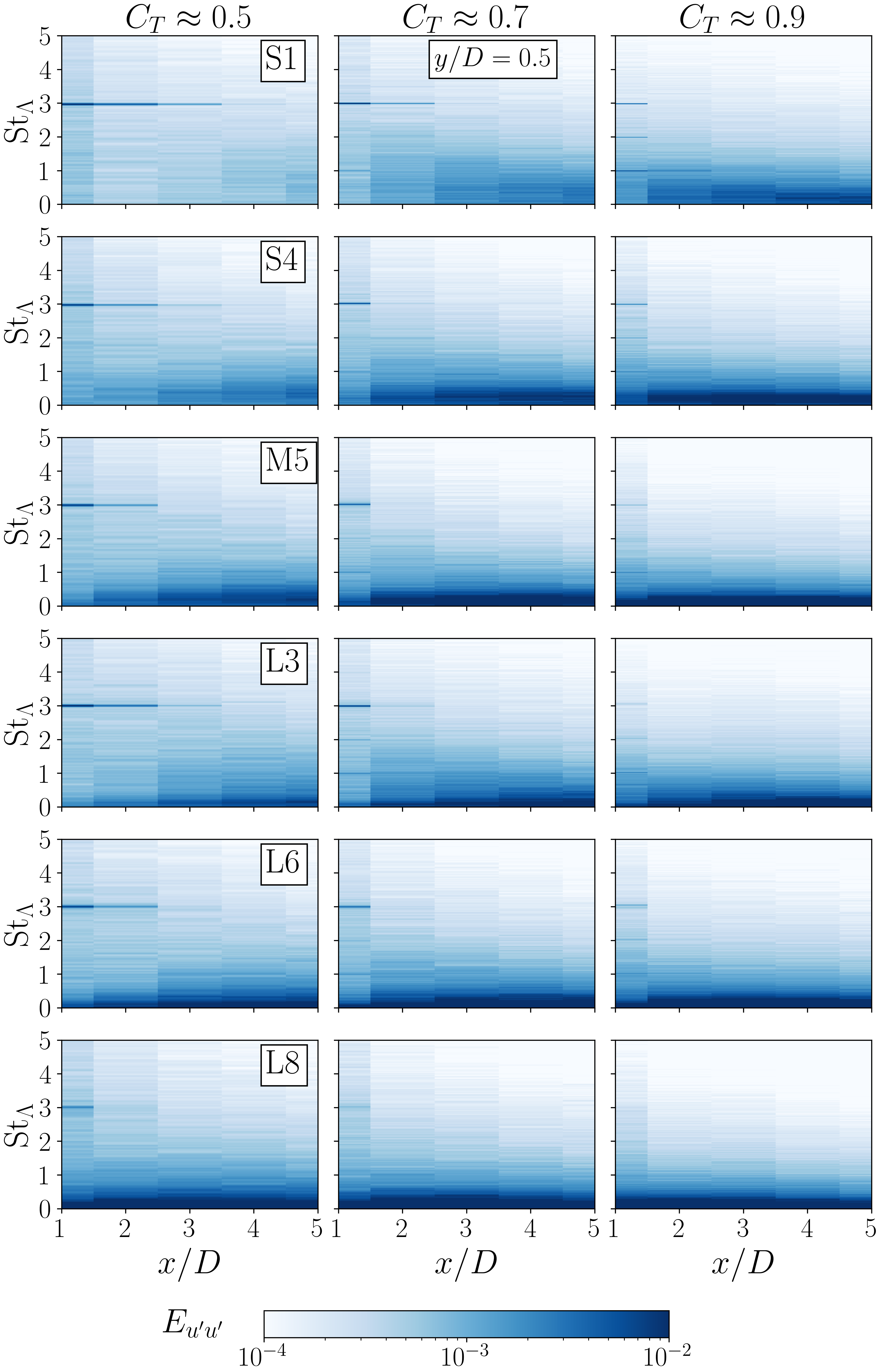}
            \subcaption{\label{fig:PSD_XD_yD0p5}}
    \end{subfigure} 
     \begin{subfigure}{0.49\textwidth}
         \centering
            \includegraphics[width=\linewidth]{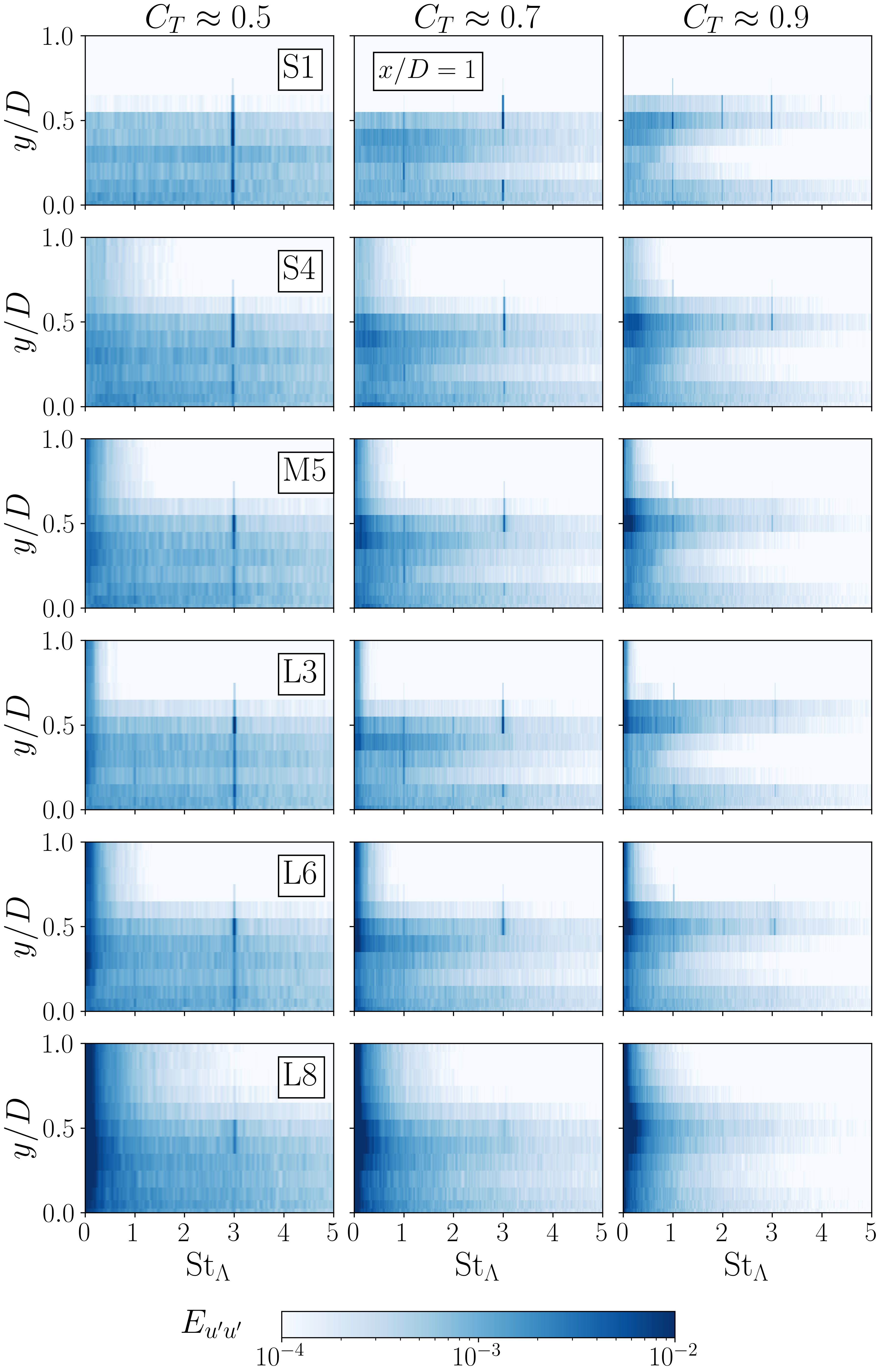}
            \subcaption{\label{fig:PSD_XD1_yD}}
    \end{subfigure}
    \caption{Spectrograms of the fluctuating velocity computed at $y/D = 0.5$ for $ 1 \leq x/D \leq 5$ (\subref{fig:PSD_XD_yD0p5}), and at $x/D = 1$ for $0 \leq y/D \leq 1$ (\subref{fig:PSD_XD1_yD}). The Strouhal number $\textrm{St}_{\Lambda}$ used is based on $L_c$. }
    \label{fig:spectrograms_NearWake}
\end{figure}

\subsection{Near wake dynamics \label{subsec:near wake dynamics}}

To gain further insight into the near wake dynamics, we computed the power spectra of the fluctuating velocity, denoted here as $E_{u'u'}$. Close to the turbine, certain dominant structures are expected to exhibit dynamics related to one of the turbine scales, such as its tip-speed ratio $\Lambda$. Therefore, instead of using the classical definition of the Strouhal number based on the turbine diameter $\textrm{St}_D = fD/U_{\infty}$, we adopt an alternative formulation, $\text{St}_{\Lambda} = fL_c/U_{\infty}$, where $L_c = \pi D/\Lambda$ \citep{Biswas2024}. The length scale $L_c$ can be interpreted as the distance travelled by a fluid particle at the freestream velocity over the time required for the turbine to complete a full rotation. Therefore, $\textrm{St}_{\Lambda} = f/f_r$, where $f_r$ is the rotor rotational frequency, and the two expressions for the Strouhal number are related by $\textrm{St}_{\Lambda} = \pi \text{St}_D/\Lambda$. 

Figure~\ref{fig:PSD_XD1_YD0p5} shows the power spectra computed in the near wake shear layer at $\{x/D,y/D\}= \{1,0.5\}$, for the 3 $C_T$ and 5 FST cases. For reference, the FST power spectra are also shown in each individual subfigure. Focusing on the first column of figures ($C_T \approx 0.5$, $\Lambda =1.7$), we observe that for the lowest tip-speed ratio, the dominant frequency for all FST cases is $\textrm{St}_{\Lambda} \approx 3$, which corresponds to the blade passage frequency, denoted here as $3f_r$. As $C_T$ and $\Lambda$ increase, additional peak frequencies—multiples of the rotor rotational frequency—appear, notably $1f_r$ ($\textrm{St}_{\Lambda}\approx1$) and $2f_r$ ($\textrm{St}_{\Lambda}=2$). In general, as $\Lambda$ increases, the spectral amplitude of the $3f_r$ peak decreases, while those of the $1f_r$ and $2f_r$ peaks become more pronounced and progressively dominate over $3f_r$. Finally, as $C_T$ and $\Lambda$ increase, the energy content of low-frequency structures ($\textrm{St}_{\Lambda} \leq 1$) increases.

Regarding the effects of FST, although no clear influence of ${\cal L}_x$ is observed, an increase in $TI_{\infty}$ leads to a significant reduction in the spectral amplitude at all three leading frequencies. In particular, for the highest $TI_{\infty}$ case (L8), no distinct spectral peaks are detected at the nearest measurement station from the turbine for the two highest tip-speed ratios. This is further emphasised in the spectrograms presented in figure~\ref{fig:PSD_XD_yD0p5}, which show that the streamwise persistence of these three rotor-related frequencies diminishes significantly as $TI_{\infty}$ increases, thereby highlighting the early dissipation of high-frequency turbine-related structures, such as the tip vortices.

The spectrogram shown in figure \ref{fig:PSD_XD1_yD} highlights the spanwise persistence of the turbine-related coherent motions at $x/D=1$. Interestingly, for the lowest tip-speed ratio ($C_T \approx 0.5$), $3f_r$ is observed across almost the entire blade span ($0 \leq y/D \leq 0.5$), with its spectral amplitude being more pronounced near the wake edge ($y/D \approx 0.5$) and close to the wake centreline ($y/D \approx 0.1$), highlighting the presence of tip and root vortices. The presence of $3f_r$ in the mid-span blade region emphasises the existence of the vortex sheet shed from the trailing edge of the rotor blades, that connects tip and root vortices \citep{Qin2021}. It is noteworthy that, in the wind turbine vortex system proposed by \cite{Okulov2007}, the tip vortices are assumed to be embedded within a helical vortex field, formed by the blade's trailing edge vortex sheet and root vortices, which appears to align with the spectrograms obtained with the turbine operating at the lowest tip-speed ratio. For $C_T \approx 0.7$ and $C_T \approx 0.9$, the presence of $3f_r$ is confined to the tip-shear layer and near the root of the blade, with no indication of the trailing edge vortex sheet, suggesting its early dissipation. As previously mentioned, increasing $\Lambda$ leads to the emergence of multiple rotor-related frequencies, with their spanwise existence varying depending on $C_T$ and the FST. Notably, at $C_T \approx 0.9$, $1f_r$, $2f_r$, $3f_r$, along with a small trace of $4f_r$ are observed in both the tip shear ($y/D \approx 0.5$) and root ($y/D \approx 0.1$) regions for the weakest turbulent background case S1. The existence of multiple coherent dynamics in wind turbine near wakes, with additional frequencies appearing as $\Lambda$ increases, has also been reported by \cite{Biswas2024,Biswas2024b}, who conducted experiments at a lower Reynolds number ($\textrm{Re} = 4 \times 10^4$) and with a similar FST intensity to that of case S1 ($TI_{\infty} \approx 1\%$). The spectral amplitude of these frequencies diminishes as $TI_{\infty}$ increases, with the coherent motions at these frequencies dissipating faster at the root region compared to the tip region. This emphasises that the root vortices are weaker \citep{Vermeer2003} and diffuse more rapidly \citep{Sherry2013,Herraez2016} than the tip vortices, with FST further enhancing their dissipation. Specifically, no traces of $\Lambda$-dependent high-frequency structures are observed for case L8 at $C_T \approx 0.9$ and $x/D = 1$, underscoring the significant and immediate influence of FST on the persistence of these structures. For low and moderate $TI_{\infty}$ cases (S1, S4, M5 and L3) at $C_T \approx 0.7$, $1f_r$ spans nearly the entire blade span, suggesting the presence of a vortex sheet analogous to the one observed at $3f_r$ for $C_T\approx0.5$, though of lower frequency and strength. 

The presence of multiple spectral peaks in the tip shear layer provide insights about the break up process of the helical vortices. Firstly, with the exception of the most turbulent case L8, the spectra shown in figure~\ref{fig:PSD_XD1_YD0p5} are qualitatively similar across the FST cases for each $C_T$, although the spectral amplitude at each frequency varies with $TI_{\infty}$. This suggests that the tip-speed ratio plays a dominant role in determining the dynamic structures introduced by the turbine into the flow, while FST primarily influences their strength and dissipation rate thereafter. Moreover, in the spectrograms displayed in figure~\ref{fig:PSD_XD_yD0p5}, multiple spectral peaks are observed at $C_T \approx 0.7$ and $C_T \approx 0.9$, while no signs of $f_r$ and $2f_r$ appear for $C_T \approx 0.5$ until $3f_r$ dissipates. Thus, at $C_T \approx 0.5$ tip vortices might primarily dissipate and mix with the flow whereas at $C_T \approx 0.7$ and $C_T \approx 0.9$, the breakdown of tip vortices seems to follow a multistage process. Specifically, the presence of spectral peaks at $1f_r$ and $2f_r$ supports a multi-step merging process, in which adjacent helical filaments become sufficiently close to interact and roll up around each other, potentially resulting in the leapfrogging phenomenon, \emph{i.e.} the pairing of consecutive vortices \citep{Felli2011,Sherri2013,Lignarolo2015,Qin2021}. This is a specific mode of instability for helical vortex filaments, which was described by \cite{Widnall_1972} as the mutual-inductance instability mode, where the grouping mechanism is driven by the mutual inductance between adjacent spirals. Recently, \cite{Biswas2024b} further emphasised the link between the tip vortex merging process and the energy exchanges between the coherent modes. In particular, at low tip-speed ratios, the larger spacing between tip vortices limits the interaction between multiple vortex filaments, leading to weaker triadic energy exchanges between the modes $f_r$, $2f_r$ and $3f_r$ \cite{Biswas2024}. Therefore, since the turbine operates at a very low tip-speed ratio ($\Lambda = 1.7$) for $C_T \approx 0.5$, this results in a large spacing between filaments, potentially inhibiting their mutual interactions. Alternatively, the energy content of the coherent motion arising from their merging might not be sufficiently high to be detected in the spectra.

\begin{figure}
    \centering
    \begin{subfigure}{\textwidth}
         \centering
            \includegraphics[width=1\linewidth]{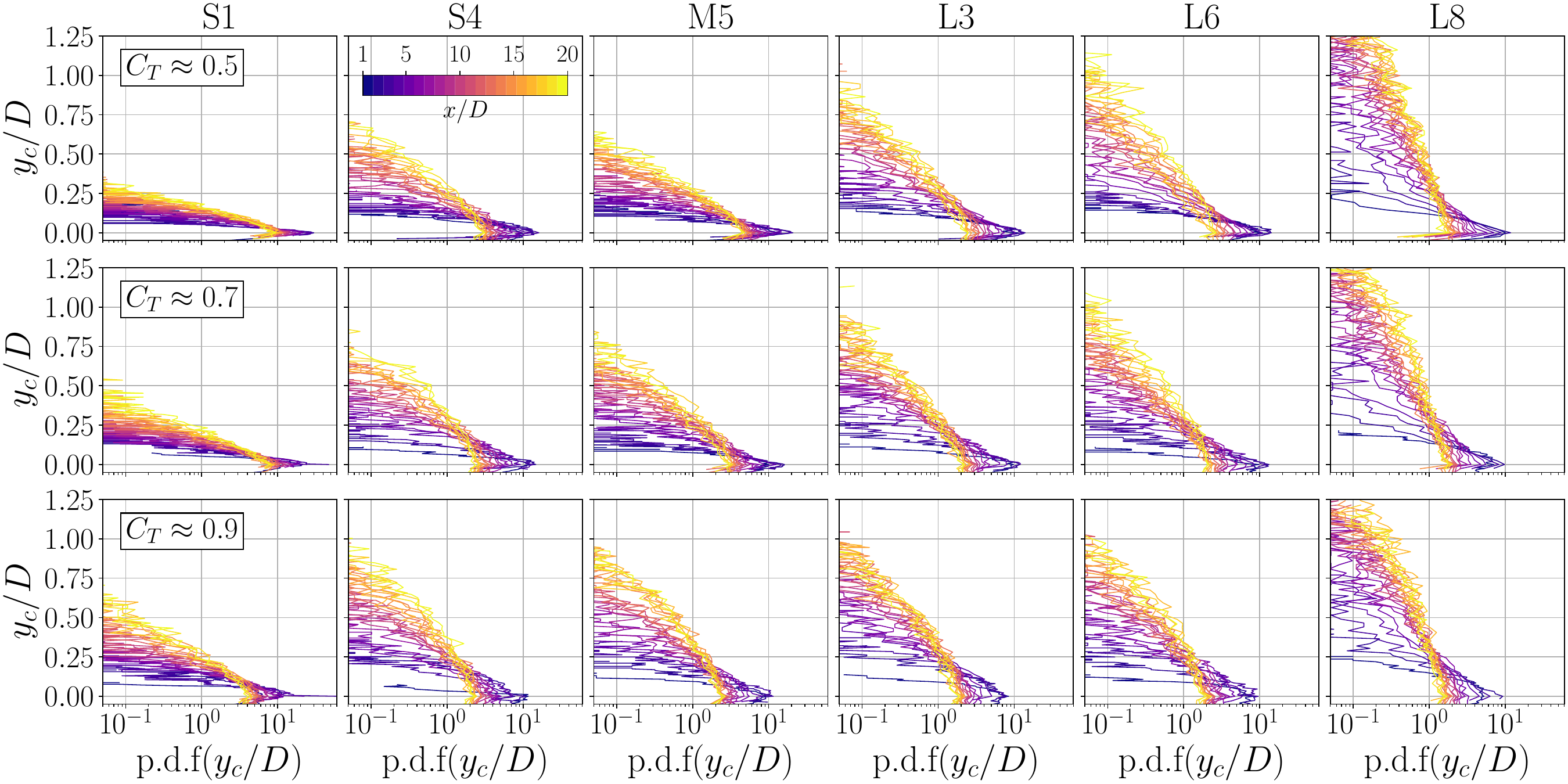}
            \subcaption{\label{fig:pdf_centreline}}
    \end{subfigure} \\
     \begin{subfigure}{\textwidth}
         \centering
            \includegraphics[width=\linewidth]{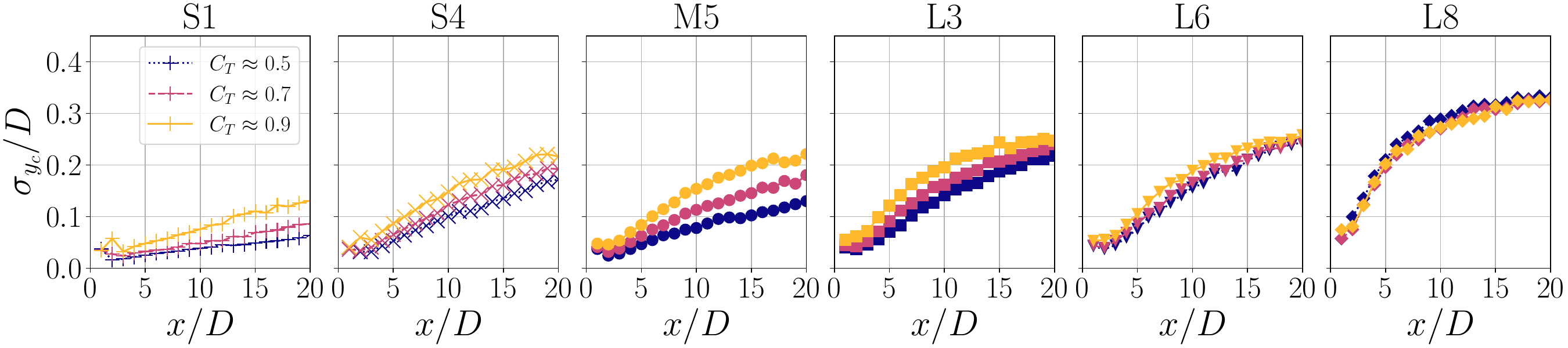}
            \subcaption{\label{fig:sigma_Ct}}
    \end{subfigure} \\
    \begin{subfigure}{\textwidth}
         \centering
            \includegraphics[width=\linewidth]{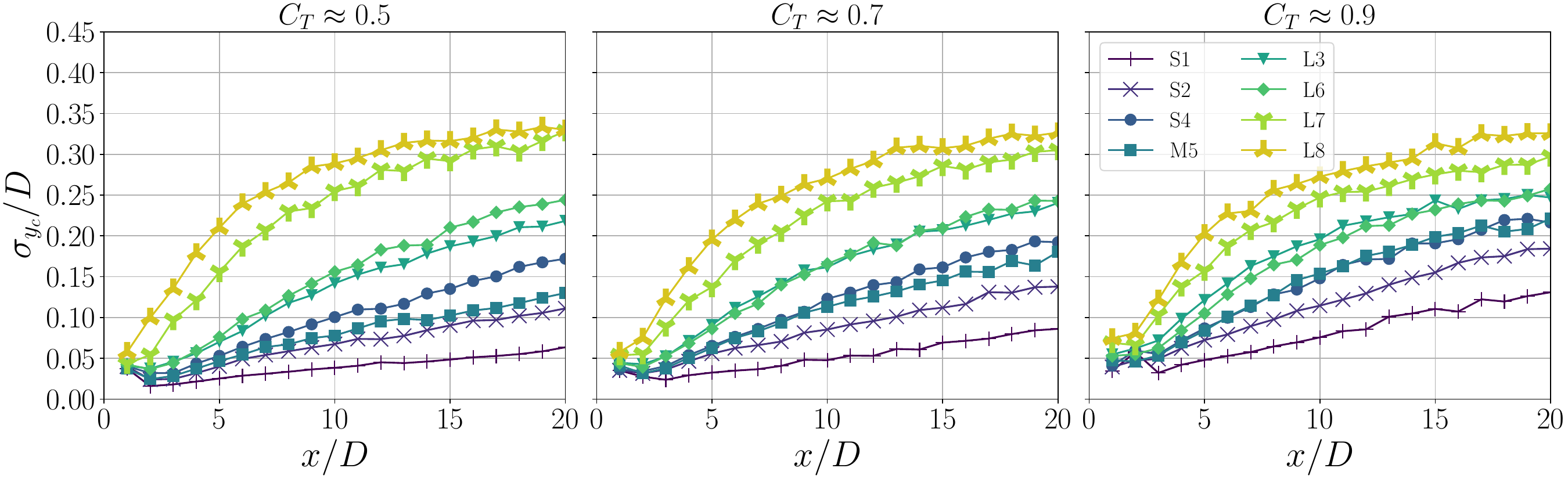}
            \subcaption{\label{fig:sigma_x}}
    \end{subfigure}    
    \caption{Probability density functions, $\textrm{p.d.f}$ $(y_c/D)$ (\subref{fig:pdf_centreline}), and standard deviations, $\sigma_{y_c}/D$ ((\subref{fig:sigma_Ct}) \& (\subref{fig:sigma_x})) of the wake centre positions. A focus is placed on the influence of $C_T$ on $\sigma_{y_c}/D$ in panel (\subref{fig:sigma_Ct}), and of FST in panel (\subref{fig:sigma_x}).}
    \label{fig:sigma}
\end{figure}

\subsection{Statistics and dynamics of wake meandering \label{subsec:wake meandering}}

In their review paper on the meandering of wind turbines wakes, \cite{Yang2019} describes wake meandering as the large-scale, low frequency motions of the wake, typically occurring in the far wake. When performing horizontal scans of wakes, it manifests through the large lateral motion of the wake centreline. This dynamic has been linked to the presence of a low-frequency broad peak in the velocity spectra, within the range, $\textrm{St}_D \in [0.1,0.4]$, which is influenced by the wind turbine's tip-speed ratio, thrust coefficient, and FST conditions  \citep{Medici2006,Medici2008,Chamorro2009,Yang2019,Biswas2024}. In the literature, wake meandering has been identified as potentially resulting from two distinct mechanisms: the passive downstream advection of the turbine wake by large eddies in the atmospheric turbulence, or a shear layer instability similar to vortex shedding from bluff and porous bodies. In this section, we analyse the existence and characteristics of wake meandering for all $\{C_T,\textrm{FST}\}$ combinations by examining the statistics of the wake centre positions $y_c$ as in \cite{Li2022,Dong2023}, along with the streamwise evolution of the power spectra of the fluctuating velocity, focusing on the low-frequency range ($\textrm{St}_D \in [0,1]$). 

At each streamwise measurement location $x/D$, the 240~s time series measured by the 21 hot-wires is divided into 2400 bins of 0.1~s each. Within each bin, the wake centre position $y_c$ is obtained by averaging the instantaneous wake centre positions, identified as the radial location where the velocity is minimal. If the instantaneous wake centre position is found in $ y < 0 $, the data is discarded.  The choice of bin size is motivated by the diameter-based Strouhal number commonly reported in the literature for wake meandering, $\textrm{St}_D \approx 0.2$, corresponding to a frequency of $\approx2$~Hz in our experiments. By applying bin averaging with a window size of 0.1~s, we ensure the sampling follows the Shannon criterion, while primarily focusing on the effects of large-scale, low frequency structures on the wake centre positions. The probability density functions of wake centre positions, $\textrm{p.d.f}(y_c/D\geq0)$, at all streamwise positions are compared for different $\{ C_T, \textrm{FST}\}$ combinations in figure~\ref{fig:pdf_centreline}. Figures~\ref{fig:sigma_Ct} \& \ref{fig:sigma_x} show the streamwise evolution of the standard deviation of the wake centre location, $\sigma_{y_c}/D$, with the first figure illustrating the influence of $C_T$ across different FST cases, and the second highlighting the effect of FST at each turbine operating point. 

Firstly, for all $\{ C_T, \textrm{FST}\}$ cases, as $x/D$ increases, the $\textrm{p.d.f}$ of the wake centre positions exhibits a reduction in the central peak, a broadening distribution, and more pronounced tails. This indicates increased variability in the wake's lateral position and a higher probability of significant deviations from the mean wake centreline. The central peak of the $\textrm{p.d.f}$ is the most pronounced for case S1 and the least for case L8, indicating that the wake in turbulent backgrounds experienced larger deviations from its time-averaged trajectory. Additionally, the flatness of the $\textrm{p.d.f}$ increases with both $TI_{\infty}$ and ${\cal L}_x$ , progressively deviating from a Gaussian distribution, with the tails of the $\textrm{p.d.f}$ becoming more pronounced. This highlights the increased likelihood of the wake centreline deviating from its mean position, larger lateral displacement and the more erratic movements of the wake over time with higher $TI_{\infty}$ and ${\cal L}_x$. We also observe a strong influence of $C_T$ on the $\textrm{p.d.f}$. For all FST cases, the central peak decreases, the $\textrm{p.d.f}$ becomes wider, and exhibit heavier tails as $C_T$ increases. The influence of $C_T$ is more pronounced in low-$TI_{\infty}$ and small ${\cal L}_x$ ambient conditions, where clear differences in the shape of the $\textrm{p.d.f}$ are observed between the $C_T$ cases across the entire measurement domain. A similar influence of $C_T$ on the $\textrm{p.d.f}$ is observed in the near wake for high-$TI_{\infty}$ and large ${\cal L}_x$ cases; however further downstream, no clear differences are observed.

The standard deviation of wake centre positions, $\sigma_{y_c}/D$, increase with $C_T$, except for the two highest $TI_{\infty}$ cases (L8 and L7, not shown for brevity), for which the influence of $C_T$ becomes negligible due to the dominant effect of background turbulence. Moreover, similar to the wake width evolution, $\sigma_{y_c}/D$ reaches a plateau in the far wake for cases M\# and L\#, whereas for S\# cases, it follows an almost linear evolution across the entire measurement domain. A clear effect of ${\cal L}_x$ is observed on $\sigma_{y_c}/D$, with FST cases L\# exhibiting higher standard deviation at all $x/D$ positions. Additionally, the increase in $\sigma_{y_c}$ is further amplified by the presence of a strong background turbulence intensity $TI_{\infty}$. In summary, wake meandering is more pronounced in cases with high ${\cal L}_x$ and high $TI_{\infty}$, with the integral length scale seemingly being the dominant parameter. This aligns with the findings of \cite{Kankanwadi2023}, who reported increased wake meandering in the wake of a cylinder as the FST integral length scale increases, as well as with the LES studies of wind turbine wakes by \cite{Hodgson2023,Vahidi2024}, which observed increased wake meandering in wind turbine wakes with both increasing ${\cal L}_x$ and $TI_{\infty}$. Moreover, as discussed in \S~\ref{subsec:WakeWidth}, the time-averaged wake width $\delta$ is a metric that captures both the width of the instantaneous wake and the lateral displacement of the wake (figure~\ref{fig:cartoon}). We have previously observed a reduction in $\delta$ in the near wake for FST cases L3 and L6, compared to S4 and M5, all of which had nearly identical FST intensity levels, but with L3 and L6 exhibiting significantly larger ILS. Since increasing the inflow ILS at a relatively fixed $TI_{\infty}$ has been shown to enhance the amplitude of lateral motion of the wake while also reducing $\delta$, it could be speculated that FST with a large ILS leads to a reduction in the instantaneous wake width $\delta_i$, which would be in line with the findings of \cite{Kankanwadi2023} for planar wakes of cylinders.

In addition, while the turbine thrust coefficient enhances the lateral displacement of the wake in weakly turbulent backgrounds, its influence becomes secondary under strongly turbulent conditions. It is well-documented that both thrust and tip-speed ratio influence the meandering of turbine wakes \citep{Medici2006, Yang2019, Biswas2024}, particularly the observed decrease in the frequency of wake meandering as the tip-speed ratio decreases. In high-$TI_{\infty}$ conditions, the ambient flow exerts a strong and immediate influence on the wake recovery, rapidly diminishing the impact of changes in the thrust coefficient on the wake recovery. As in such cases, wake recovery is primarily governed by the ambient turbulence, it reduces the sensitivity of $\sigma_{y_c}$ to variations in $C_T$. Conversely, under low FST intensity conditions, where wake recovery is more influenced by the turbulence within the wake itself than by ambient turbulence, the lateral displacement of the wake becomes more sensitive to changes in turbine operating conditions.

\begin{figure}
    \centering
    \begin{subfigure}{0.48\textwidth}
         \centering
            \includegraphics[width=\linewidth]{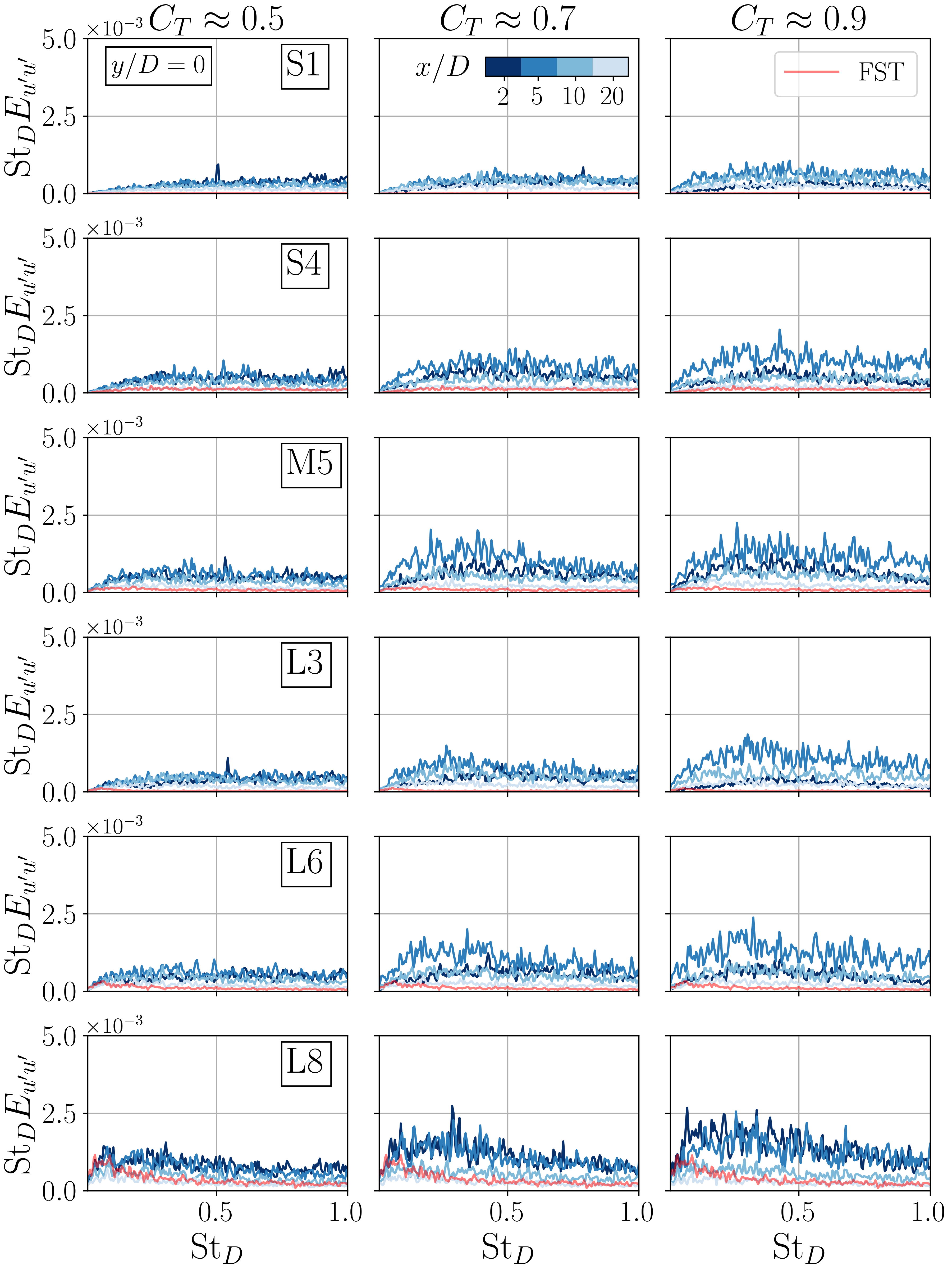}
            \subcaption{$y/D=0$\label{fig:psd21}}
    \end{subfigure} \hfill
     \begin{subfigure}{0.48\textwidth}
         \centering
            \includegraphics[width=\linewidth]{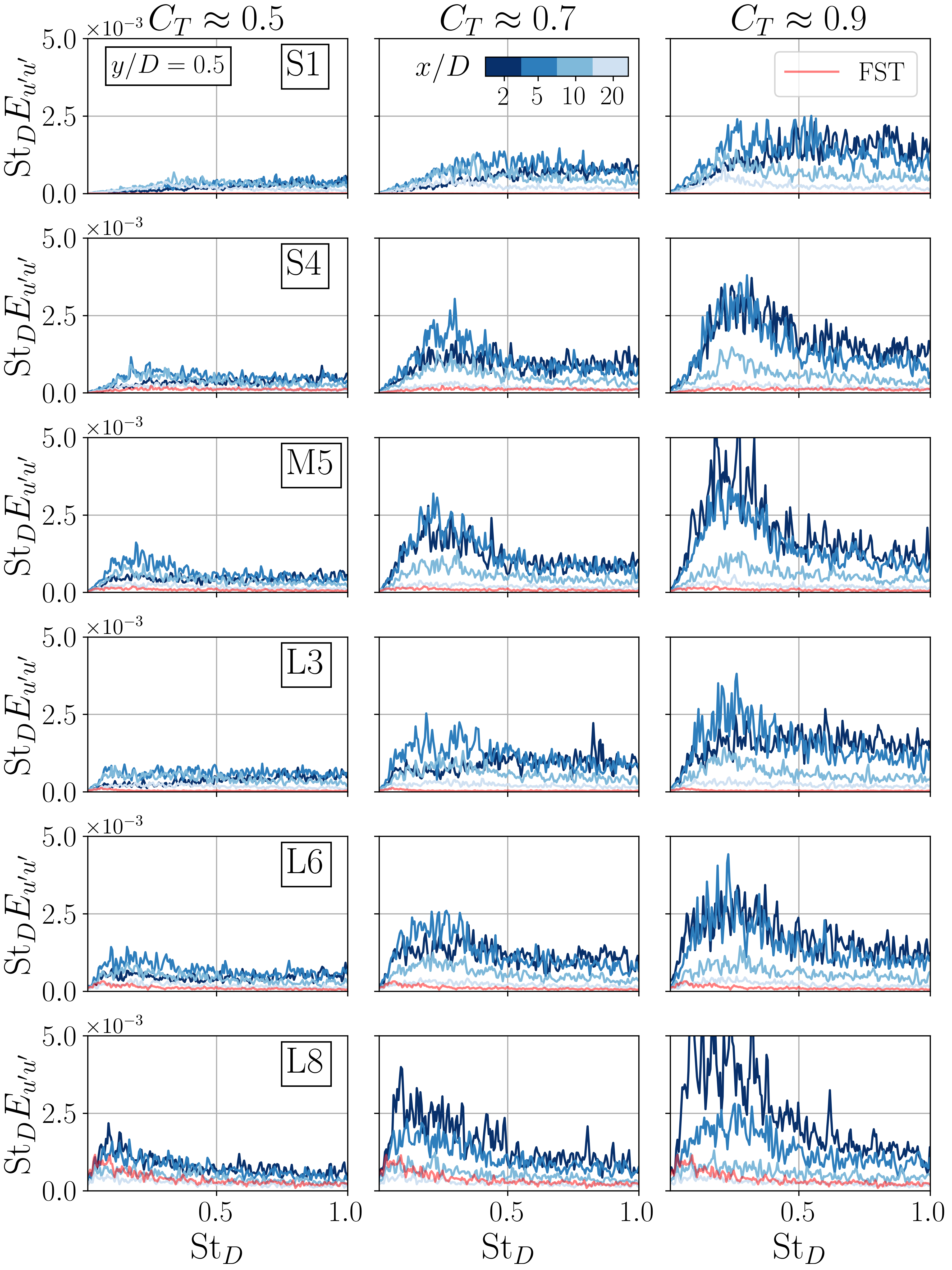}
            \subcaption{$y/D=0.5$\label{fig:psd22}}
    \end{subfigure} \\
    \begin{subfigure}{0.48\textwidth}
         \centering
            \includegraphics[width=\linewidth]{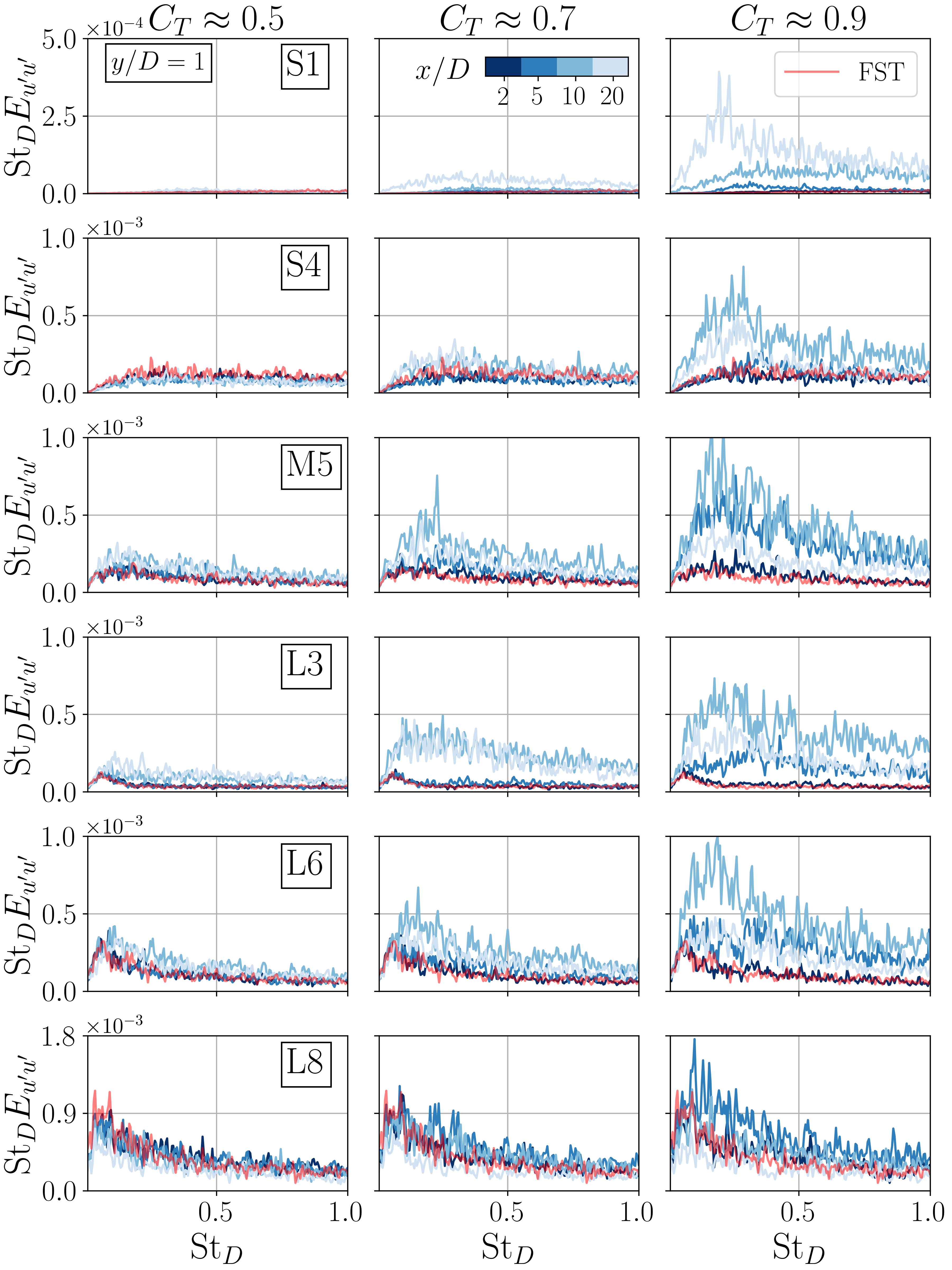}
            \subcaption{$y/D=1.0$\label{fig:psd23}}
    \end{subfigure}  \hfill
     \begin{subfigure}{0.48\textwidth}
         \centering
            \includegraphics[width=\linewidth]{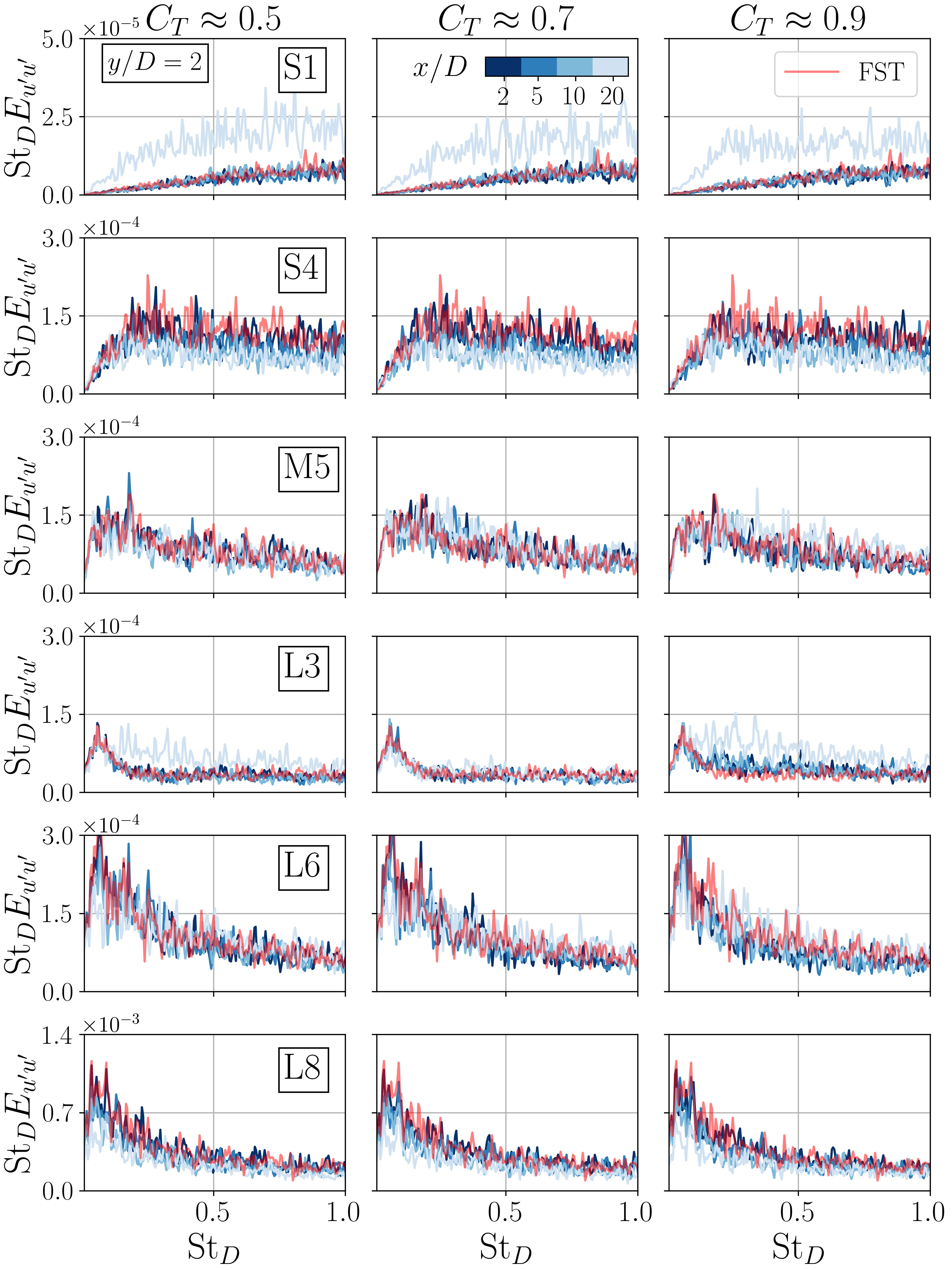}
            \subcaption{$y/D=2.0$\label{fig:psd24}}
    \end{subfigure}
    \caption{Premultiplied power spectra of the fluctuating velocity $u'$, computed at $y/D=0$ (\subref{fig:psd21}), $y/D=0.5$ (\subref{fig:psd21}), $y/D=1$ (\subref{fig:psd23}), and $y/D=2$ (\subref{fig:psd24}) and for 4 streamwise positions $x/D= 2 - 5 - 10 - 20$, with the abscissa representing the Strouhal number based on the turbine diameter. FST power spectra, measured without the turbine at $\{x/D , y/D\} = \{0,0\}$, are shown in {\textcolor{red}{red}}. Note the change in the ordinate axis scale between the figures.}
    \label{fig:powerspectra}
\end{figure}

\begin{figure}
    \centering
    \begin{subfigure}{\textwidth}
         \centering
            \includegraphics[width=0.45\linewidth]{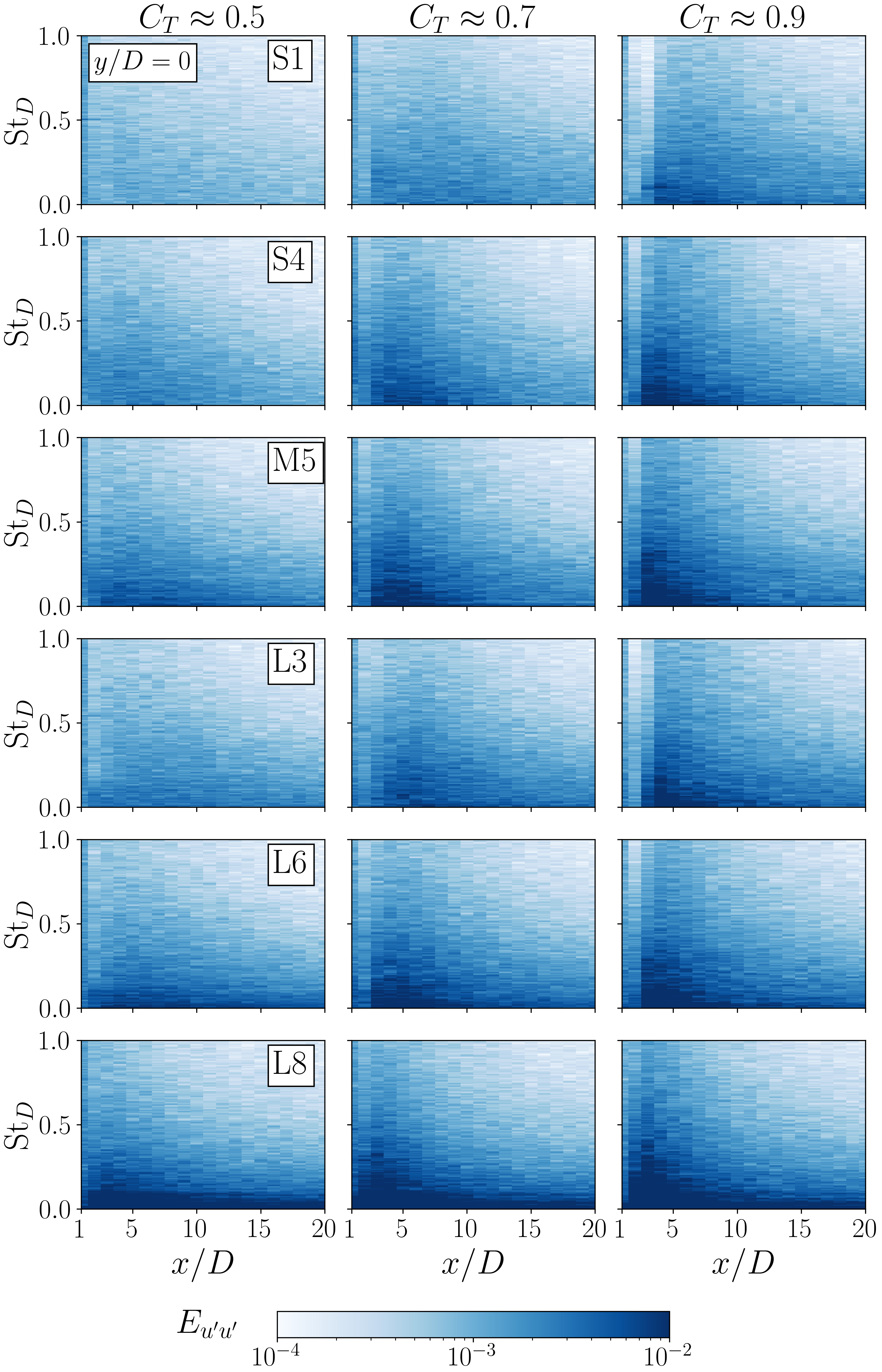} \hfill
            \includegraphics[width=0.45\linewidth]{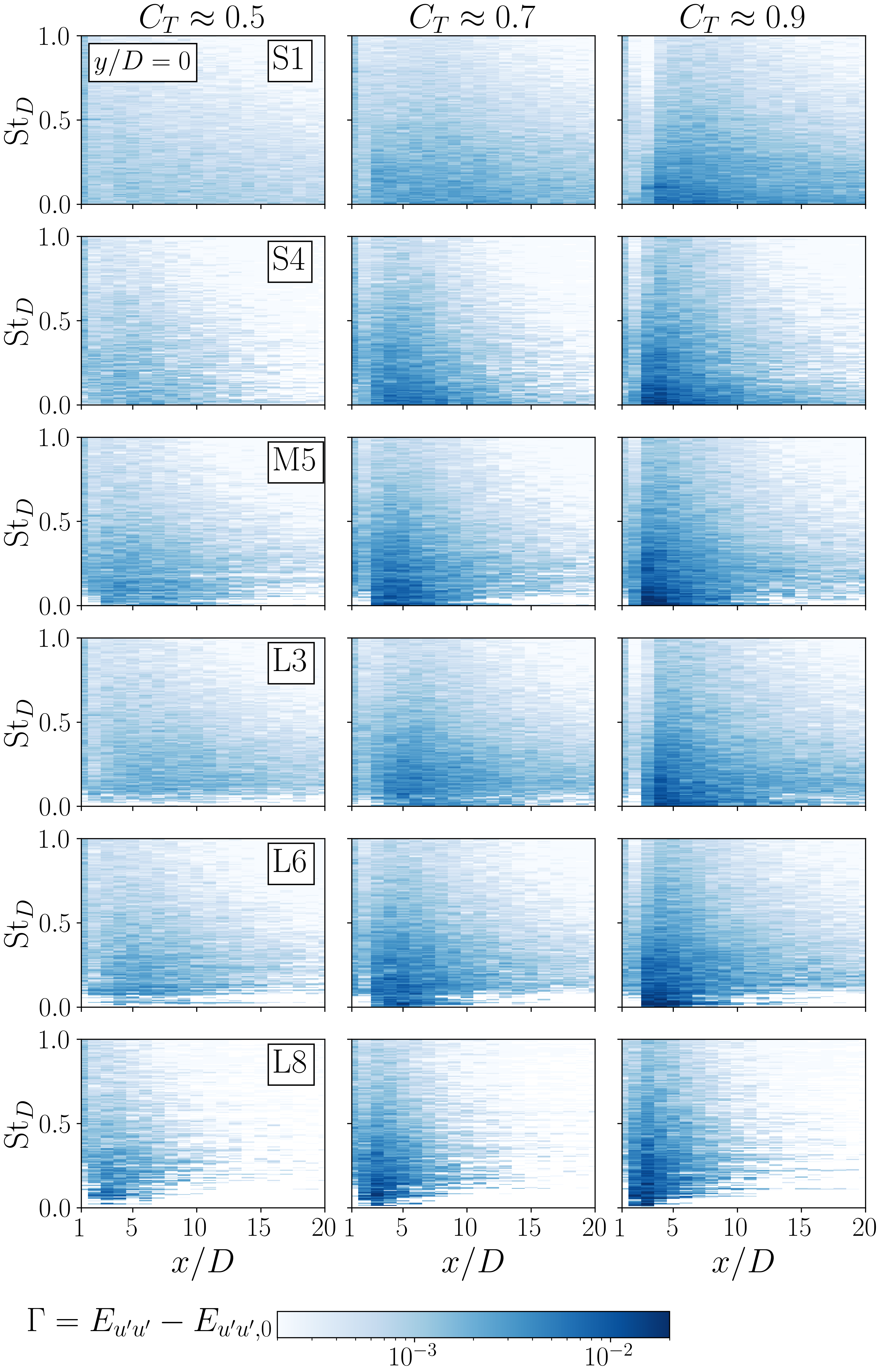}
            \subcaption{$y/D=0$\label{fig:PSD_yD0}}
    \end{subfigure}\\
    \begin{subfigure}{\textwidth}
         \centering
            \includegraphics[width=0.45\linewidth]{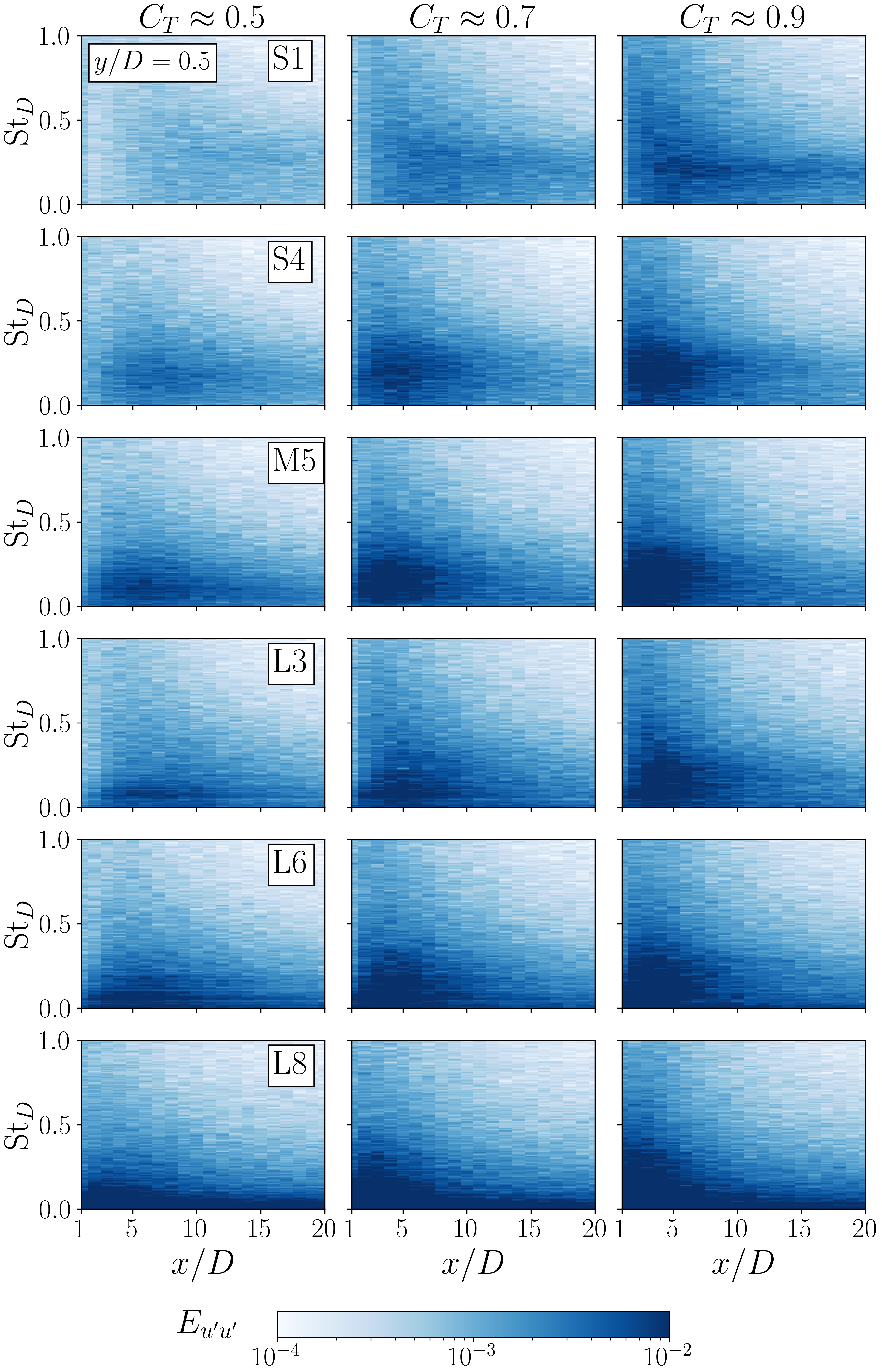} \hfill
            \includegraphics[width=0.45\linewidth]{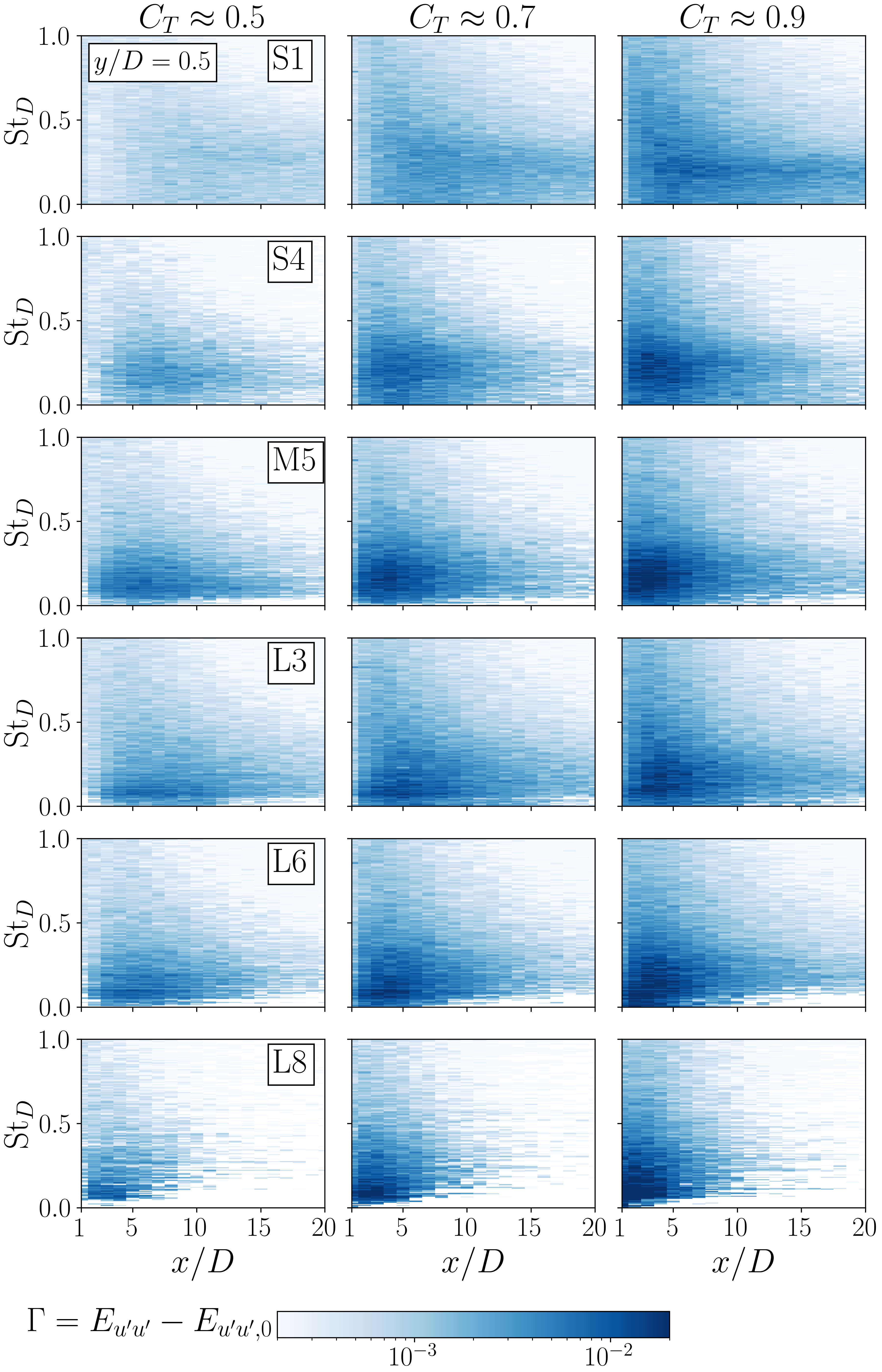}
            \subcaption{$y/D=0.5$\label{fig:PSD_yD0.5}}
    \end{subfigure}
    \caption{Spectrograms of the fluctuating velocity computed at $y/D = 0$ (\subref{fig:PSD_yD0}) and $y/D = 0.5$ (\subref{fig:PSD_yD0.5}). In the figures on the right, the freestream turbulence spectra, $E_{u'u',0}$, measured prior to the turbine experiments at $\{x/D , y/D\} = \{0,0\}$, have been subtracted from the energy spectra, $E_{u'u'}$, shown in the left figures. The ordinate represents the Strouhal number based on the turbine diameter, $\textrm{St}_{D}$.}
    \label{fig:spectrograms2}
\end{figure}

\begin{figure}
    \centering
    \begin{subfigure}{\textwidth}
         \centering
            \includegraphics[width=0.45\linewidth]{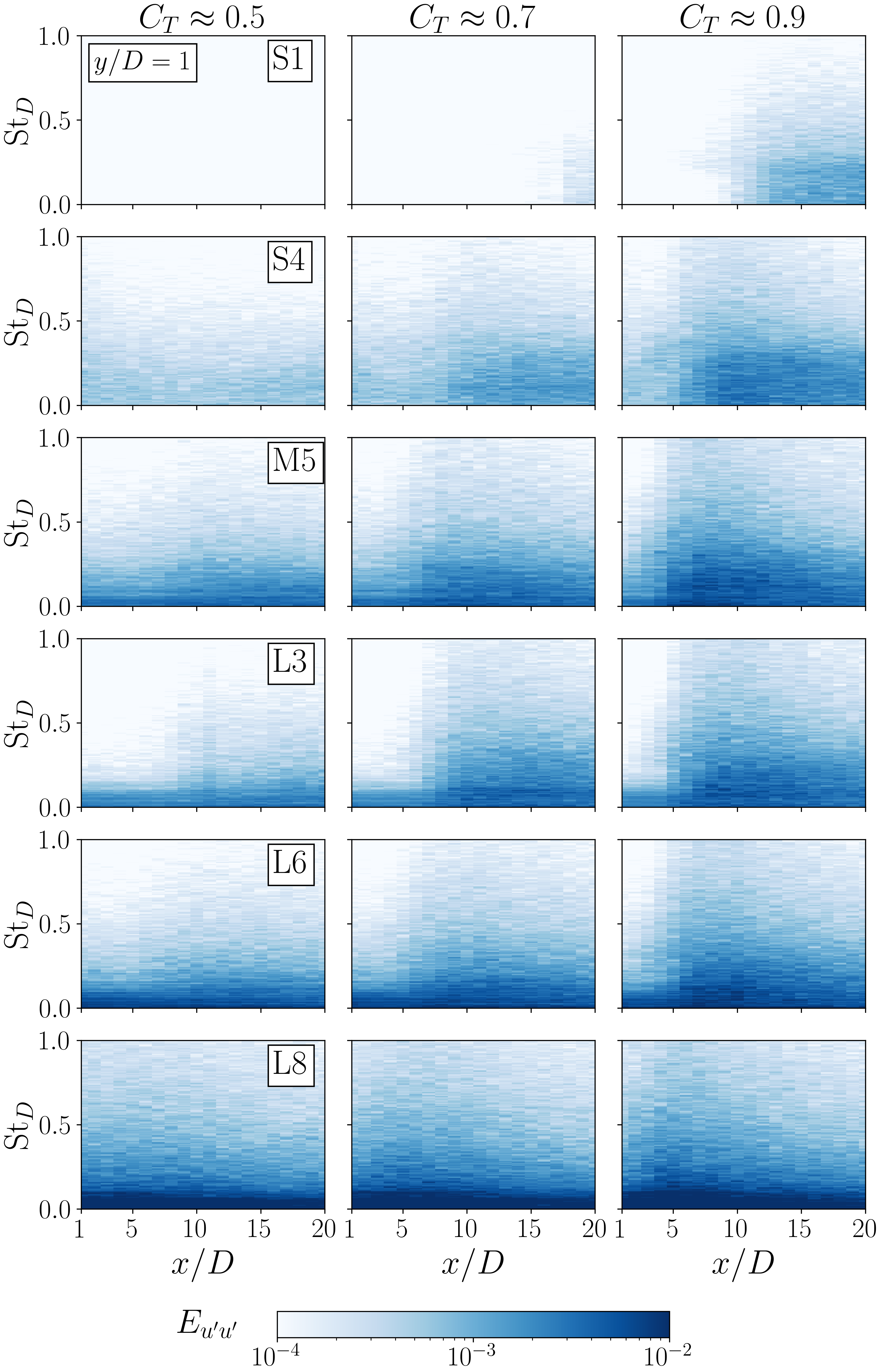} \hfill
            \includegraphics[width=0.45\linewidth]{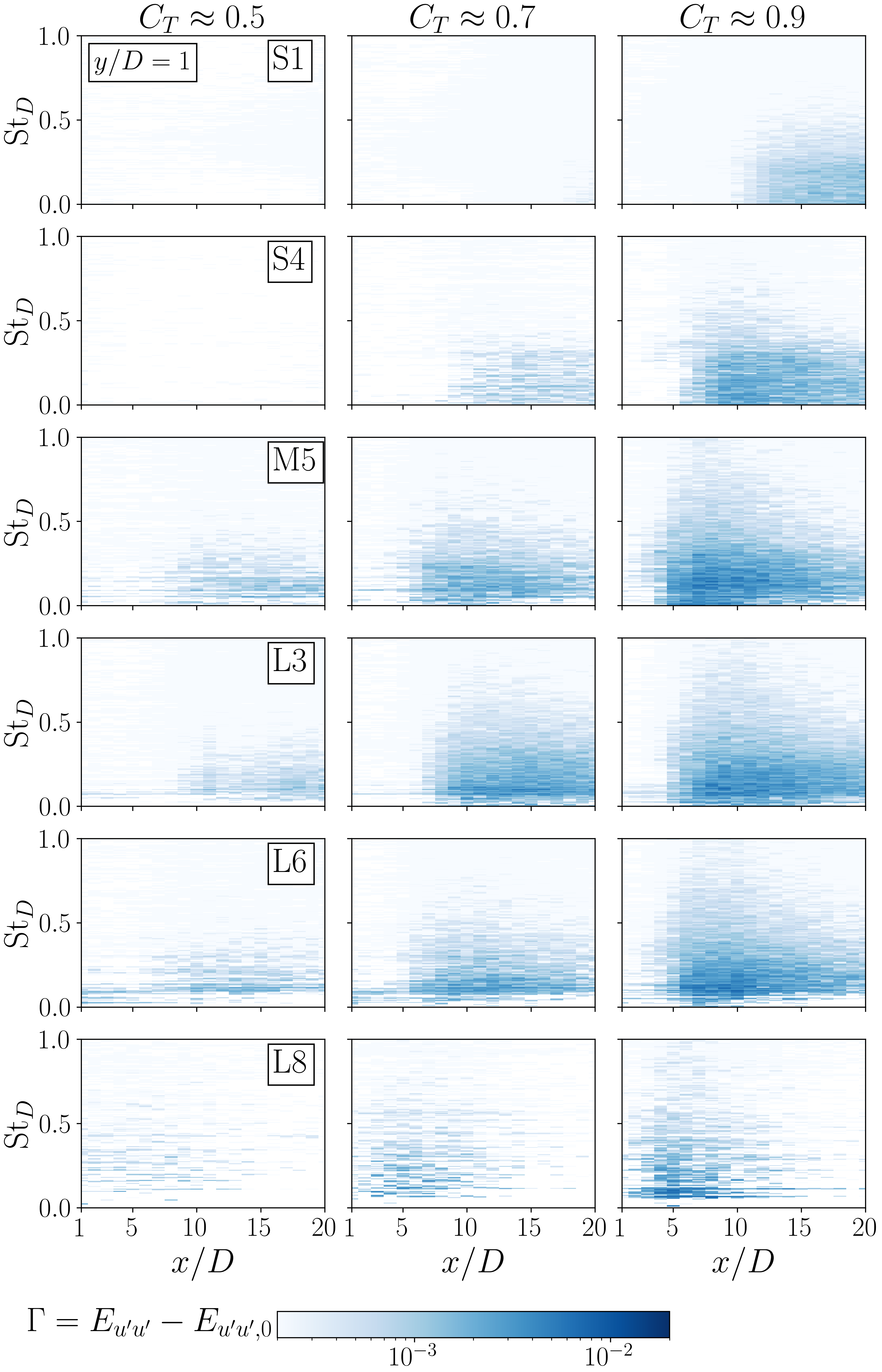}
            \subcaption{$y/D=1.0$\label{fig:PSD_yD1}}
    \end{subfigure}\\
    \begin{subfigure}{\textwidth}
         \centering
            \includegraphics[width=0.45\linewidth]{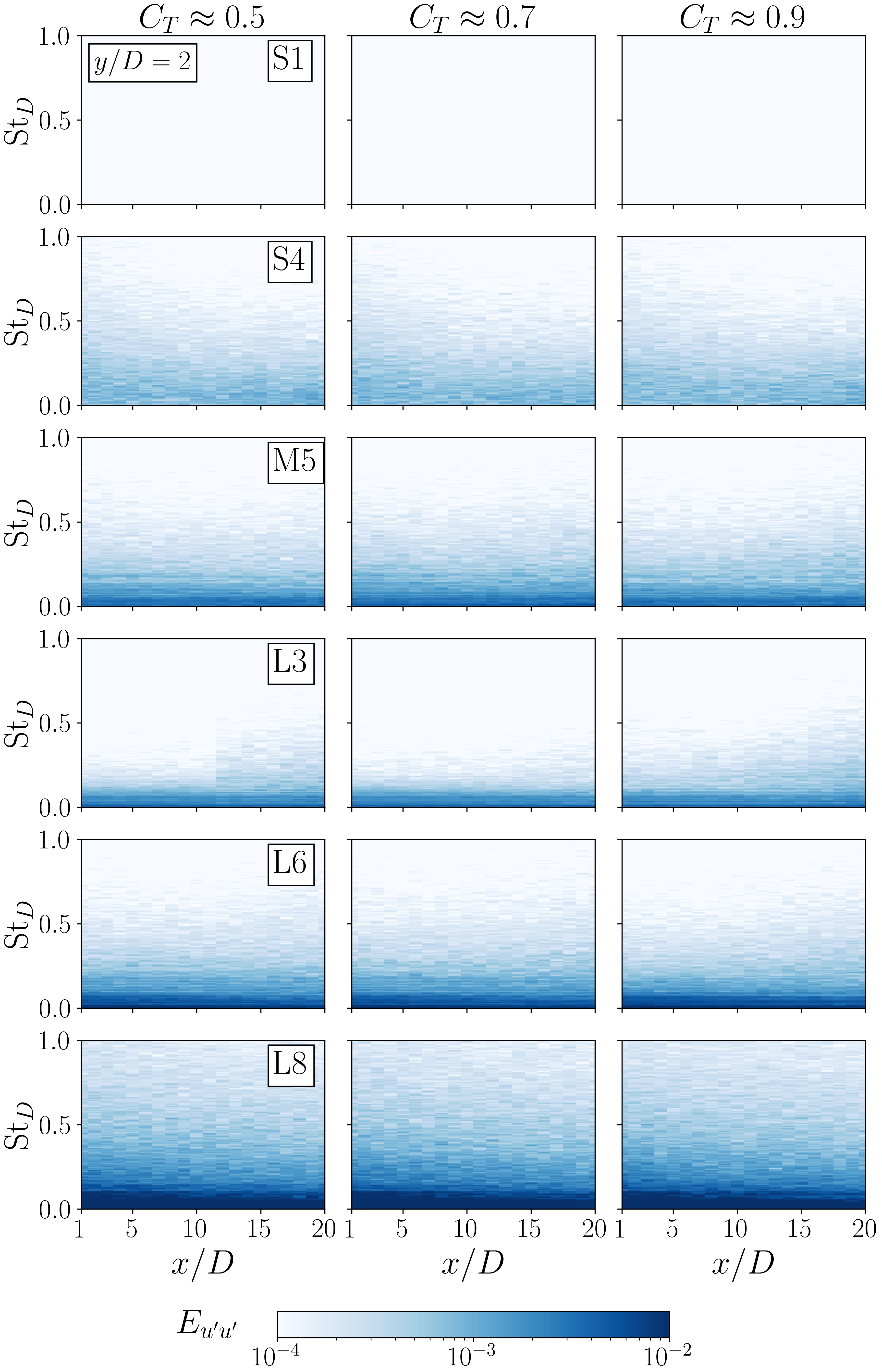} \hfill
            \includegraphics[width=0.45\linewidth]{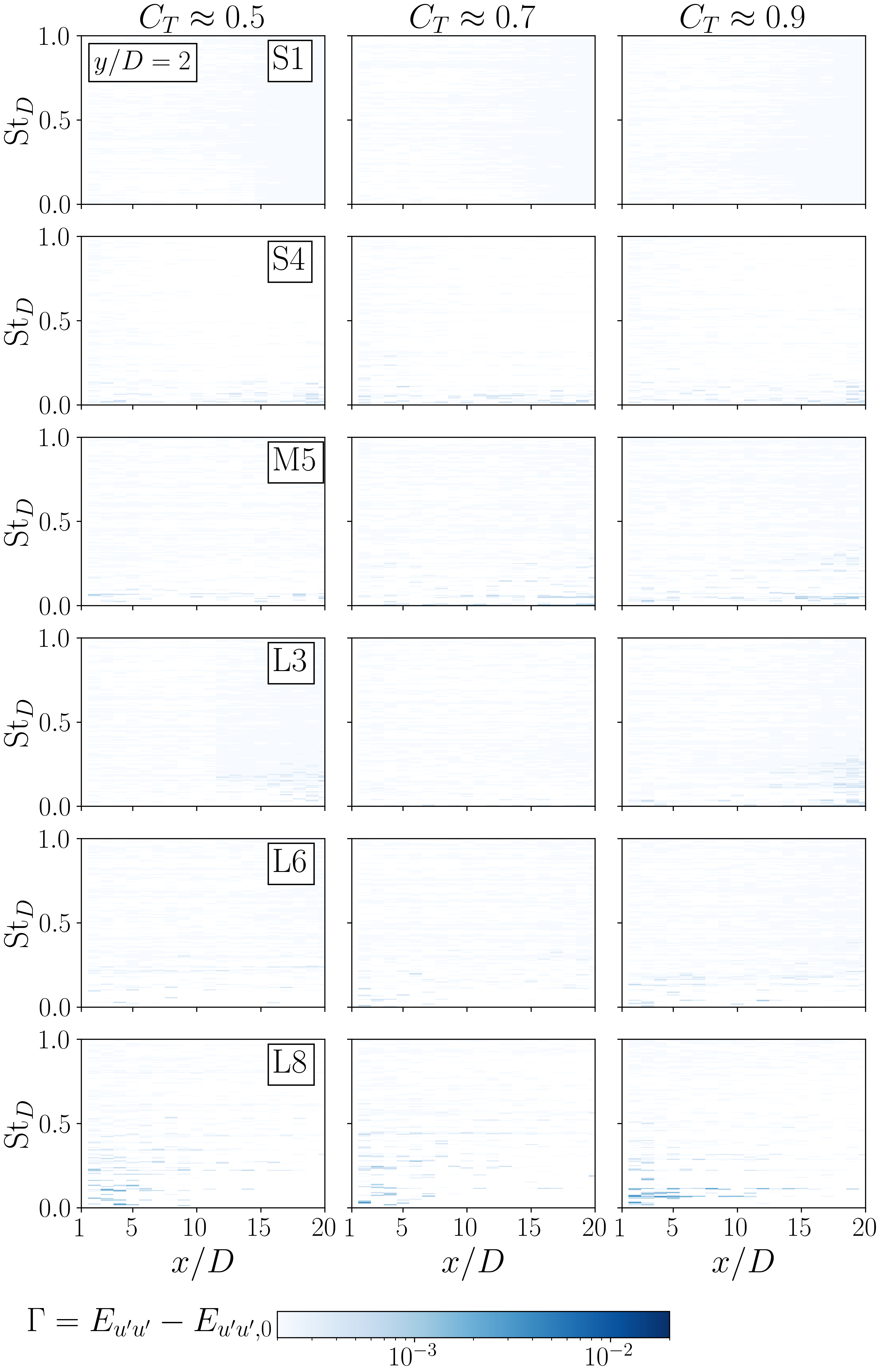}
            \subcaption{$y/D=2.0$\label{fig:PSD_yD2}}
    \end{subfigure}
    \caption{Spectrograms of the fluctuating velocity computed at $y/D = 1$ (\subref{fig:PSD_yD1}) and $y/D = 2$ (\subref{fig:PSD_yD2}). In the figures on the right, the freestream turbulence spectra, $E_{u'u',0}$, measured prior to the turbine experiments at $\{x/D , y/D\} = \{0,0\}$, have been subtracted from the energy spectra, $E_{u'u'}$, shown in the left figures. The ordinate represents the Strouhal number based on the turbine diameter, $\textrm{St}_{D}$.}
    \label{fig:spectrograms2b}
\end{figure}

We will now investigate wake meandering from the perspective of a low-frequency dynamic. Figure~\ref{fig:powerspectra} compares the premultiplied power spectra at 4 specific radial locations: $y/D=0$ (\ref{fig:psd21}), $y/D=0.5$ (\ref{fig:psd22}), $y/D=1$ (\ref{fig:psd23}) and $y/D=2$ (\ref{fig:psd24}), for 4 streamwise positions $x/D= 2 - 5 - 10 - 20$, across different $\{ C_T, \textrm{FST}\}$ cases. The freestream turbulence spectra, detailed in the Appendix, are superimposed in red on each figure, to distinguish the turbine-induced structures from the background turbulence structures. Given that wake meandering is a low-frequency dynamic, the focus in these figures is on $\textrm{St}_D \leq 1$. The spectrograms shown in figures~\ref{fig:spectrograms2} \& \ref{fig:spectrograms2b} illustrate the streamwise evolution of the energy content of the low frequency dynamics. For each of the 4 radial locations, the left figures show the turbulence power spectra $E_{u'u'}$, while the right figures present $\Gamma =  E_{u'u'} -E_{u'u',0}$, where the power contribution from background turbulence, $E_{u'u',0}$, to the total spectrum has been subtracted. $\Gamma$ provides additional insights into the turbine's contribution to the energy content of low-frequency structures by distinguishing it from the background turbulence. It is important to note that, since $E_{u'u',0}$ is computed at the turbine hub location, $\Gamma$ also accounts for the evolution of the energy of background turbulence structures along the wind tunnel. However, the power spectra obtained from the hot-wire at $y/D =2$, located outside the wake up to several turbine diameters, closely overlap with those of the FST, with $\Gamma \approx 0$, suggesting limited evolution of the background turbulence structures' frequency and energy content along the tunnel (figures~\ref{fig:psd24} \& \ref{fig:PSD_yD2}).

A common broad spectral peak between $\textrm{St}_{D} \approx 0.1$ and $\textrm{St}_{D} \approx 0.4$ emerges in the tip-shear layer at $y/D = 0.5$ for all FST cases (figures~\ref{fig:psd22} \& \ref{fig:PSD_yD0.5}). This frequency interval coincides with the Strouhal number range commonly reported in the literature for wake meandering, and is distinct from the low-frequency spectral content of the different turbulent backgrounds, as evidenced by the comparison with the FST spectra in figure~\ref{fig:psd22}, and the positive values for $\Gamma$ in the spectrograms (figure~\ref{fig:PSD_yD0.5}). Focusing on the hot-wire at $y/D=0.5$, we observe a clear increase in $E_{u'u'}$ and $\Gamma$ within that frequency band as the thrust coefficient increases. At the lowest thrust coefficient $C_T \approx 0.5$, the energy content is so low that the existence of wake meandering can be questioned, particularly in the lowest turbulence cases, S1 and S2 (omitted here for brevity). This behaviour mirrors vortex shedding from porous bodies, which occurs only above a certain porosity threshold and whose strength increases with reduced porosity and higher thrust coefficient \citep{Cannon1993}. Furthermore, the observation that the highest energy around $\textrm{St}_D \approx 0.2$ occurs in the tip-shear layer and in the near the wake, followed by a decay of the structures, aligns with the dynamics of vortex shedding downstream of porous bodies \citep{Biswas2024, Bourhis2024}.

In addition, freestream turbulence plays a pivotal role in the dynamics of wake meandering. Both $E_{u'u'}$ and $\Gamma$ within $\textrm{St}_D \in [0.1 -0.4]$ increase with both $TI_{\infty}$ and ${\cal L}_x$, aligning with the previous observation of an increased standard deviation in the wake centre positions. The streamwise location $x/D$ where these quantities are maximised shifts closer to the turbine as $TI_{\infty}$ and ${\cal L}_x$ increase. Moreover, while a clear broad peak is observed for the Group 2 and Group 3 FST cases, it is less distinct for Group 1 cases, highlighting the influence of freestream turbulence in either triggering or at least enhancing wake meandering. Comparing Group 2 FST cases (S4, M5, L3 and L6), increasing ${\cal L}_x$ appears to lead to a slight downstream shift in the streamwise location where $E_{u'u'}$ and $\Gamma$ are maximised. Moreover, although the initial energy of this low-frequency structure is enhanced with increasing $TI_{\infty}$, its decay is accelerated, as illustrated in the spectrograms at $y/D =0.5$, where $\Gamma$ is higher for cases L3 than L6 for $x/D \geq 10$. A parallel can be drawn between this low-frequency dynamic and vortex shedding observed in porous and bluff bodies regarding the influence of FST on both. Notably, our observations mirror the findings in \cite{Oliveira2024} who reported that the presence of FST leads to a broadening of the frequency band associated with regular vortex shedding downstream of a cylinder, an increase in the energy associated with this flow structure immediately downstream, and a faster energy decay further downstream. Similar trends have also been reported for porous discs in \cite{Bourhis2024}.

Interestingly, we previously reported that for small ${\cal L}_x$ FST cases (S1, S2, and S4), a peak of ${\cal L}_x$ in the profiles was found in the tip-shear layer, which matches the observation that the maximum energy content of the low-frequency structure is found at $y/D=0.5$ (see the profiles of ${\cal L}_x$ in figure~\ref{Fig:LProfiles}). Hence, the largest structure in the flow are those introduced by the turbine, originating from tip-shear instability, and can give rise to wake meandering. However, at small ${\cal L}_x$ and low $TI_{\infty}$, both the strength of the broad energy peak and the amplitude of wake meandering, \emph{i.e.} characterised by $\sigma_{y_c},$ are reduced. For large ${\cal L}_x$ and high $TI_{\infty}$ FST cases, wake meandering still appears to originate in the tip-shear layer, but freestream turbulence amplifies it, as evidenced by the higher energy content and increased amplitude of the wake centreline position. Moreover, for large ${\cal L}_x$ FST cases, the peak energy content at the frequency $\textrm{St}_D \approx 0.2 $ is observed immediately downstream of the turbine, after which the structure rapidly decays, and becomes negligible in comparison to the low-frequency turbulent background structures. As the wake develops, the energy content of the frequencies in $\textrm{St}_D \in [0.1 -0.4]$ at the radial location $y/D=1$ progressively increases (figures \ref{fig:psd23} \& \ref{fig:PSD_yD1}). Both $E_{u'u'}$ and $\Gamma$ are smaller than at $y/D=0.5$ and the maximum of these quantities is observed further downstream, highlighting the spreading and diffusion of these low-frequency structures throughout the wake. Finally, at the hot-wire position furthest from the wake centreline ($y/D =2$), the spectra for each FST are qualitatively similar across the streamwise measurement locations, and $\Gamma \approx 0$, indicating that ambient structures dominate at this location, except for case S1, where the increase in $E_{u'u'}(x/D=20)$ emphasises the presence of the turbine's wake (figures \ref{fig:psd24} \& \ref{fig:PSD_yD2}). 

In summary, from a purely spectral perspective, wake meandering seems to originate from a tip-shear layer instability, as a broad spectral peak in the Strouhal number range characteristic of wake meandering is found in that region and close to the turbine. Additionally, this dynamic exhibits similarities to vortex shedding observed in the wakes of bluff and porous bodies, particularly in low-turbulence-intensity inflows, where it becomes pronounced only above a certain thrust coefficient and tip-speed ratio.  However, wake meandering is strongly modulated in terms of energy content and wake motion amplitude by the freestream turbulence. Large integral length scales and turbulence intensities in the freestream substantially amplify the large scale motion of the wake, as evidenced by the increased standard deviation of the wake centreline position. It also results in a higher energy content around $\textrm{St}_D \approx0.2$. When inflow turbulence intensity and length scale are low (\emph{e.g.}, S1, S2), wake meandering results solely from shear-layer instability. Shear-layer-induced wake meandering leads to limited lateral motions of the wake, though it is sensitive to variations in $C_T$. In highly turbulent freestreams, the influence of $C_T$ becomes secondary, likely due to the dominance of ambient turbulence over shear-layer-induced wake meandering. In such conditions, wake meandering is primarily governed by inflow turbulence, leading to more intense wake motion. Finally, wake meandering cannot be solely attributed to the passive advection of the wake by large structures in the ambient flow, as hypothesised in the DWM model, or shear-layer instability. Instead, it arises from a combination of both factors, with their respective contributions and resulting dynamics' significance in the flow depending on the FST conditions and the turbine's operating point.

\section{Summary and conclusions \label{Conclusion}}

An in-depth experimental investigation was conducted to assess the influence of freestream turbulence on wind turbine-generated wakes over an extended streamwise distance and at high Reynolds numbers. Freestream turbulence intensity ($1\% \lesssim TI_{\infty} \lesssim 11\%$) and integral length scale ($0.1 \lesssim {\cal L}_x/D \lesssim 2$)  along with the turbine thrust coefficient ($C_T \approx [0.5, 0.7,0.9]$), were independently varied across a broad range representative of typical real-world wind turbine operating conditions.

First, it was observed that a wake defined as a flow region of increased turbulence intensity and TKE vanishes earlier than a wake defined as a flow region of reduced momentum. Indeed, while a clear time-averaged velocity difference between the wake and the background persists throughout the entire measurement range for all FST cases, this distinction was not always observed for the $TI$ and TKE. Specifically, for high $TI_{\infty}$ inflows, a flow region within the wake was identified several diameters downstream of the turbine, where the velocity deficit remains, despite turbulence intensity and TKE having already homogenised with the freestream, challenging conventional wake definitions established for non-turbulent freestream. Beyond this location, the decay of the velocity deficit slows, and a pronounced reduction in the wake growth rate is observed. This faster homogenisation of TKE emphasises the enhanced entrainment rate of TKE relative to mass and momentum, particularly in the near wake— a phenomenon also observed in the near wakes of bluff bodies exposed to a turbulent background \citep{Buxton_Chen_2023}. While Gaussian-like velocity deficit profiles, and saddle-shape TKE profiles, are observed in the far wake for low $TI_{\infty}$ cases, typical of a self-similar evolution for axisymmetric wakes, the presence of high ambient $TI_{\infty}$ appears to suppress the possibility of such self-similar evolution. A similar behaviour has also been reported by \cite{Rind2012} in the far wake of solid discs. The turbine acts as a high-pass filter to the large structures present in the inflows, with a higher $C_T$ leading to an increased blocking of these structures, mirroring the behaviour of porous discs with decreasing porosity \citep{Bourhis2024}. For inflows with a large ILS, the ILS within the wake recovers and gradually aligns with that of the background flow. For inflows with a small ILS, the largest structures in the flow are those introduced by the turbine, originating from the tip-shear layer and expanding radially as the wake develops. Moreover, akin to porous bodies, the size and strength of these structures increases with the thrust coefficient, \emph{i.e.}, as the turbine's porosity decreases. Hence, when $TI_{\infty}$ and ${\cal L}_x$ are large, the turbulence within the wakes tends to adjust to the background turbulence, with the turbulence intensity and structures introduced by the turbine—and, consequently, its operating point—having a a diminished role in the turbulence evolution. In contrast, for low FST intensity and ILS conditions, the turbulence evolution in the wake is primarily driven by the turbulence generated by the turbine itself.

Second, a novel and noteworthy finding is the significant change in the slope of the wake width evolution, especially for wakes exposed to high FST intensity and integral length scale. For all $\{C_T,\textrm{FST}\}$ combinations, the wakes initially grow linearly, as typically reported for wind turbine wakes. However, this linear expansion does not persist throughout the entire wake but transitions into a plateau phase, characterised by a significant deceleration in wake expansion. The plateau phase begins closer to the turbine, with a more abrupt transition as $TI_{\infty}$ increases. In particular, a very clear turning point in the wake width evolution was observed for highly turbulent cases (L7,L8), beyond which the wake growth rate significantly diminishes. In the near wake region ($x/D \lesssim 7$), the wake width increases with $C_T$, $TI_{\infty}$ and ${\cal L}_x$. While the wake growth rate also increases with $C_T$ and $TI_{\infty}$, we found that the ILS has a negligible effect on it. In this region, these findings align with the current state of the art on wind turbine wakes \citep{PorteAgel2019}. However, in contrast to the near wake, sufficiently far from the turbine ($x/D \gtrsim 15$), a significant reduction in the wake growth rate is observed as $TI_{\infty}$ increases. Although these results contrasts with the current state of the art on wind turbine wakes, they simultaneously bridge the gap with the entrainment behaviours observed in bluff and porous body wakes. Specifically, it aligns with prior studies showing reduced entrainment rates in the far wake of porous bodies \citep{Vinnes2023, Bourhis2024} and bluff bodies \citep{Kankanwadi2020, Chen2023,Chen_Buxton_2024} exposed to FST as $TI_{\infty}$ increases. This also highlights a key limitation of standard wind turbine wake growth rate models, which assume continuous wake expansion and an increasing growth rate with $TI_{\infty}$ throughout the entire wake. In particular, these models do not account for streamwise variations in how FST influences wake evolution, despite such spatial dependence being previously observed in bluff and porous body wakes and, as demonstrated in the present study, also in wind turbine wakes. As a result, they appear valid only within a limited region downstream of the turbine ($\lesssim 7D$), where wake growth remains linear, with the extent of this region depending on $C_T$, $TI_{\infty}$ and ${\cal L}_x$. Although this range can be considered as the ``technologically useful'' range, given that inter-turbine spacing in wind farms typically ranges from $5D$ to $10D$, it could, however, have significant implications for farm-to-farm spacing.

Third, based on radially and temporally averaged velocity and TKE, it was found that $TI_{\infty}$ accelerates the wake recovery, whereas larger ${\cal L}_x$ delay its onset. Under low turbulence intensity inflows, the influence of $C_T$ persists throughout the wake, with the initial differences induced by a change in the turbine's operating conditions remaining noticeable further downstream. For high turbulence intensity inflows, the differences induced by a change in $C_T$ diminishes rapidly.

Fourth, the different dynamics present in the wakes were investigated. Multiple dynamics exist in the near wake and are strongly dependent on the turbine's operating point ($\Lambda$ and $C_T$). Specifically, the presence, absence, and strength of certain dynamics in the near wake are largely determined by the turbine's tip-speed ratio. However, while these dynamics persist in the wake under low $TI_{\infty}$ conditions, they dissipate rapidly under high $TI_{\infty}$ conditions.

Fifth, the statistical and dynamic characteristics of wake meandering were examined. With the exception of the two lowest-$TI_{\infty}$ case (S1 \& S2) at low thrust coefficient, for most FST cases and $C_T$, a broad spectral peak around $\textrm{St}_D \approx 0.2$, characteristic of wake meandering, was found in the tip-shear layer. However, the amplitude of lateral wake motion is minimal under weakly turbulent backgrounds. In contrast, inflows with large FST integral length scales and intensities significantly amplify the wake's lateral motions, with the ILS playing a significant role, as reported for bluff bodies \citep{Kankanwadi2023}. Hence, it appears that wake motion and meandering are only significant when the inflow turbulence ILS and intensity are large, although a spectral signature of wake meandering is also found for low turbulence intensity inflows. Based on these results, we hypothesised that wake meandering in weakly turbulent backgrounds is driven solely by shear layer instability, resulting in minimal wake motion, whereas in strongly turbulent backgrounds, wake meandering is dominated by inflow turbulence, leading to more significant wake displacement. This finding aligns with the LES of \cite{Li2022}, which showed that wake meandering was dominated by the inflow turbulence for high inflow turbulence intensity, whereas for low inflow turbulence intensity, wake meandering was induced by shear layer instability. Finally, while increasing $C_T$ leads to enhanced wake meandering in weakly turbulent backgrounds—consistent with previous observations with porous discs \citep{Bourhis2024}—$C_T$ has a negligible effect in strongly turbulent backgrounds, highlighting the interplay between ambient flow turbulence characteristics and the turbine's operating point in driving this dynamic.

To conclude, to the best of the authors' knowledge, this is one of the first experimental studies to examine wind turbine wakes over such long distances and across a broad range of freestream turbulence conditions and thrust coefficients, covering a large parametric space with 24 $\{C_T,\textrm{FST}\}$ configurations. Notably, most studies on the impact of turbulence length scales on wakes have been carried out using LES simulations or with smaller bluff bodies—such as cylinders and discs—due to physical constraints in generating large turbulence length scales in wind tunnels. The distinct roles of the thrust coefficient, turbulence intensity, and integral length scale in wake turbulence statistics and dynamics—both near and far from the turbine—are emphasised, also highlighting the spatial dependence of these effects on the streamwise location within the wake. Our findings align with some of the LES results reported, bridging the gap between numerical and experimental outcomes regarding the effect of FST on wind turbine wakes. Finally, several parallels were drawn between the wakes of wind turbine surrogates, such as porous discs, revealing that similar physical mechanisms govern the wakes of both bluff bodies and wind turbines. Notably, the observation of reduced wake growth rates in the far wake in the presence of background turbulence intensity, along with similarities in wake dynamics within the tip shear layer, helps bridges the gap between the current body of knowledge of wind turbine wake and bluff body wakes.

Finally, some comments on future studies are worth making. From a fundamental perspective, while experimentally challenging, capturing the interface that demarcates the wind turbine wake from the background flow to measure entrainment fluxes and analyse its dynamical and geometrical features would be of significant interest (\emph{cf.} similarly to \cite{Kankanwadi2020,Kankanwadi2023,Chen2023,Chen_Buxton_2024}). Notably, this would allow for the evaluation of the contributions of the different entrainment mechanics-nibbling and engulfment \citep{daSilva2014}- on the recovery of wind turbine wakes. Additionally, determining whether, as reported for bluff body far wakes, strong detrainment events are responsible for reduced entrainment rates in wind turbine far wakes exposed to FST would provide a deeper understanding of the universality of FST’s influence on entrainment mechanics across different wake-generated objects. From a ``technological'' perspective, further investigations in different wind tunnels, employing various wind turbine models, operating conditions, and inflow conditions, are needed to assess the universality of the present results and expand the current dataset to improve the modelling of wind turbine wakes. For instance, \cite{Dong2023} have shown that blade geometry still influences the far wake, suggesting that the hypothesis that far-wake statistics depend solely on a global turbine parameter, such as the thrust coefficient, is insufficient. Moreover, the individual and separate effects of $C_T$ and $\Lambda$ have yet to be fully examined, to refine wake empirical models that currently use only $C_T$ as the turbine's input. Additionally, a wind farm can, in the first approximation, be modelled as a porous body—or a sum of porous bodies; therefore, whether a reduction in entrainment and growth rates occurs in wind farm far wakes is an interesting topic for further research, with potentially significant implications for engineering applications.

\backsection[Supplementary data]{\label{SupMat}Supplementary material and movies are available at ...}

\backsection[Acknowledgements]{The authors would like to acknowledge J. Jüchter and L. Neuhaus for their valuable help with the experiments.}

\backsection[Funding]{The authors gratefully acknowledge the Engineering and Physical Sciences Research Council (EPSRC) for funding this work through grant no. EP/V006436/1.}

\backsection[Declaration of interests]{The authors report no conflict of interest. For the purposes of open access, the authors have applied a Creative Commons Attribution (CC BY) licence to any Author Accepted Manuscript (AAM) version arising.}

\backsection[Author ORCIDs]{Martin Bourhis, https://orcid.org/0000-0003-3107-032X; Thomas Messmer, https://orcid.org/0009-0007-7758-6387; Michael Hölling, https://orcid.org/0000-0003-4736-8526; Oliver R.H. Buxton, https://orcid.org/0000-0002-8997-2986.}

\appendix

\section{Characteristics of the different turbulent inflows \label{AppendixA}}

\begin{figure}
    \centering
    \begin{subfigure}{\textwidth}
        \centering
        \includegraphics[width=\linewidth]{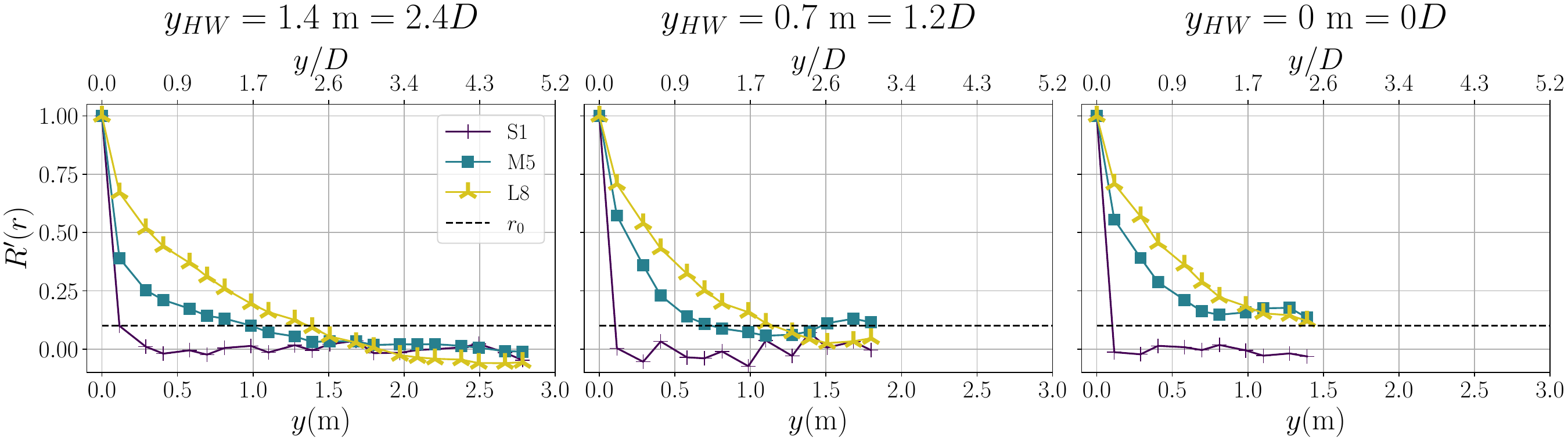}
        \subcaption{}
        \label{fig:Correlation_a}
    \end{subfigure} \\ \vspace{0.2cm}
    \begin{subfigure}{\textwidth}
        \centering
        \includegraphics[width=\linewidth]{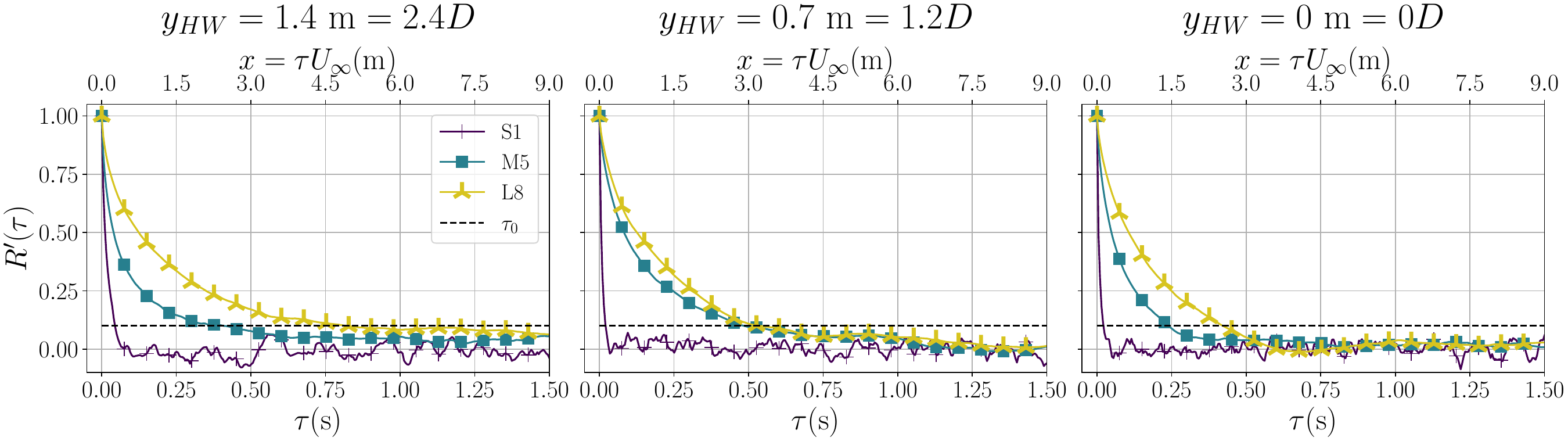}
        \subcaption{}
        \label{fig:Correlation_b}
    \end{subfigure}
        \caption{Spatial $R'(r)$ (\subref{fig:Correlation_a}) and temporal $R'(\tau)$ (\subref{fig:Correlation_b}) normalised autocorrelation functions of the fluctuating velocity $u'$ computed at $\{x/D,z/D\}=\{ 0,0\}$ for three different radial positions $y/D$ along the wind tunnel width and for three FST cases.}
        \label{Fig:Correlation}
\end{figure}

\begin{figure}
    \centering
    \begin{subfigure}{0.49\textwidth}
        \centering
        \includegraphics[width=\linewidth]{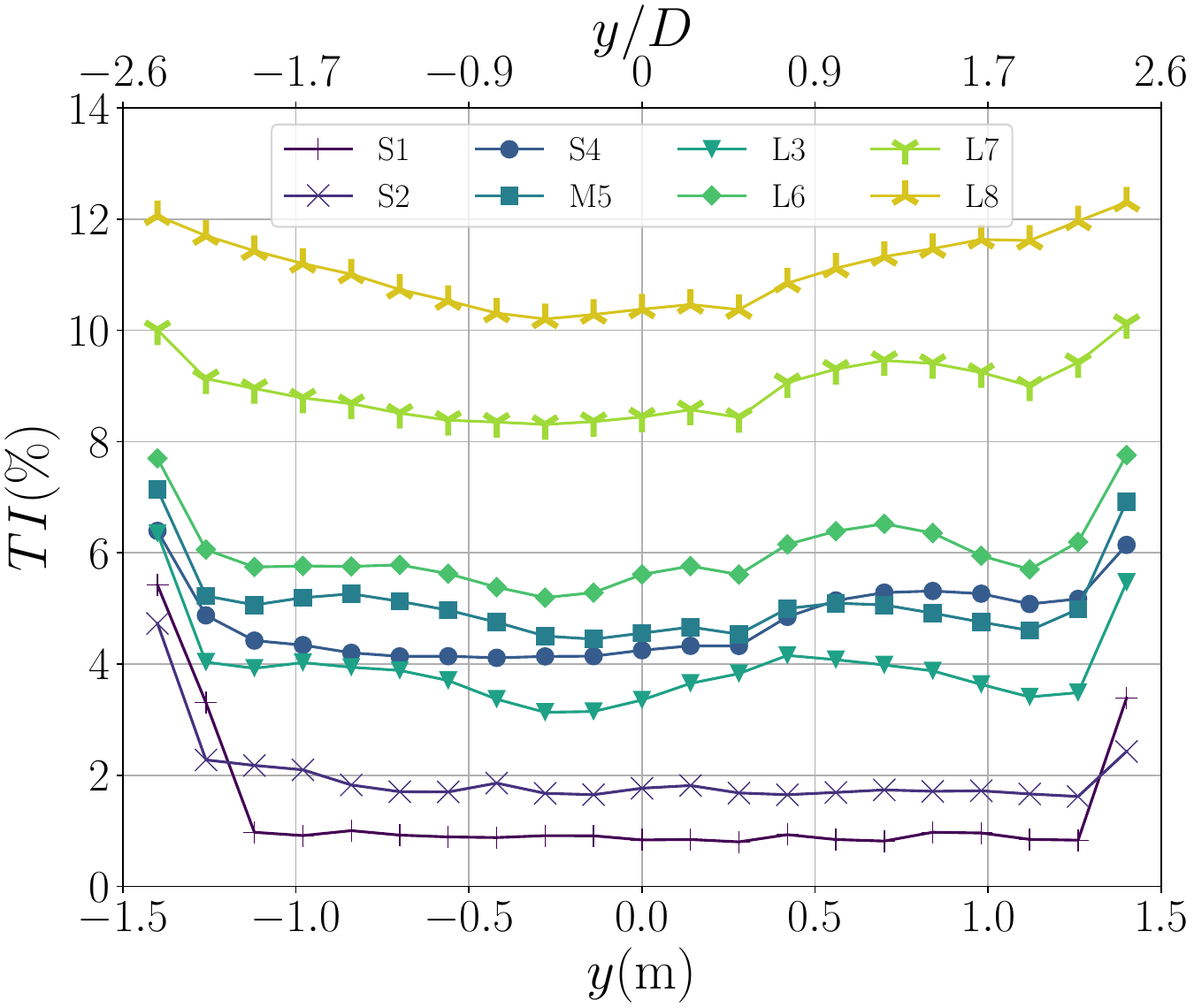}
        \subcaption{}
        \label{fig:FST_Horizontalprofiles_TI}
    \end{subfigure}
    \\
    \begin{subfigure}{0.49\textwidth}
        \centering
        \includegraphics[width=\linewidth]{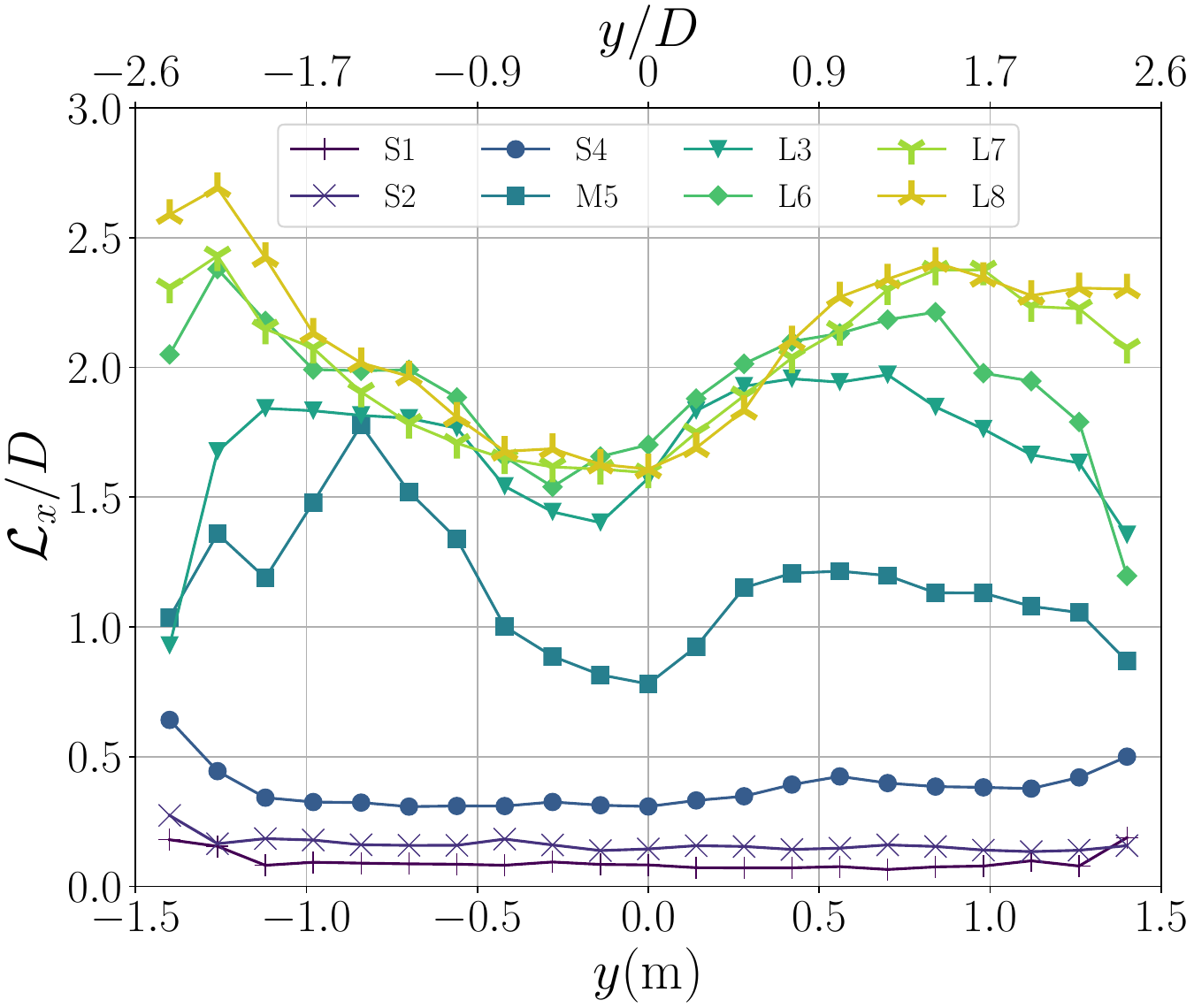}
        \subcaption{}
        \label{fig:FST_Horizontalprofiles_Lx}
    \end{subfigure}
    \hfill
    \begin{subfigure}{0.49\textwidth}
        \centering
        \includegraphics[width=\linewidth]{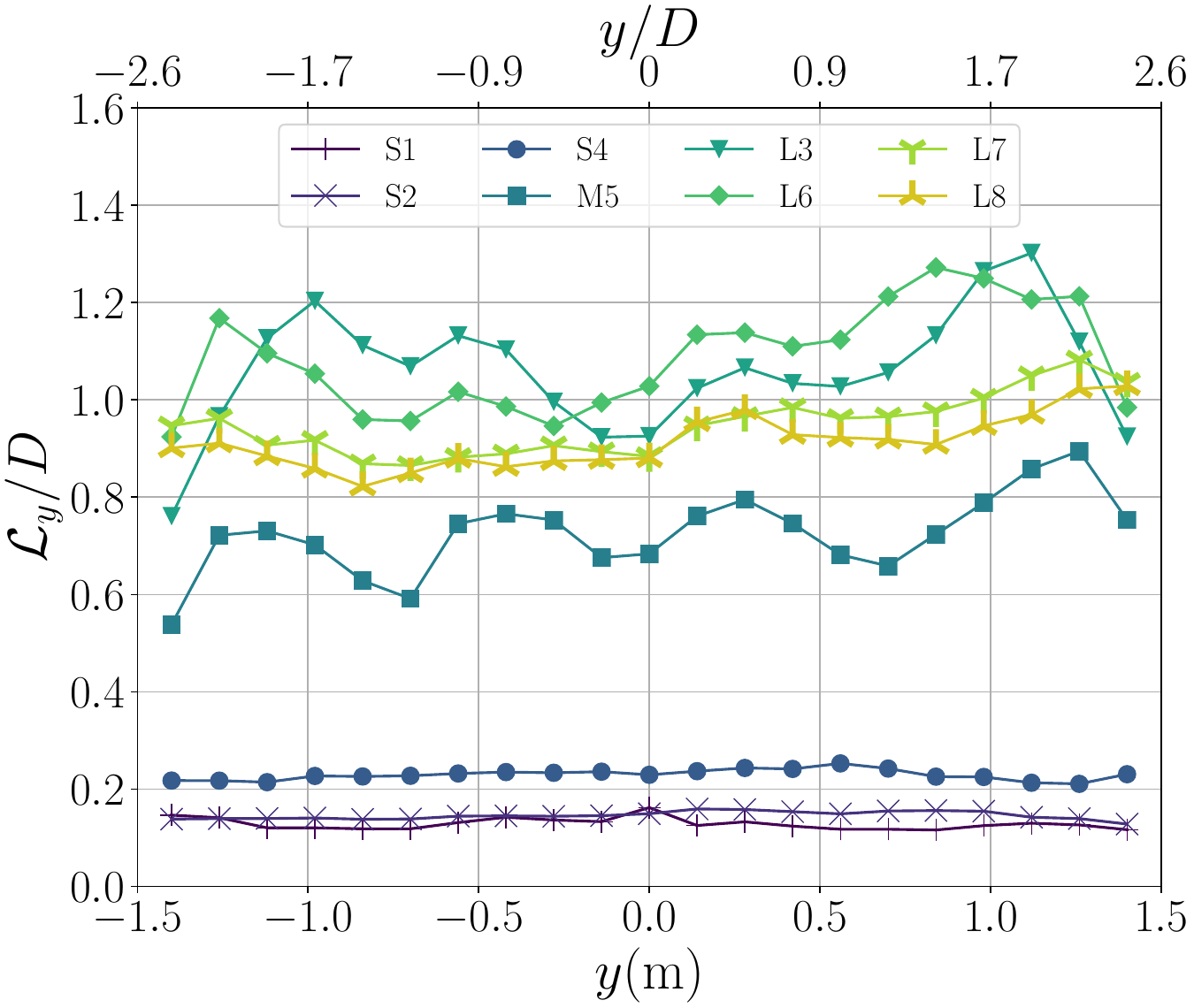}
        \subcaption{}
        \label{fig:FST_Horizontalprofiles_Ly}
    \end{subfigure}
        \caption{Profiles of turbulence intensity $TI (\%)$ (\subref{fig:FST_Horizontalprofiles_TI}), streamwise integral length scale ${\cal L}_x/D$ (\subref{fig:FST_Horizontalprofiles_Lx}), and spanwise integral length scale ${\cal L}_y/D$ (\subref{fig:FST_Horizontalprofiles_Ly}), computed at $ \{ x/D,z/D \} = \{ 0,0\}$. The abscissae are given in metres (bottom) and as multiples of the turbine diameter $D$ (top).}
        \label{fig:FST_Horizontalprofiles}
\end{figure}

\begin{figure}
    \centering
    \begin{subfigure}{0.48\textwidth}
        \centering
        \includegraphics[width=\linewidth]{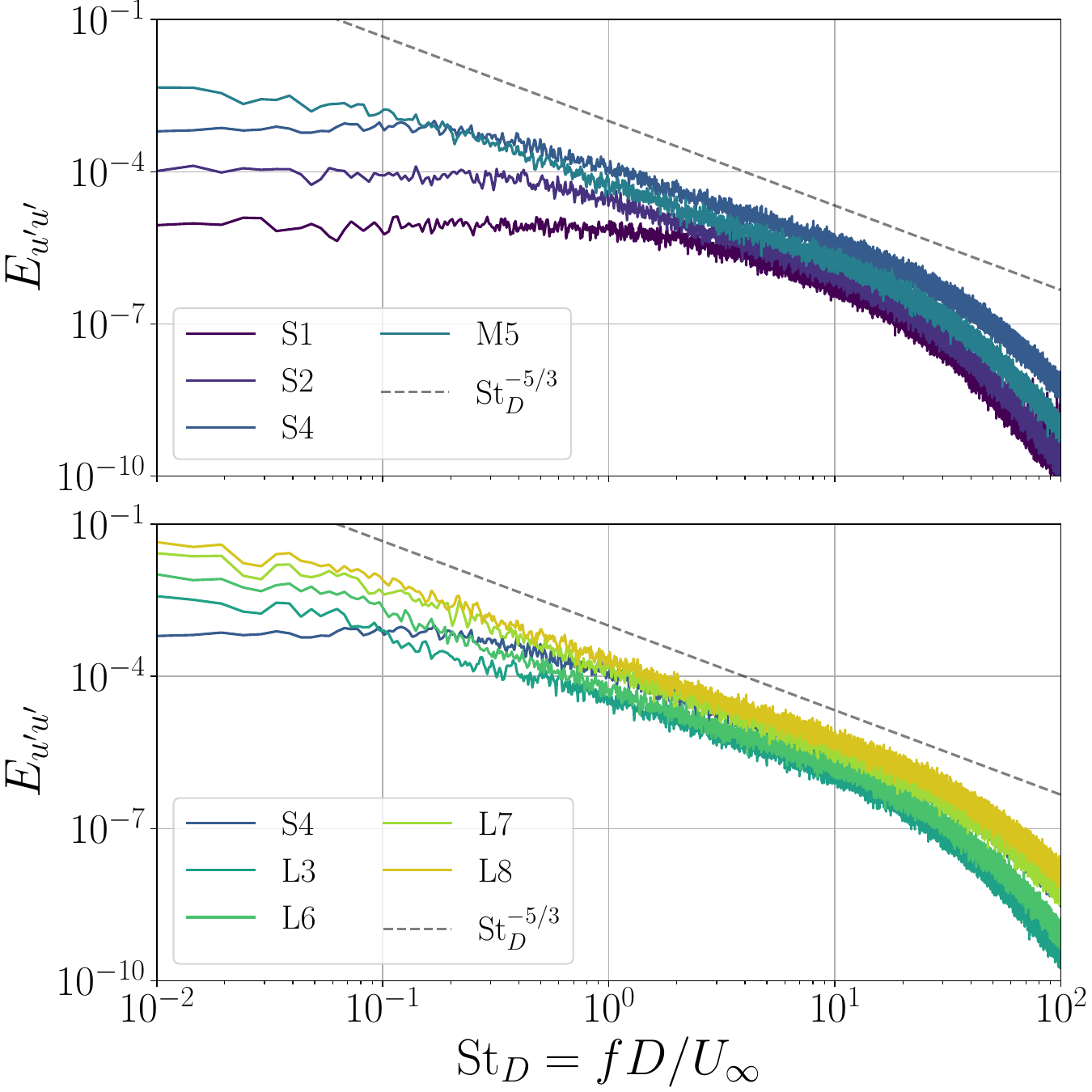}
        \subcaption{}
        \label{Fig:FST_Spectra_0}
    \end{subfigure} \\
    \begin{subfigure}{0.48\textwidth}
        \centering
        \includegraphics[width=\linewidth]{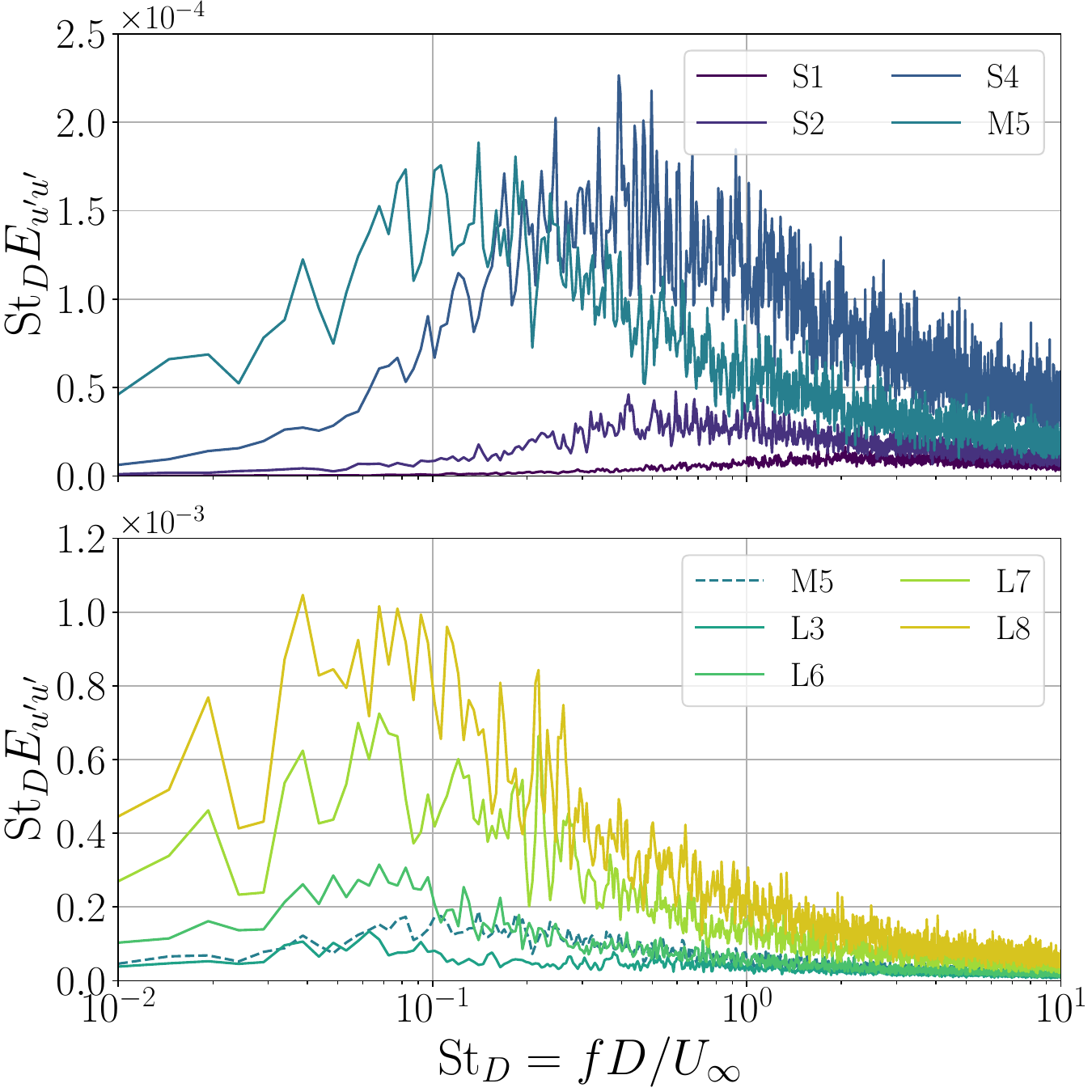}
        \subcaption{}
        \label{Fig:FST_Spectra_1}
    \end{subfigure} \hfill
     \begin{subfigure}{0.48\textwidth}
        \centering
        \includegraphics[width=\linewidth]{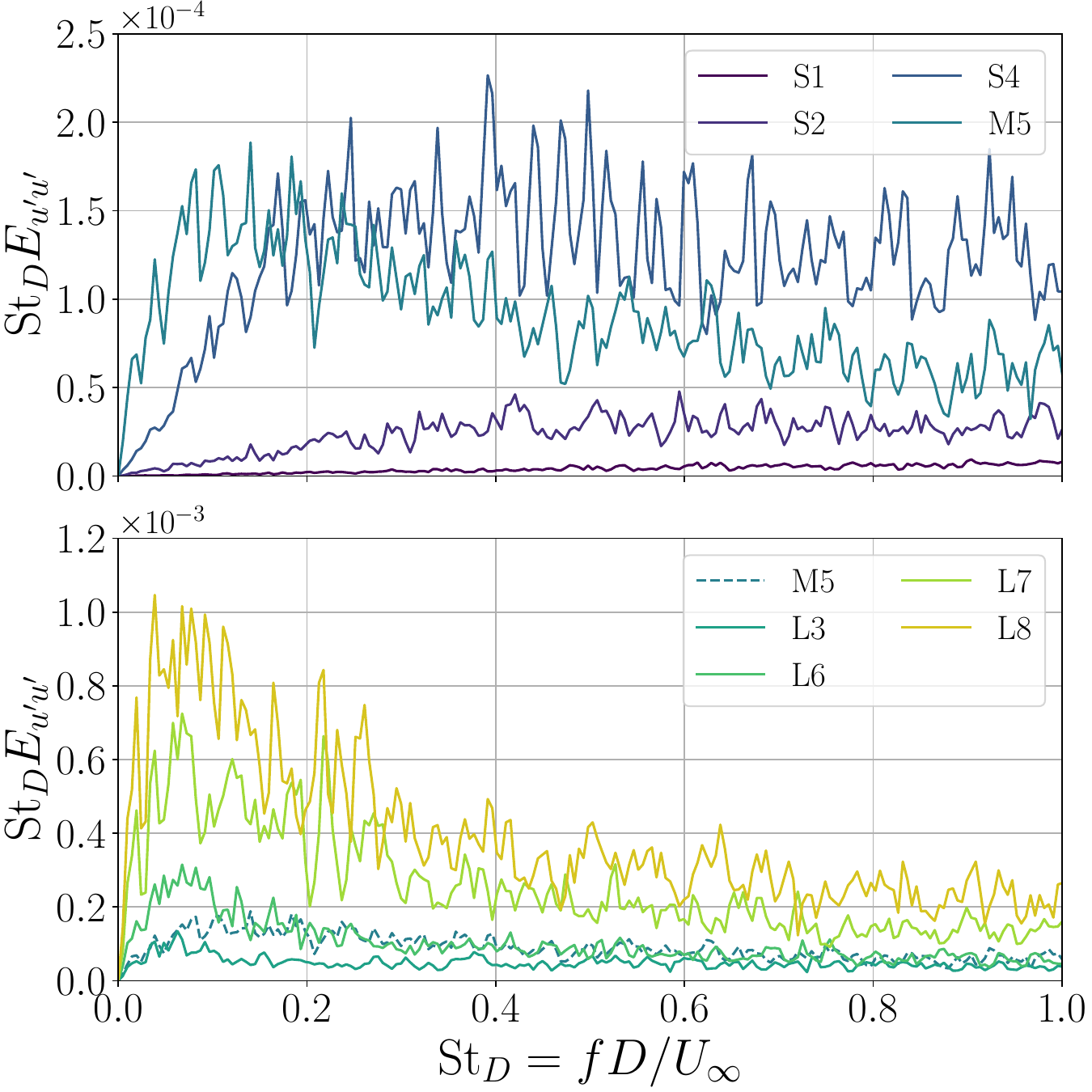}
        \subcaption{}
        \label{Fig:FST_Spectra_2}
    \end{subfigure}
        \caption{Power spectra, $E_{u'u'}$, (\subref{Fig:FST_Spectra_0}) and premultiplied power spectra, $\textrm{St}_DE_{u'u'}$, (\subref{Fig:FST_Spectra_1} \& \subref{Fig:FST_Spectra_2}) of the fluctuating velocity $u'$, as a function of the diameter-based Strouhal number $\textrm{St}_D=fD/U_{\infty}$. Spectra are computed at the hub centre location $\{x/D,y/D,z/D\} = \{ 0,0,0\}$. A particular focus is placed on low-frequency dynamics in the premultiplied spectra, with the only difference between the two figures being the abscissa scale, logarithmic in (\subref{Fig:FST_Spectra_1}) and linear in (\subref{Fig:FST_Spectra_2}).}
        \label{Fig:FST_Spectra}
\end{figure}

This appendix provides further details on the characteristics of the different turbulent inflows generated by the active grid. The 8 inflows were characterised across the width of the wind tunnel at the turbine's location, \emph{i.e.}, along the horizontal axis $y$ at $\{x/D,z/D\}=\{ 0,0\}$.
Figure~\ref{Fig:Correlation} presents the normalised spatial and temporal autocorrelation functions of the fluctuating velocity $u'$, for different horizontal positions and FST cases. Figure~\ref{fig:FST_Horizontalprofiles} shows the profiles of turbulence intensity and integral length scales. Finally, the FST spectra and premultiplied spectra of the fluctuating velocity are provided in figure~\ref{Fig:FST_Spectra}.

\bibliographystyle{jfm}
\bibliography{jfm}

\end{document}